\documentclass{aa}  
\usepackage[hidelinks]{hyperref}
\usepackage{placeins}
\usepackage[dvipsnames]{xcolor}
\usepackage{tabularx}
\hypersetup{colorlinks, linkcolor={blue!70!black}, citecolor={blue!70!black}, urlcolor={blue!70!black}}
\usepackage{graphicx}

\usepackage{txfonts}
\usepackage{array}
\begin{document}
\setlength{\extrarowheight}{5pt}

\title{Chronology of our Galaxy from Gaia Colour-Magnitude Diagram-fitting (ChronoGal). I. The formation and evolution of the thin disk from the Gaia Catalogue of Nearby Stars}
\author{Carme Gallart\inst{1, 2} \and Francisco Surot\inst{1} \and Santi Cassisi$^{\dagger,}$ \inst{3,4} \and Emma Fernández-Alvar$^{\dagger,}$ \inst{1, 2} \and David Mirabal$^{\dagger,}$ \inst{ 2} \and Alicia Rivero $^{\dagger,}$\inst{2} \and Tomás Ruiz-Lara$^{\dagger,}$ \inst{5, 6} \and Judith Santos-Torres$^{\dagger,}$ \inst{2, 7} \and Guillem Aznar-Menargues \inst{2} \and Giuseppina Battaglia\inst{1, 2} \and Anna B. Queiroz\inst{1, 2} \and Matteo Monelli\inst{1,2} \and EugeneVasiliev\inst{8} \and Cristina Chiappini\inst{9} \and Amina Helmi\inst{10} \and Vanessa Hill\inst{11} \and Davide Massari\inst{12} \and Guillaume F. Thomas\inst{1,2} }

\titlerunning{Chronogal: The formation and evolution of the Milky Way thin disk from the GCNS}
\institute{Instituto de Astrof\'isica de Canarias, E-38200 La Laguna, Tenerife, Spain
   \and 
   Departamento de Astrof\'isica, Universidad de La Laguna, E-38205 La Laguna, Tenerife, Spain 
   \and
 INAF -- Astronomical Observatory of Abruzzo, via M. Maggini, sn, 64100 Teramo, Italy
   \and
   INFN, Sezione di Pisa, Largo Pontecorvo 3, 56127 Pisa, Italy 
\and
   Universidad de Granada, Departamento de F\'isica Te\'orica y del Cosmos, Campus Fuente Nueva, Edificio Mecenas, E-18071, Granada, Spain.
   \and 
   Instituto Carlos I de F\'isica Te\'orica y computacional, Universidad de Granada, E-18071 Granada, Spain
   \and
   Isaac Newton Group of Telescopes
   \and 
   Institute of Astronomy, Madingley road, Cambridge, CB3 0HA, UK
   \and 
   Leibniz-Institut f\"ur Astrophysik Potsdam (AIP), An der Sternwarte 16, 14482 Potsdam, Germany
   \and 
   Kapteyn Astronomical Institute, University of Groningen, P.O. Box 800, 9700 AV, Groningen, The Netherlands
   \and
   Universit\'e Côte d'Azur, Observatoire de la Côte d'Azur, CNRS, Laboratoire Lagrange, Bd de l'Observatoire, CS 34229, 06304
Nice cedex 4, France
       \and 
   INAF - Osservatorio di Astrofisica e Scienza dello Spazio di Bologna, Via Gobetti 93/3, I-40129 Bologna, Italy
}
   \date{}

\abstract
{The study of the Milky Way is living a golden era thanks to the enormous high-quality datasets delivered by Gaia, and space asteroseismic and ground-based spectroscopic surveys. However, the current major challenge to reconstruct the chronology of the Milky Way is the difficulty to derive precise stellar ages for large samples of stars. The CMD-fitting technique offers an alternative to individual age determinations to derive the star formation history (SFH) of complex stellar populations.} 
{We aim to obtain a detailed "dynamically evolved" SFH (deSFH) of the solar neighbourhood, and the age and metallicity distributions that result from it. We define deSFH as the amount of mass transformed into stars, as a function of time and metallicity, in order to account for the population of stars contained in a particular volume of the MW.}
{We present a new package to derive SFHs from CMD-fitting tailored to work with Gaia data, called CMDft.Gaia and we use it to analyse the CMD of the Gaia Catalogue of Nearby Stars (GCNS), which contains a complete census of the (mostly thin disk) stars currently within 100 pc of the Sun.}
{We present an unprecedented detailed view of the evolution of the Milky Way disk at the solar radius. The bulk of star formation started between 11-10.5 Gyr ago at metallicity around solar and continued with a slightly decreasing metallicity trend until 6 Gyr ago.  Between 6 and 4 Gyr ago, a notable break in the age-metallicity distribution is observed, with three stellar populations with distinct metallicities (sub-solar, solar, and super-solar), possibly indicating some dramatic event in the life of our Galaxy. Star formation then resumed 4 Gyr ago with a somewhat bursty behaviour, metallicity near solar and average star formation rate higher than in the period before 6 Gyr ago. The derived metallicity distribution closely matches precise spectroscopic data, which also show stellar populations deviating from solar metallicity. Interestingly, our results reveal the presence of intermediate-age populations exhibiting both a metallicity typical of the thick disk, approximately ${\rm[M/H]}\simeq-0.5$, and supersolar metallicity.}
{The many tests performed indicate that, with high precision photometric and distance data such as that provided by Gaia, CMDft.Gaia is able to achieve a precision of $\lesssim$ 10 \% and an accuracy better than 6\% in the dating of stellar populations, even at old ages. The comparison with independent spectroscopic metallicity information shows that metallicity distributions are also determined with high precision, without imposing any a-priory metallicity information in the fitting process. This opens the door to obtaining detailed and robust information on the evolution of the stellar populations of the Milky Way over cosmic time. As an example we provide in this paper an unprecedented detailed view of the age and metallicity distributions of the stars within 100 pc of the Sun.}


   \maketitle

   \def\thefootnote{$\dagger$}\footnotetext{These authors contributed in an equally crucial way to this work}\def\thefootnote{\arabic{footnote}}

\section{Introduction}\label{intro}

The Milky Way (MW) is the galaxy that we can study with the greatest detail, over its whole history, by characterising its stellar content on a star-by-star basis. The low mass stars that formed since the first star formation events in the Universe inform us of the rate of star formation as a function of time, and how metals build up in successive stellar generations.

The study of the MW is living a golden era. On the one hand, the impressive data-sets delivered by the ESA mission {\it Gaia} \citep{Prusti2016TheGaiaMission, GaiaDR2_2018_Brown2018, GaiaDR3_2023Vallenari} are revolutionising our current view on our Galaxy \citep{Brown2021ARAA}. On the other hand, several ground based spectroscopic surveys, ongoing or planned, (e.g. LAMOST, \citealt[][]{2020ApJS..251...27W, 2019RAA....19...75L}; RAVE, \citealt[][]{2020AJ....160...82S, 2020AJ....160...83S};  GALAH, \citealt[][]{2021MNRAS.tmp.1259B}; APOGEE, \citealt[][]{2020ApJS..249....3A}; WEAVE, \citealt{WEAVE_2023_presentation_general}; 4MOST, \citealt{4MOST2019MWDHR_messenger, 4MOST2019MWDLR_messenger}; or DESI \citealt{DESI_MW2023}), are complementing the {\it Gaia} mission by obtaining spectroscopy with higher resolution and/or down to a fainter limiting magnitude. Finally, asteroseismic missions such as Kepler \citep{Borucki2010Sci_Kepler} and K2 \citep{Howell2014_K2} have shown the potential of adding seismic information to derive ages.

As stated in the {\it ''Gaia Red Book''}\footnote{{\it Gaia} report on the Concept and Technology Study (sometimes referred to as the "Gaia Red Book"), see https://www.cosmos.esa.int/documents/29201/297049/report-science.pdf/}  {\it ''A primary scientific goal of the {\it Gaia} mission is the determination of the star formation histories, as described by the temporal evolution of the star formation rate, and the cumulative numbers of stars formed, of the bulge, inner disk, Solar neighbourhood, outer disk and halo of the Milky Way''}. The most difficult part to accomplish this goal, once the necessary data is available, is the determination of stellar ages, since they cannot be directly measured: they need to be inferred by comparing the observed properties with the predictions of stellar evolution models.

The most suitable method for determining the age of a given star depends on its mass and/or evolutionary stage, and thus deriving homogeneous ages for a broad range of stellar types or for the full age range is virtually impossible \citep[][for detailed reviews]{soderblom2010, SalarisCassisi2005}. In the most widely used method in Galactic archaeology, a set of physical parameters derived from spectra (and/or photometry), such as effective temperature, surface gravity and metallicity (or colours and luminosities), are compared to a set of stellar evolution models, which predict age as a function of these parameters \citep[][]{sahlholdt_2019_gaia_benchmark}.  Isochrone fitting is in practice prone to large uncertainties, both due to the difficulty of accurately deriving the needed stellar parameters, and due to the biases introduced by the isochrone interpolation. Even in the favourable case of stars with well measured distances and accurate spectroscopic parameters, typical age errors of {\textit{individual}} stars are around 25\% \citep{SandersDas2018, Mints2018Isofit, Queiroz2018Starhorse, Kordopatis2023GaiaAges}, and only estimated lower in particularly exquisite instances \citep[e.g.][]{haywood2013}. In spite of this, the wealth of data from large spectroscopic surveys, and the increasing availability of distances (initially from {\it Hipparcos} and currently from {\it Gaia}), has led to works exploiting advanced statistical methods, in particular based on Bayesian statistics \citep[][]{PontEyer2004, JorgensenLindegren2005}, to derive ages for large stellar samples \citep[e.g.][]{Holmberg2009, Feuillet2016, SandersDas2018,Mints2019spectroSurveys, Frankel2019InsideOut, Sahlholdt2022MNRAS.510.4669S, Xiang_Rix_2022Natur.603..599X, Queiroz2023_starhorse}.

Asteroseismology has turned into the big hope to derive precise stellar ages \citep{Miglio2017_Plato}. When combined with spectroscopy, it provides solid constraints on stellar mass, radius and evolutionary state \citep[see][]{Mathur2012, Chaplin2020NatAs...4..382C}, particularly for bright red giants, which allows to obtain ages for distant stellar samples.  A number of stellar catalogues have already exploited this combination \citep[][]{martig2016RGBages, Ness_2016AgesRGB, Anders2017Corogee, rendle2019_GalacticDisk}. However, age errors are still large \citep[see Fig 22 in][]{Pinsonneault2018}. A best case scenario expected to provide an age precision of 10\% is discussed by \citet{Miglio2017_Plato} for long duration observations like those planned with the Plato satellite. 

These {\it individual stellar age} determinations require detailed and costly observations, only possible for a tiny fraction of MW stars. This results in very complicated selection functions. These samples allow inferring information such as age-metallicity or age-velocity trends, but it is basically impossible to retrieve from them the much sought holy grail of galaxy evolution, that is the star formation history (SFH), or to produce unbiased age and metallicity distributions directly comparable with predictions from galaxy models. 

The Colour-Magnitude Diagram fitting (CMD-fitting) technique offers a highly complementary way to approach the problem of deriving the SFH of a composite stellar population.  In the case of the MW, because only {\it Gaia} CMDs reaching the old main sequence turnoffs (oMSTO) are required, SFH derivation is possible for unbiased and huge stellar samples. With current and forthcoming {\it Gaia} data releases it will be possible, using this methodology, to obtain SFHs and precise age distributions out to distances of several kpc from the Sun, thus exploring all Galactic components and addressing major questions of Galactic astronomy. In fact, the {\it Gaia Red Book} proposes the CMD-fitting methodology as the best method to derive the SFH of the MW\footnote{(Gaia Red Book, p. 34)}, after the early successes of this methodology in providing SFHs of Local Group galaxies from deep CMDs obtained from ground based or {\it Hubble Space Telescope} (HST) imaging. 

Indeed, in extra-galactic archaeology, deep CMDs reaching the oMSTO with good photometric accuracy and precision, analysed with the robust technique of CMD-fitting \citep{Tosi1991, Bertelli1992_SFH_LMC, Tolstoy1996_method,  Aparicio1997LGS3,  Dolphin1997_1st_method, gallart1999, dolphin2002, apariciogallart2004, gallart2005, apariciohidalgo2009, cignonitosi2010}, are regarded as the most direct and reliable observable to obtain a detailed SFH and stellar age distributions of a galaxy, from its earliest epochs to the present time.  CMD-fitting has been, over 25 years, the standard to determine detailed SFHs for Local Group galaxies, from nearby MW satellites using ground based data, to more distant members using large allocations of HST time. This has provided insight on a number of important topics of near-field cosmology, such as the role of reionization on the early SFH of dwarf galaxies \citep{cole2007leoa, monelli2010cetus, monelli2010tucana, Weisz2014reio, ruiz-lara2018leoa}, the origin of their different morphological types \citep{Gallart2015}, the differences between the MW and M31 satellite systems \citep{monelli2016, skillman2017}, the spatial gradients \citep{Noel2009_SMC, Cignoni2013_SMC, meschin2014, Rubele2018_SMC} and synchronised SFHs of the Magellanic Clouds \citep{Ruiz-Lara2020LMC, Massana2022SMC-LMCdance} and the SFHs of the M31 halo, spheroid, and outer disk \citep{brown2008m31, bernard2012m31}. CMD-fitting has flourished in the context of the study of Local Group galaxies, because they are sufficiently close to resolve their individual stars, yet far enough that all their stars can be considered to be at the same distance, which can be obtained accurately using various well calibrated distance indicators \citep{benedict2007, beaton2016}. This is fundamental to transform the measured apparent luminosities to absolute magnitudes that can be compared with the predictions of stellar evolution models.

In the case of the MW, precise and accurate distances for each individual star are necessary, and thus the first examples of CMD-fitting to derive the SFH of stellar samples in the very close solar vicinity (within approximately 50 pc) used Hipparcos data \citep{hernandez2000, bertellinasi2001, cignoni2006_Hipparcos}. However, the real breakthrough came with the availability of {\it Gaia} data, which allowed to extend the study of the solar neighbourhood to 100-250 pc, first with {\it Gaia} DR1 \citep{bernard2018proc} and then with {\it Gaia} DR2 \citep{Alzate2021_Gaia, DalTio2021_200pc}.  Additionally, samples within 2 kpc allowed to date the accretion time of Gaia-Enceladus and the early SFH of MW thick disk and halo \citep{Gallart2019Gaia}, and the possible repeated influence of the Sagittarius dSph on the SFH of the MW disk, since its accretion some 6 Gyr ago \citep{Ruiz-Lara2020Sgr}, as well as SFH gradients as a function of distance from the MW plane \citep{Gallart2019gaiaproc, Mazzi2023_cylinder}. 

In this paper, we describe in detail our current implementation of the CMD-fitting technique to derive detailed SFHs from {\it Gaia} CMDs, which has been upgraded in several aspects (e.g. the adopted stellar evolution library, and the CMD-fitting code itself) with respect to the procedures used in \citet{Gallart2019Gaia} and \citet{Ruiz-Lara2020Sgr}. One salient characteristic of our methodology is that no {\it a priori} assumptions are made on the age-metallicity relation, on the metallicity distribution, or on the functional form of the star formation rate as a function of time, SFR(t). We apply this methodology to derive a first detailed SFH of the Solar neighbourhood, using the exquisite {\it Gaia Catalogue of Nearby Stars (GCNS)} dataset \citep{gcns}, therefore presenting an unprecedented detailed view of the evolution of the Milky Way disk at the solar radius.

This paper is organised as follows: Section~\ref{data} summarises the contents of the original GCNS dataset relevant for this study, which are complemented with Gaia DR3 data on chemical abundances and radial velocities; Section~\ref{sec3:method} presents CMDft.Gaia, a new suite of procedures for CMD-fitting specially tailored for the analysis of Gaia CMDs; Section~\ref{sec4:deriving} describes the particular application of CMDft.Gaia to the CMD of the GCNS while Section~\ref{results} presents the derived deSFH and age-metallicity distributions and the robustness of these results, and compared them with literature spectroscopic metallicity distributions and age-metallicity relations; Section~\ref{discussion} discusses the evolutionary history of the Milky Way (thin) disk in the light of the derived SFH; finally Section~\ref{summary} summarised the main results and conclusions, both regarding the evolution of the Milky Way disk and the performance of CMDft.Gaia. A number of Appendix present complementary information on various aspects of this work.

\section{The data} \label{data}

We base our analysis on the \textit{Gaia} Catalogue of Nearby Stars \citep[GCNS,][]{gcns}, which comprises 331312 stars residing within a sphere of 100 pc radius centred on the Sun, selected from the full \textit{Gaia} EDR3 catalogue. It is a volume-complete sample of all objects with spectral type earlier than M8 down to the nominal $G=20.7$ magnitude limit of \textit{Gaia}. 

For details on this catalogue we refer the reader to the original paper \citep{gcns}. Here suffices to say that it originates from a selection of all sources in {\it Gaia} EDR3 with measured parallaxes $\omega > 8 \, \rm{mas}$ (corresponding to a maximum distance of 125 pc), from which objects with spurious astrometric solutions were removed with a random forest classifier \citep{Breiman2001}. Subsequently, posterior probability densities for the true distance of each source were inferred with a simple prior not dependent on the sky position or type of star, based on the distance distribution of stars in GeDR3mock \citep{Rybizki2020PASP} with observed parallax greater that 8 mas\footnote{For this sample of very nearby, well measured stars, the dist$_{50}$ distances obtained in this way are practically identical to those that would be obtained by simply inverting the parallax. However, for consistency we adopted dist$_{50}$ from the GCNS as the distance, and ($\rm{dist_{84}-dist_{16})/2}$ as its error.}.  Finally, all the stars with a non-zero probability of being within 100 pc were retained in the catalogue.

In terms of photometric information, apart from Gaia eDR3 data, the original GCNS also included magnitudes from external optical and infrared catalogues. We checked the resulting CMDs in a number of combinations of the available filters and concluded that, for the purposes of SFH derivation, the CMD in the {\it Gaia} bands [$G_{BP}-G_{RP}, G$] was definitely superior. This is therefore the combination we will use in the rest of the article. 

\citet{gcns} do not discuss the amount of extinction affecting the stars in the sample. We have used the \citet{Green2019} 3D dust map to derive extinctions at the 3D position of each star and have verified that they are negligible.

In this section we provide a bird's-eye view of the GCNS stellar content in terms of the CMD, kinematic and global chemical information, and how it is globally split between the Milky Way thin and thick disc and stellar halo components. Given that Gaia DR3 data have become available since the publication of the GCNS, here we complement the original data-set with Gaia DR3 line-of-sight velocities for 174221 stars, as well as metallicity ([M/H]) and [$\alpha$/Fe] abundance measurements for 23629 stars from the $Gaia$ Radial Velocity Spectrometer (RVS) as measured by the General Stellar Parametriser-spectroscopy module, \textit{GSP-Spec} \citep[][]{RecioBlanco2023_GaiaDR3}. We refer the reader to Appendix~\ref{app:GDR3} for details on how these quantities were assembled and on how we assign probabilities of membership for a star to belong to the thin disc, thick disc or stellar halo.

\begin{figure*}
\centering
\includegraphics[scale=0.42]{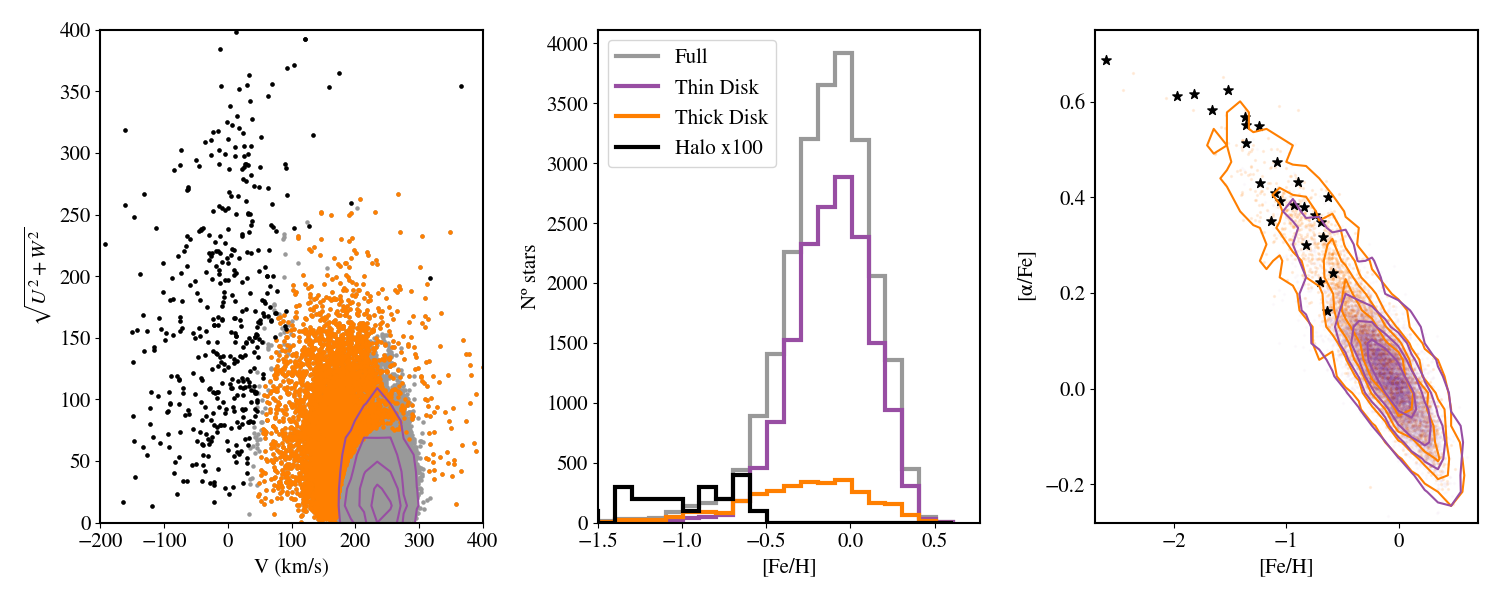}
\includegraphics[scale=0.5]{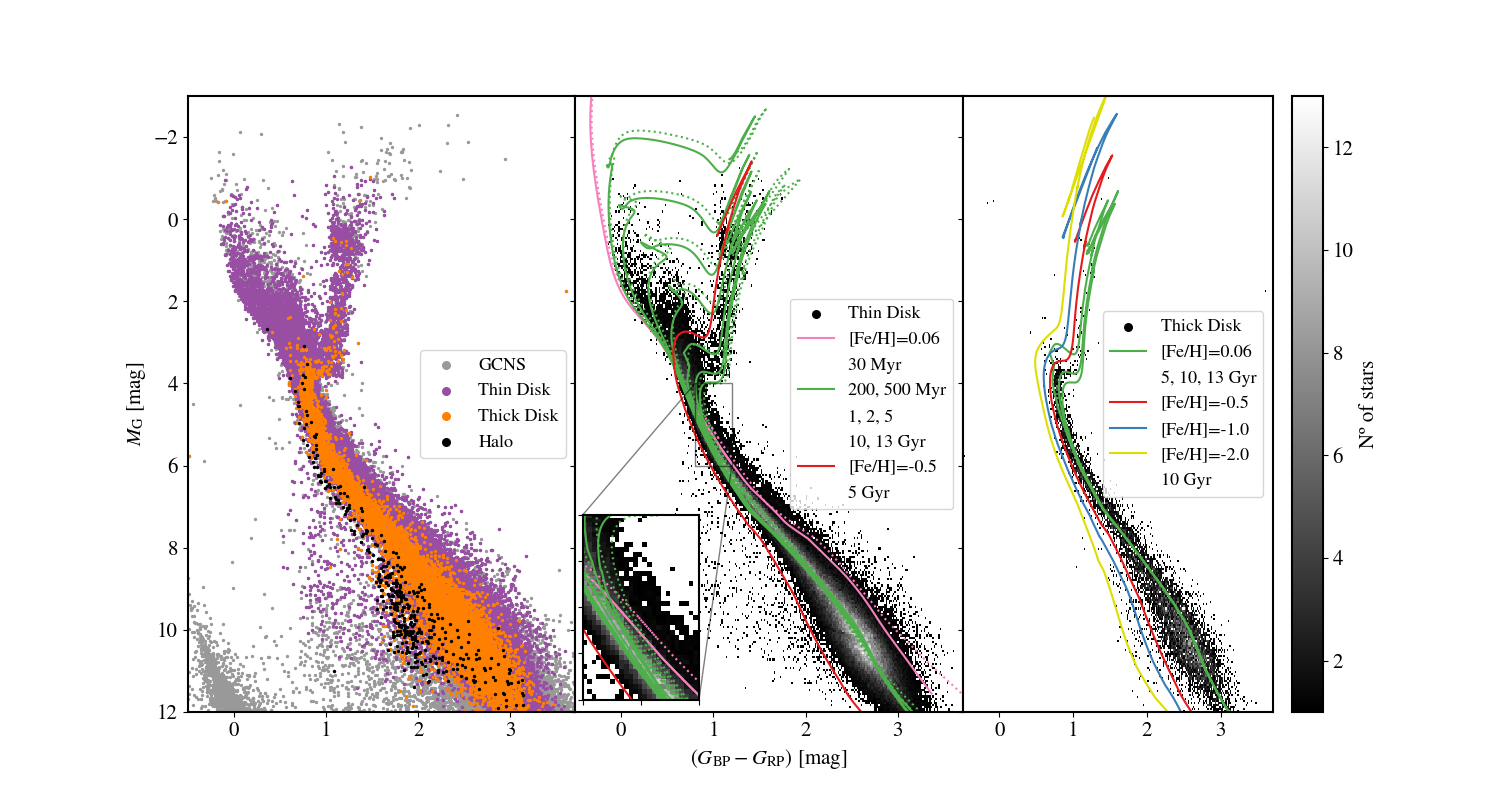}
\caption{Summary of the content of the updated GCNS. Upper left panel: Toomre diagram of the stars with line-of-sight velocities. The thin disk stellar distribution is represented with violet contours and the thick disk and halo stars as orange and black dots, respectively. The grey points in the background represent the whole GCNS sample with kinematic data from DR3.
Middle upper panel: [Fe/H] distribution for the global (grey), thin disk (purple), thick disk (orange) and halo (black; multiplied by 100). Upper right panel: [$\alpha$/Fe] distribution of halo stars (black stars), thick disk stars (orange dots and contours) and thin disk stars (purple dots and contours). In these two panels, only the stars with abundance information have been represented. Lower left panel:  CMD of the three kinematically selected components within 100 pc of the Sun, superimposed to the whole original GCNS sample (in grey): thin disk (purple), thick disk (orange) and halo (black). Middle panel: CMD of the kinematic thin disk with superimposed isochrones of solar metallicity with a range of ages (0.03, 0.2, 0.5, 1, 2, 5, 10 and 13 Gyr) from the BaSTI-IAC (solid lines) and PARSEC \citep[][dashed lines]{Bressan2012} stellar evolution libraries. A [M/H]$=-0.5$, 5 Gyr old metallicity BaSTI isochrone is also plot. Lower right panel: CMD of the kinematic thick disk with superimposed isochrones from the BaSTI-IAC library: [Fe/H]=0.06 and 5, 10 and 13 Gyr; [Fe/H]=$-0.5, -1.0, -2.0$ and 10 Gyr.} 
\label{description_catalogue}
\end{figure*}

Figure~\ref{description_catalogue} summarises the content of this updated GCNS. In the upper panels and in the bottom left, the thin disc stars are indicated by purple symbols or contours, the thick disc in orange and the halo in black. The whole kinematic, chemical and photometric samples are shown in grey in the upper left, upper middle and lower left panels, respectively. The upper left panel represents the Toomre diagram; note that the different components overlap slightly in this space. The stars classified as halo mostly have retrograde orbits and follow the distribution that has been associated to the remnants of Gaia-Enceladus-Sausage \citep{belokurov2018_sausage,helmi2018}.
The middle panel presents the metallicity distribution [Fe/H] of each component and that of the whole sample (that of the halo has been multiplied by 100 to make it visible). Note that the thick disk distribution is only slightly shifted to lower metallicity values compared to the thin disk distribution (mean [M/H]$_{thin}=-0.11$ dex, $\sigma_{thin}=0.2$ dex; mean [M/H]$_{thick}=-0.29$ dex, $\sigma_{thick}=0.4$ dex). Some contamination from the thin disk may be responsible of this slightly higher metallicity compared to the 'canonical' metallicity of the thick disk. Finally, the upper right panel displays the [$\alpha$/Fe] versus [Fe/H] distribution. Both disk components, kinematically selected, have a similar range of [$\alpha$/Fe] with thick disk stars somewhat more extended to high [$\alpha$/Fe] values and lower [Fe/H]. The reason of the relatively similar distribution of thin and thick disk stars in this plane is twofold: first, $\alpha$ abundances from {\it Gaia} GSP-Spec come primarily from Calcium measurements, and this element is produced both by SNIa and SNII, thus not resulting in such a clear-cut separation for different populations \citep[this is confirmed in the CNN analysis by][where the break in $\alpha$/Fe becomes more evident after combining Gaia data with APOGEE]{Guiglion2024CNN}; second, a separation of thin and thick disk stars with kinematic (or geometric) criteria, does not necessarily reproduce a chemical separation \citep{kawata_chiappini2016}. The kinematically selected halo stars do have a distinct distribution, with all of them having [$\alpha$/Fe] $\gtrsim$ 0.2 and [Fe/H]<-0.5.

The lower left panel shows the CMD of the global GCNS catalogue and the three kinematically selected components. One can appreciate that the dominant thin disk component (146108 stars) reaches very bright absolute magnitudes and blue colours in the main sequence, indicative of a very young population; the thick disk (13153 stars) is clearly much older, and its low main sequence is located in the blue part of that of the thin disk, indicating a lower metallicity. Finally, the few halo stars (415 stars, in black) are all located in the blue ridge of the other two populations, reflecting their even lower metallicity. 

The lower middle and right panels of Fig.\ref{description_catalogue} display the CMD of the kinematic thin and thick disk, respectively, with isochrones superimposed. In the case of the thin disk, solar metallicity isochrones ([Fe/H]=0.06)\footnote{Note that the chosen solar metallicity [Fe/H]=+0.06 is the initial metallicity of the Sun so that at its present age of $\simeq4.5$ Gyr and as a consequence of atomic diffusion, its measured photospheric metallicity is [Fe/H]=0.}  with a range of ages (0.03, 0.2, 0.5, 1, 2, 5, 10 and 13 Gyr) from the BaSTI-IAC\footnote{The BaSTI-IAC isochrones in this figure differ from the official release in the colours of the faint main sequence portion of the isochrone, below $M_G \simeq 8$. The reason for this corrections and how it is computed is discussed in Appendix \ref{app:lowmscor}.} \citep{Hidalgo2018} and PARSEC \citep{Bressan2012} stellar evolution libraries have been selected. The stars in this CMD are well matched by solar metallicity isochrones within the selected age range, from the youngest to the oldest. The youngest 30 Myr isochrone, while not obviously needed to match very bright young stars in the main sequence, seems to match very well the less populated red part of the broad low main sequence (fainter than M$_G$=8). The blue side of this low main sequence is well matched by a lower metallicity population (see 5 Gyr isochrone with [Fe/H]= -0.5, in red). Finally note that there are basically no stars in the subgiant branch between the 10 and the 13 Gyr old isochrones, hinting to an scarcity of a very old population in this sample.

The comparison between the BaSTI-IAC and PARSEC isochrones in this panel reveals some slight differences between the two stellar evolution libraries: the PARSEC isochrones are systematically redder in the main sequence (see inset) and red-giant branch, while they are brighter in the sub-giant branch compared to the BaSTI-IAC isochrones \citep[we refer to][for a more detailed comparison between the two independent isochrone libraries]{Hidalgo2018}. These differences between isochrones from different stellar evolution libraries indicate that also systematic differences may be expected between models and data. In the derivation of the SFH, we will quantify this effect by allowing small shifts between the whole observed CMD and the synthetic CMD to which it will be compared (see  section \ref{sec3:shifts}). The differences between the predictions of stellar evolution libraries lead to somewhat different SFHs and age and metallicity distributions (see Section~\ref{robustness}).
 
In the case of the thick disk, for clarity only isochrones from the BaSTI-IAC library, in a range of metallicities, are displayed ([Fe/H]=0.06 and 5, 10 and 13 Gyr; [Fe/H]$=-0.5$, $-1.0$, $-2.0$ and 13 Gyr). The thick disk population is basically matched by the 10 Gyr old solar metallicity isochrone, with a minority of stars scattered around the 5 Gyr old isochrone, and very few stars fainter (thus older) than 10 Gyr. This is similar to what was found by \citet{MiglioChiappini2021_Kepler} using isochrone ages for a Kepler sample observed by APOGEE. These authors also found an almost coeval (age scatter of around 1 Gyr) and old thick disk. Old ages for the thick disk were also found by \citet{Queiroz2023_starhorse} using data from different spectroscopic surveys together with Gaia DR3. The old, lower metallicity isochrones show that there is room for a minority lower metallicity population which, at faint magnitudes in the main sequence (M$_G < 8$) becomes bluer than the bulk of the population. 

This was a first qualitative assessment of the age and metallicity ranges of the Milky Way components present in the volume covered by the GCNS. In Section~\ref{results}, we will present a quantitative description of these stellar populations.

\section{Derivation of SFHs: CMD fitting methodology} \label{sec3:method}

In this series of papers, we will define the SFH as the amount of mass that has transformed into stars, as a function of time (t) and metallicity (Z), in order to account for the population of stars contained in a particular volume of the MW. Because the defined volumes will typically be small compared to the MW size, and stars are expected to move away from their birth position  \citep[owing to diffusive or dynamical processes induced by the spiral arms or the bar, see e.g.][]{MinchevFamaey2010, Halle2015churblur, Hayden2018churning, Feltzing2020churblur}, these SFHs may be more appropriately referred to as {\it dynamically evolved SFHs} (deSFH), and the existence of stellar migration will need to be taken into account in the interpretation of the results. In any case, from these deSFHs, local age and metallicity distributions can be derived and this is a fundamentally important information in Galactic Archaeology.

The SFH can be expressed as a combination of simple stellar populations (SSPs) with a small range of age and metallicity. A convenient way to obtain these SSPs is from a synthetic CMD computed assuming that stars are born with a constant probability for all ages and metallicities within a given age and Z range
and adopting other stellar population characteristics such as an initial mass function (IMF) and a parameterisation of the binary star population. The errors affecting the absolute colours and magnitudes of the stars, as well as the completeness function across the CMD need to be simulated in the synthetic CMD to make it comparable to the observed CMD. {\it We will call this specific type of synthetic CMD from which we derive the SSPs, mother CMD.}

Once these SSPs have been defined, any arbitrary SFH (and its associated {\it model CMD}) can be obtained as a linear combination of SSPs:

\[\Psi(t,Z)= \sum_{i} \alpha_i \Psi_i \]
where $\Psi_i$ refers to SSP $i$ and $\alpha_i$ is the strength attributed to that SSP for that arbitrary SFH. 

The best fit SFH (for a given mother CMD) is then obtained by comparing the distribution of stars across the observed CMD and in an arbitrary number of {\it model CMDs}, constructed from different combinations of SSPs defined through sets of $\alpha$, until the best possible match is found. 

\begin{figure*}
\centering

\includegraphics[width=1\textwidth]{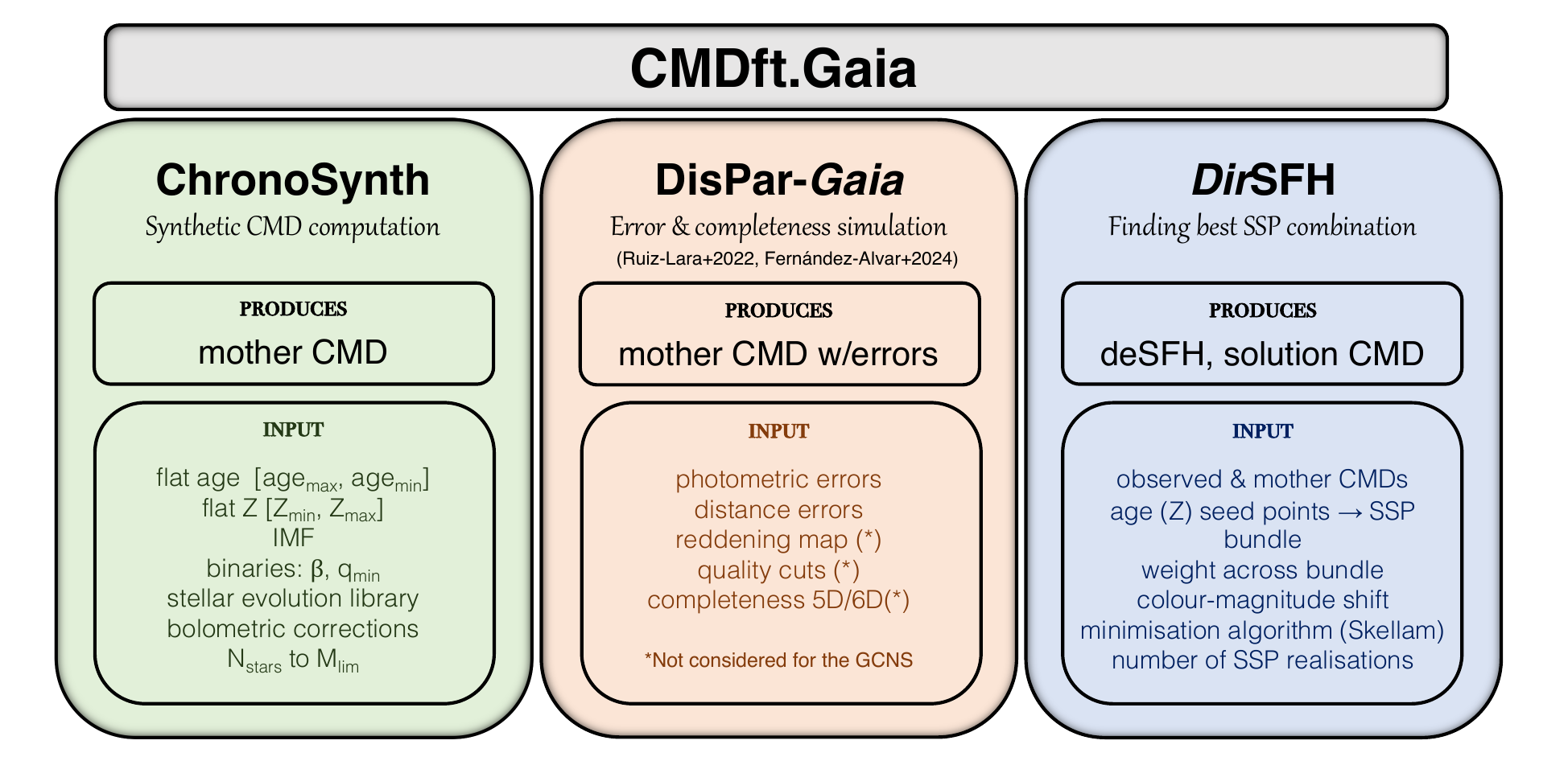}

\caption{Summary of the steps involved in the SFH derivation using CMDft.Gaia. A detailed discussion of ChronoSynth and $Dir$SFH is provided in this paper, while Dispar-$Gaia$ was presented in \citet{Ruizlara2022_HS} and will be discussed in more detail in Fernández-Alvar et al. 2024, in prep}
\label{cmdftgaia}
\end{figure*}

\begin{table*}

\caption{Glossary of acronyms and specific terms}
\centering

\begin{tabularx}{\textwidth}{lX}
\hline\hline
Term & Definition/description \\
\hline
\hline
{\bf Age (Z) bins} & Difference in age (Z) between consecutive age (Z) seed points \\
{\bf  Age (Z) seed points} & Array of age (Z) values that define a typical "size" of the SSPs reflecting the varying age (Z) resolution across the whole age (Z) interval \\
{\bf  Bundle} &  Regions of the CMD containing the stars that are used for the CMD-fitting. The bundle(s) are divided in colour-magnitude boxes (or 'pixels') where the stars (real and synthetic) are counted  \\
{\bf deSFH} & Mass, per unit time and metallicity, that has been transformed into stars  somewhere in the galaxy to account for the stars that are today in the studied volume\\
{\bf deSFR(t)} & Marginalisation over metallicity of the deSFH. It gives the mass per unit time that has been transformed into stars  somewhere in the galaxy to account for the stars that are today in the studied volume \\
{\bf MDF$_M$} & Metallicity distribution function of the mass transformed into stars \\
{\bf MDF$_S$}& Metallicity distribution function of the stars currently populating the studied volume \\
{\bf Model CMD}  & CMD resulting from an arbitrary linear combination of SSPs \\
{\bf Mother CMD} & Specific type of synthetic CMD computed assuming that stars are born with a constant probability for all ages and metallicities within a given age and metallicity range and adopting other stellar population characteristics such as an initial mass function (IMF) and a parameterisation of the binary star population, and with observational errors and completeness simulated in it. The SSPs are derived from a mother CMD \\
{\bf Solution CMD} & CMD obtained with the best fit SFH, and matching the number of stars of the observed CMD inside the bundle \\
{\bf SSP} & Simple Stellar Population, population of synthetic stars in a small range of age and metallicity \\
{\bf Synthetic CMD} & Any CMD obtained from the predictions of stellar evolution models \\
  \hline
  \end{tabularx}
  \label{terms}
\end{table*}

In the following we will discuss in detail all the steps involved in our implementation of the SFH derivation procedure, which we call {\it CMDft.Gaia} (first introduced in \citealt{Ruizlara2022_HS}). It has been considerably updated compared to previous works such as \citet{Gallart2019Gaia} and \citet{Ruiz-Lara2020Sgr}.  

\FloatBarrier

CMDft.Gaia is a suite of procedures specifically designed to deal with {\it Gaia} data that includes i) the computation of synthetic CMDs in the {\it Gaia} bands (Sec~\ref{sec3:Synthetic_CMD_Computation}; ii) the simulation in the synthetic CMDs of the observational errors and completeness affecting the observed CMD after quality and reddening cuts (Section~\ref{sec3:ErrorSimulation}; and iii) the derivation of the SFH itself (Section~\ref{sec3:FitSFH}). We explain these steps in detail below, while a summary is presented in Figure~\ref{cmdftgaia}.

\subsection{Synthetic CMD computation} \label{sec3:Synthetic_CMD_Computation}

All mother CMDs adopted in the present investigation have been computed with our own synthetic CMD code, that results from a deep evolution/update of the code presented in \cite{pietrinferni2004} and \cite{cordier07} and that we will call ChronoSynth. Due to the need of computing mother CMDs hosting a huge number of synthetic stars, the current version of the code has been parallelised in order to speed up the whole computational procedure. The code provides magnitudes and colours of stars belonging to a synthetic stellar population with an arbitrary SFH, as well as the total mass that has been transformed into stars associated to that population, which includes that of the stars already dead and thus not present in the synthetic CMD. For this purpose, the code relies on a grid of isochrones in a wide age and metallicity range, which depends on the adopted stellar model library.

The code has some flexibility in the types of SFH that can be adopted. However, for the purpose of SFH derivation, the synthetic mother CMDs are computed adopting a flat probability distribution for star to be born within the whole defined age and metallicity range (the latter can be defined flat in Z or log(Z)). The lower and upper limits both in age and metallicity have to be specified in the input file, as well as the adopted IMF\footnote{The current version of the code allows to choose between a power law (with exponent chosen by the user), and the \citet{kroupa1993} IMF.} and the characteristics of the binary population, which are parameterised as a function of the binary frequency $\beta$ and minimum mass ratio q$_{min}$.

To create the synthetic stellar population, with a desired number of stars (N$_{\rm{stars}}$) down to a given limiting magnitude (M$_{\rm{max}}$), according to the adopted SFH, a random value of the stellar age and metallicity are drawn from the whole age/metallicity range. Then a stellar initial mass, M, is randomly selected following the adopted IMF. These values of age, stellar mass, and metallicity are used to interpolate in the isochrones of the selected grid to determine the bolometric luminosity, effective temperature, and current value of the mass\footnote{The current mass of the star, that can be different than the initial one as a consequence of mass loss, is needed in order to properly evaluate the actual surface gravity that is -- together with the effective temperature and chemical composition -- the quantity needed to evaluate the bolometric correction for any selected photometric passband. The occurrence of mass loss is taken into account by adopting a stellar model grid computed by accounting for a selected value of the mass loss efficiency \citep[we refer to][for a more detailed discussion on this topic]{Hidalgo2018}.} of the synthetic star. These properties are then used to predict its absolute magnitudes in the various {\it Gaia} DR3 photometric passbands \footnote{Many other photometric systems are also available in the code.} on the basis of the bolometric correction tabulations\footnote{These bolometric correction tables are used for all the libraries of stellar models than can be selected as input in the synthetic CMD code. This means that they are adopted also when selecting the PARSEC library; however it is worth noting that both the BaSTI/BaSTI-IAC library and the PARSEC one adopt very similar spectral libraries in the regime appropriate for main sequence and sub-giant branch stars as discussed by \cite{Hidalgo2018} and \cite{Chen2019}.} provided by \cite{Hidalgo2018}.

To include a given population as unresolved binary systems, the binary fraction $\beta$ and the minimum mass ratio $q_{\rm min}$ between secondary and primary stars have to be specified. Then, for each generated synthetic star, an additional random number is used to determine if it is a component of an unresolved binary. If this is the case, the mass of the secondary star is randomly selected from the distribution given by \cite{woo03}, that is, the mass of the secondary of a primary star with mass M$_p$ is randomly chosen - according to a flat distribution - between $q_{\rm min} \times \rm{M_p}$ and  $\rm{M_p}$. The predicted properties (luminosity, effective temperature, magnitudes) of this unresolved binary system are calculated properly adding the fluxes of the two unresolved components. No evolution of the binary system itself is considered.
 
Finally, in order to investigate the impact of the choice of different stellar model libraries on the  properties derived for the studied stellar populations, we have implemented different model grids in ChronoSynth. The current version of the code allows to use the following libraries: \footnote{In addition to the two libraries quoted in the main text, the first release of the BaSTI isochrone grid (hereinafter BaSTI) for both the solar scaled \citep{pietrinferni2004} and $\alpha-$enhanced \citep{pietrinferni2006} mixture is also kept in the code in order to allow -- if needed -- some comparison with previous works. Indeed, this library is the one used in \citet{Gallart2019Gaia} and \citet{Ruiz-Lara2020Sgr}. The selected model set from the former library is the one accounting for convective core overshooting during the central H-burning stage, and mass loss according to the Reimers law with the free parameter $\eta$ fixed at a value equal to 0.4.}

-the BaSTI-IAC grid both for solar-scaled \citep{Hidalgo2018} and $\alpha-$enhanced heavy element mixture \citep[][]{pietrinferni21}. The selected grid is that accounting for diffusive processes, core convective overshooting and mass loss (with efficiency fixed by selecting a value for the free parameter $\eta$ equal to 0.3);

-the PARSEC stellar model library for a solar-scaled mixture as provided by \cite{Bressan2012}.

\subsection{Simulation of the observational errors and completeness} \label{sec3:ErrorSimulation}

Prior to the comparison between the observed and synthetic star distribution across the CMD, it is necessary to simulate the observational errors and the completeness in the synthetic CMD. In the case of the GCNS, since it is basically complete and the photometric and distance errors are really small, we adopted a simplified error simulation procedure (see Section~\ref{sec4:ErrorSimulationGCNS}). A comprehensive description of the general method we will adopt in this series of papers to simulate the observational errors and completeness in the synthetic CMD, called DisPar-$Gaia$ will be provided in Fernández-Alvar et al. (2024, in prep), while a summary is provided in \citet{Ruizlara2022_HS} (see also Figure~\ref{cmdftgaia}).

\subsection{Determination of the best fit SFH} \label{sec3:FitSFH}

To determine the best fit SFH through comparison of the observed CMD with model CMDs computed from combinations of SSPs, we use a new CMD-fitting package that we call $Dir$SFH\footnote{Because an important feature of the code is the use of a $Dir$ichtlet (or Voronoi) tessellation algorithm to define the SSPs}. This package is a sophisticated evolution of previous CMD-fitting software such as IAC-pop \citep{apariciohidalgo2009} and TheStorm \citep{bernard2018MNRAS, Rusakov2021Fornax}. We will discuss the novel approach followed by $Dir$SFH to the different steps involved in the CMD-fitting process, which lead to extremely robust solutions providing a large amount of details in the age-metallicity plane. We will mention the main differences with respect to IAC-pop and TheStorm, the latter being the code used in our two first works delivering SFHs of the MW with {\it Gaia} CMDs, namely \citet{Gallart2019Gaia} and \citet{Ruiz-Lara2020Sgr}.

\subsubsection{Definition of the simple stellar populations (SSPs)} \label{sec3:SSPdefinition}

An innovative aspect of $Dir$SFH lays in the way SSPs are defined. The mother CMD is dissected as a function of age and metallicity in a semi-random Dirichlet-Voronoi tessellation, based on a grid of seed points in both age and metallicity, which allows the user to define a typical "size" of the SSPs reflecting the varying age and metallicity resolution across the whole age and metallicity interval. A minimum number of stars is required in each SSP, and if this number is not reached, a SSP may be merged with neighbour ones until it has the minimum required number of stars. Therefore, a large enough mother diagram is necessary to avoid degrading the resolution in age and metallicity of the derived SFH.

\begin{figure*}
\centering
\includegraphics[scale=0.37]{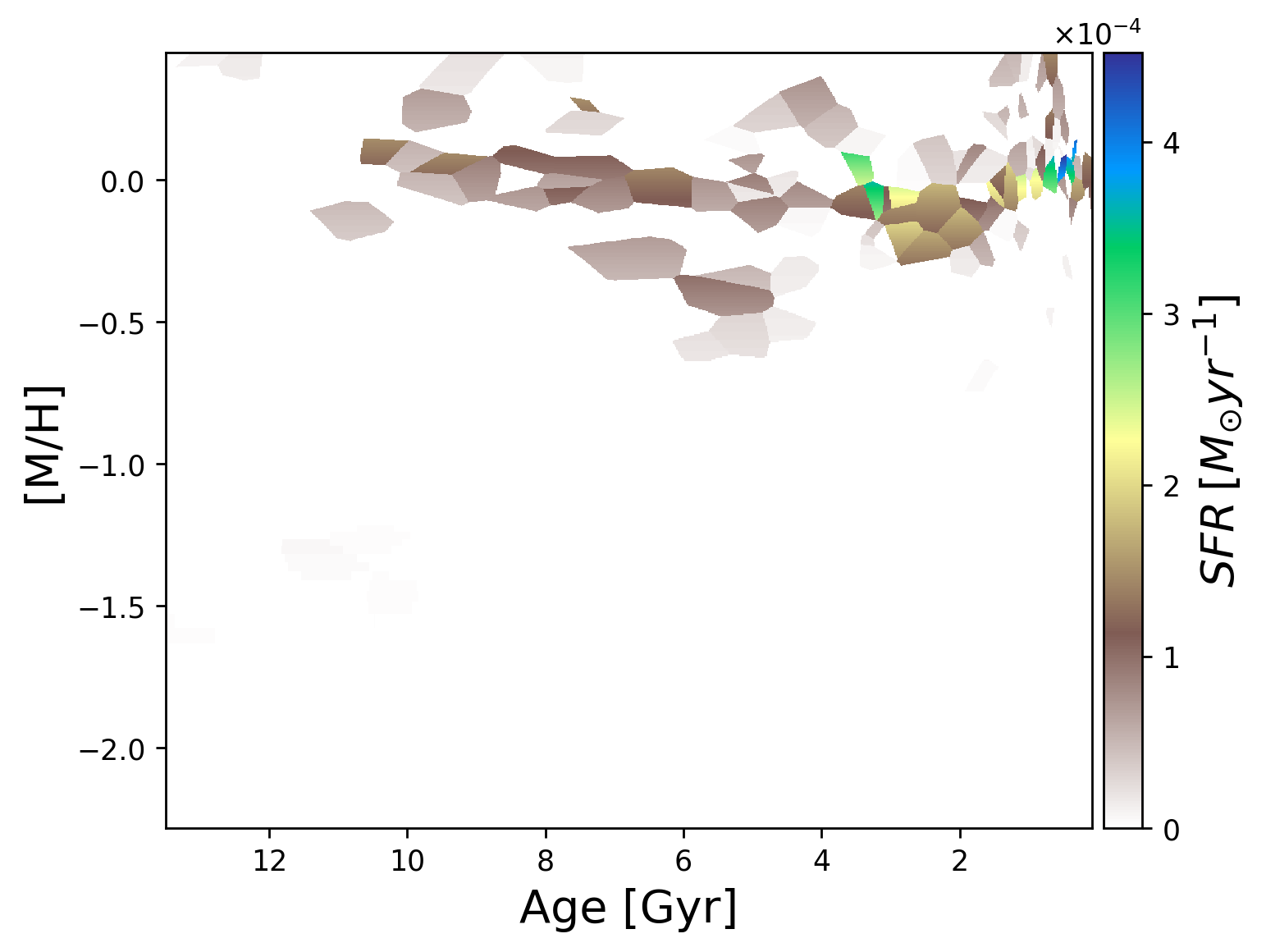}
\includegraphics[scale=0.37]{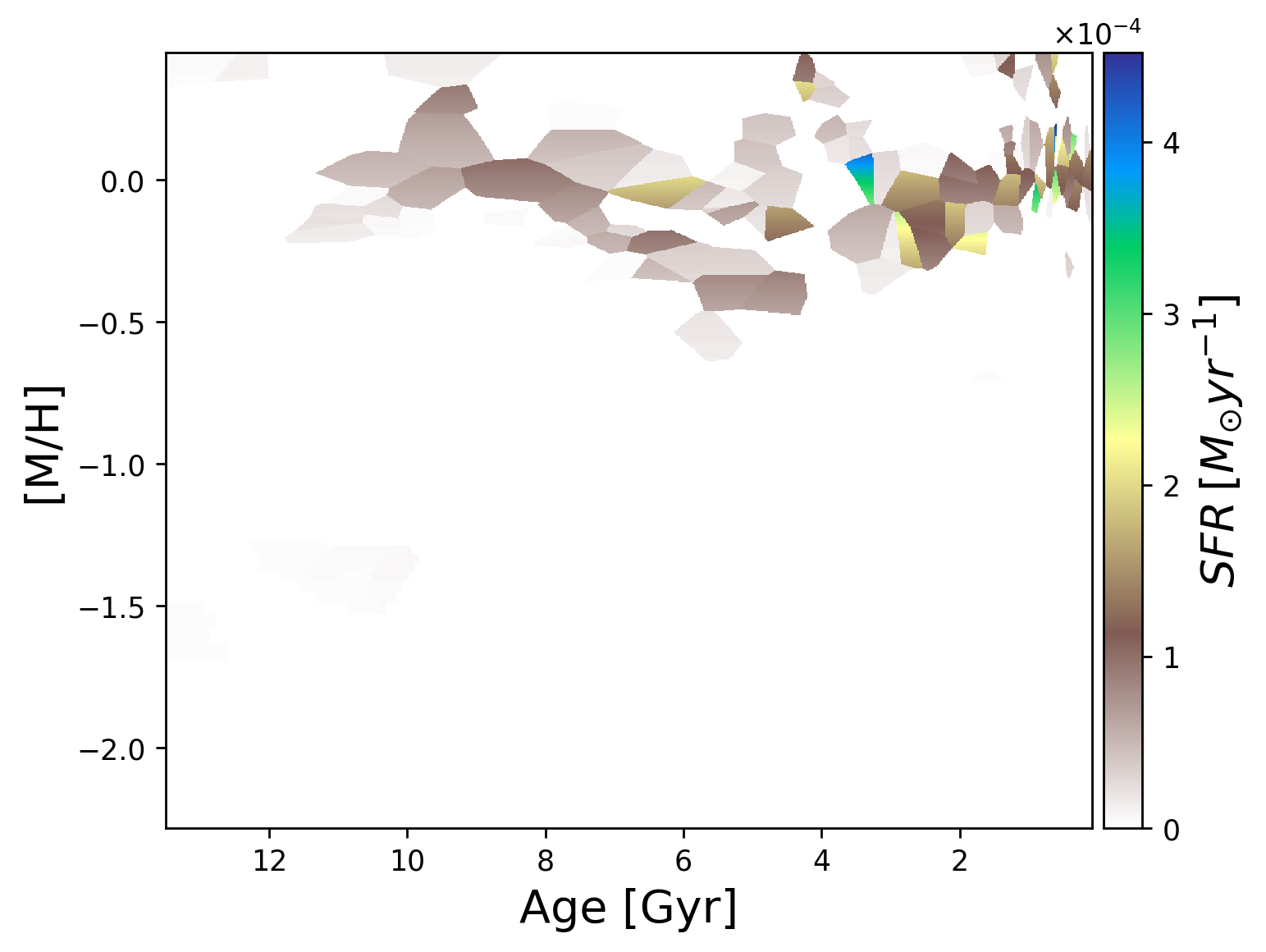}
\includegraphics[scale=0.37]{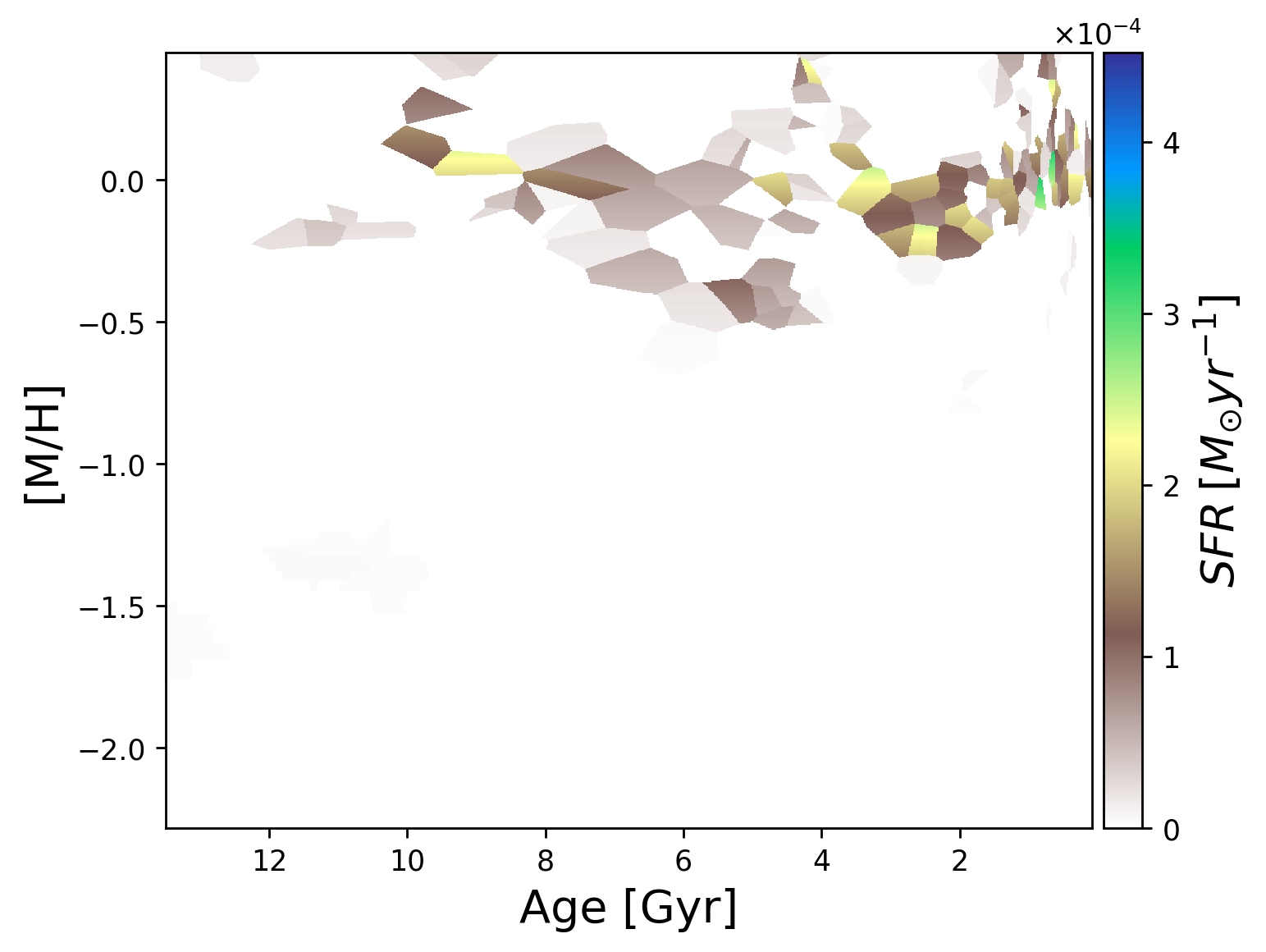}
\caption{Three individual solutions of the GCNS SFH are shown. The Dirichlet-Voronoi tessellations of the mother CMD, different for each solution, can be appreciated, while the overall shape and characteristics of the derived SFH is preserved.}
\label{voronois}
\end{figure*}

A large number N of 'individual' SFHs are derived with different sets of SSPs constructed with the above method, and the final SFH is the average of all of them. Figure~\ref{voronois} shows three individual SFH solutions in which the Dirichlet-Voronoi tessellation of the mother CMD can be appreciated. Note that each solution is somewhat different from another, even though they share general trends, including a low metallicity old population with very low signal, as well as some high metallicity population.

\subsubsection{Sampling the star's distribution across the CMD and weights of the fit} \label{sec3:samplingCMD} 

As mentioned in the introduction of this section, the distribution of stars in the observed CMD and in the model CMDs resulting from each combination of SSPs, need to be compared. For this, observed and model CMDs are binned as a function of colour and magnitude in the same way. 

\begin{figure}
\centering
\includegraphics[scale=0.55]{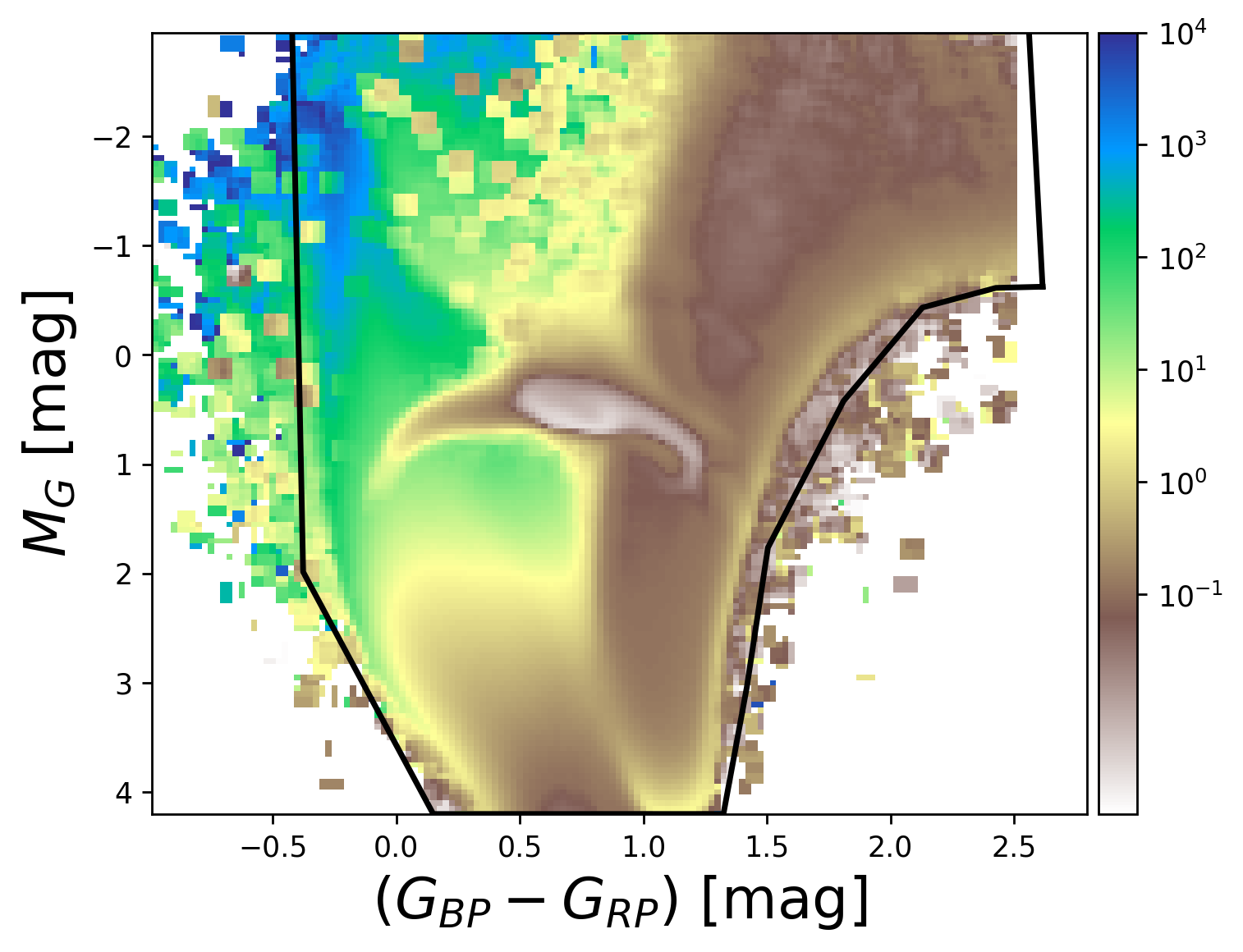}
\caption{Example of the weights applied across the CMD for the q01b03\_120M\_M5 mother CMD. The weight of each 'pixel' in the CMD is calculated as the inverse of the variance of the stellar ages in that 'pixel'. The bundle defining the area where observed and model CMDs are compared is also shown. Note that the bundle tightly delimits the mother CMD; a few 'pixels' containing synthetic stars outside the bundle correspond to stars that have been scattered in the simulation of the observational errors.}
\label{weights}
\end{figure}

The amount of information on the SFH that a particular region of the CMD can provide depends mainly on a) how separated in colour and magnitude the populations of different ages and metallicities are in that region, b) the accuracy with which stellar evolution models are able to predict the positions and lifetimes of the stars~\citep{gallart2005} in the corresponding evolutionary phase, and c) the number of stars populating it. The main sequence (and particularly the region around the turnoffs) and the sub-giant branch are thus the CMD regions that provide most information on the SFH. In contrast, the (shorter lived) red giant branch, the red clump and the horizontal branch phases are affected both by larger uncertainties in the stellar evolution model predictions (including more uncertain bolometric corrections) and by the superposition in a small colour-magnitude region of stars that encompass almost the whole range of possible ages and metallicities. The latter leads to a poor age and metallicity resolution which is exacerbated by a larger age-metallicity degeneracy compared to the rest of the CMD. To take this into account, in IAC-Pop or TheStorm, several {\it bundles} in the CMD were typically defined, each in turn divided in colour-magnitude boxes (or 'pixels') of different sizes where the number of stars are counted for the comparison between the observed and the {\it model} CMD. These {\it bundles} could exclude, or sample more coarsely, CMD regions corresponding to certain stellar evolutionary phases, in order to modify their overall weight in the fit. This approach has the disadvantage of a certain subjectivity in the bundle definition, which however, was shown to have little effect on the final SFH \citep[see, for example][]{Ruiz-Lara2021LeoI}. $Dir$SFH uses a single bundle defined by the user to tightly include both the observed and the mother CMD down to a given limiting magnitude, which is used only as a delimiter of the fitting space. Within this bundle, a weight matrix is calculated from the mother CMD based on how precisely a given 'pixel' of that CMD is unique in terms of age (see next paragraph for a discussion of how these 'pixels' are defined). In particular, in the current implementation, the weight of each 'pixel' in the CMD is calculated as the inverse of the variance of the stellar ages in that 'pixel', such that a 'pixel' populated by a small range of age will be given more weight (see Figure~\ref{weights} for an example of the weights applied across the CMD for the q01b03\_120M\_M5\footnote{From now on, this will be the typical naming convention for mother CMDs. In this particular case, q01b03\_120M\_M5 means a  CMD computed  adopting q$_{min}$=0.1, $\beta$=0.3, and with 120 million stars down to $M_G$=5.0. The Basti-IAC stellar evolution library is used unless stated otherwise} mother CMD). This occurs, for example, in the bright, blue parts of the CMD, which are populated exclusively by young stars. Giving more weight to those areas defines the populations therein (mostly young SSPs) with higher accuracy, imposing a strong constrain on the presence of young stars in areas where the CMD has severe overlap of ages. The option of a uniform weight across the whole CMD is also possible, and we have verified that the results are not affected in a significant way by the weighting scheme applied (see Figure \ref{amr_wei_uni}).

The way the CMD 'pixels' are defined in $Dir$SFH takes into account the fact that the number of stars in each pixel is subject to statistical fluctuations. In TheStorm this was taken into account by shifting the limits of the boxes by a fraction of their size and calculating new SFHs, which would at the end be averaged together and used to calculate statistical errors. In $Dir$SFH, a 'coarse' grid of pixels is defined (typically of size 0.2 magnitudes in M$_G$ and 0.1 magnitudes in $(G_{BP}-G_{RP})$) and used to calculate a colour-magnitude histogram. Then, the grid is shifted a number of times (25 times, for 5x5 steps in colour and magnitude in the current implementation) and new histograms are calculated for each shift. This results in an over-sampled representation of the CMDs, maintaining the number statistics of the coarse grid, but adding the information of the variation of the number of stars across the CMD in much finer steps (effectively increasing the 'resolution' by 25 times). This preserves fine details in the stellar distribution across the CMD that would have otherwise been destroyed by the simple coarse grid. 

\subsubsection{Search for systematic shifts between theory and observations}\label{sec3:shifts}

Uncertainties in the effective temperature scale, and/or in the bolometric corrections adopted to transfer stellar evolution predictions from the H-R diagram to the observational plane (in this case the {\it Gaia} photometric system), as well as residual uncertainties in the photometric calibration, may lead to slight overall systematic shifts between the observed and the synthetic CMDs. An evidence of the existence of such shifts can be observed in the data shown in the lower middle panel of Figure~\ref{description_catalogue}, where PARSEC and BaSTI-IAC isochrones have been superimposed to the {\it Gaia} CMD in the absolute plane: in most evolutionary phases, small but noticeable systematic shifts do exist between isochrones of identical age and metallicity belonging to the two model sets (as also discussed in \cite{Hidalgo2018}). Similar systematic shifts may be expected between data and models.

In order to derive the size and direction of these systematic offsets, several SFHs are calculated with the mother CMD shifted in colour-magnitude space within a maximum specified range. Then, the residuals of these fits are analysed, and an appropriate weighted average of the colour and magnitude shifts leading to the smallest residuals is adopted as the best shift for the final SFH calculation. This is a similar, but slightly more sophisticated procedure compared to that adopted in TheStorm or IACpop \citep[see, for example, Figure~5 in][]{Rusakov2021Fornax}.

\subsubsection{The minimisation algorithm}\label{sec3:minimizacion} 

In $Dir$SFH, the goodness of the coefficients in the linear combination that defines each model CMD is measured through a Skellam distribution (as opposed to simple Poisson statistics in TheStorm). This statistic considers both the observed and model CMD histograms to be stochastic in nature. For any given colour-magnitude pixel, the difference in the number-counts between observed and model is evaluated, and the probability of this result, considering both inputs to be Poisson distributed, is used to calculate the goodness-of-fit of the ensemble. This implies a critical difference with TheStorm approach: in $Dir$SFH, cases where there are observed stars but no model CMD stars are treated completely equivalent to the cases where there are no observed CMD stars but model stars are present. In particular, $Dir$SFH minimises:

\[ S = \sum_i{\frac{(O_i-M_i)^2}{O_i+M_i}} \]

Where $O_i$ and $M_i$ are the number counts for each pixel in the ensemble for the observed CMD and the model CMD, respectively.

\subsubsection{Calculation of the final SFH and associated errors} \label{sec3:errors}

The CMD-fitting procedure is repeated for a large number ($N \simeq 100$) of different realisations of the SSPs and the final SFH is a weighted average of the resulting $N$ SFHs, with the error being the dispersion of the distribution. 

\subsubsection{The outputs of $Dir$SFH} \label{sec3:dirSFHoutput}

$Dir$SFH produces two main outputs:

\noindent i) The {\it SFH}, that is, the mass transformed into stars as a function of lookback time (age) and metallicity [M/H] in units of M$_\sun$ Gyr$^{-1}$ dex$^{-1}$, in the form of a 800$\times$800 grid of star formation rate values and their uncertainties as a function of age and metallicity. From this information, the SFR(t) (in units of M$_\sun$ Gyr$^{-1}$) and the metallicity distribution function of the astrated mass (MDF$_M$) (in units of M$_\sun$ dex$^{-1}$) are calculated by marginalising over metallicity and age, respectively. From the SFR(t), the cumulative mass function is calculated.

\noindent ii) A {\it solution CMD}, which is obtained by sampling the mother CMD according to the derived SFH, until the same number of stars in the observed CMD is obtained. In the solution CMD, each star has information of its age and metallicity, in addition to its colour and magnitude. The solution CMD, therefore, allows analysing the current stellar content of a given stellar population, in terms of the number of stars currently present as a function of their age and metallicity. In particular, the metallicity distribution function of the stars in the sample (MDF$_S$ hereafter) can be obtained, which is an important observable that can be compared with that derived spectroscopically. Note that no information is obtained about the age of {\it individual stars} in the observed CMD.

\section{Deriving the SFH of the GCNS} \label{sec4:deriving}

In this section we will discuss the particular application of CMDft.Gaia to the GCNS \citep{gcns}.

\subsection{Synthetic CMDs used to derive the deSFH within 100pc}\label{sec4:syntheticCMD}

As discussed in \ref{sec3:Synthetic_CMD_Computation}, in addition to the adopted stellar evolution library, a number of choices have to be made to calculate the synthetic CMDs that will be used to derive the SFH. Along the paper we will explore the impact on the SFH of the most relevant of these choices. In particular, the extraordinary depth of the GCNS CMD will allow us to check different assumptions on the binary star population using the distribution of stars in colour and magnitude in the low main sequence. For a summary of the synthetic CMDs used in this work, see Table \ref{mothers}. The choices made to compute this set of synthetic CMDs are the following:

i) Stellar evolution library. The reference stellar evolution library used is BaSTI-IAC with the solar scaled mixture \citep{Hidalgo2018}. Several synthetic CMDs with a number of choices on the other parameters have been calculated with this library. Two additional synthetic CMD has been computed using the PARSEC solar-scaled library \citep{Bressan2012}.

ii) Age and metallicity distribution, and number of stars in the synthetic CMDs: all synthetic CMDs have been computed with a flat age and metallicity distribution (flat in Z) within the age and [M/H] range indicated in Table~\ref{mothers}, and with a specified total number of stars down to a given limiting magnitude M$_{G, max}$. 
 
For each synthetic CMD, the age and metallicity limits  as well as the number of stars brighter than $M_G$=5 (which allows a homogeneous comparison of the size of each CMD in the approximate portion used to calculate the SFH, above the oMSTO) are given in Table~\ref{mothers}. 

iii) Binary star population: we have explored different populations of unresolved binary systems, parameterized as a function of the fraction of unresolved binaries $\beta$ and minimum mass ratio $q_{\rm min}$, as discussed in Section~\ref{sec3:Synthetic_CMD_Computation}. The $\beta$ and $q_{\rm min}$ adopted combinations are specified in Table~\ref{mothers}. 

iv) IMF: the \cite{kroupa1993} IMF has been used, except in one case, for which the Salpeter IMF was adopted. 

\begin{table*}
 \centering
 \caption{Characteristics of the synthetic mother CMDs used for the analysis of the SFH of the GCNS. }
 \begin{tabular}{lrcrcrcrc}
  \multicolumn{9}{c}{Synthetic Diagrams} \\
  \hline \hline
Name & Age range  & [M/H] & q$_{\rm min}$ &   $\beta$ & Library & M$_{G, max}$ &  N* ($\times 10^6$)  & shift \\ 
 & (Gyr) & range &  &   &  &  &  (M$_G\leq 5$) & (col,mag) \\ 
 \hline
q01b03\_60M\_MG10\ &13.5-0.02 & -2.2--0.45& 0.1  &  0.3 &  BaSTI-IAC & 10 &  7.0 &(-0.036, 0.041) \\ 
q01b03\_112M\_MG6\ &13.5-0.08 &-2.2--0.45 & 0.1  &  0.3 &  BaSTI-IAC & 6 &  68.2 & (-0.036, 0.040)\\ 
q01b03\_120M\_MG5\ & 13.5-0.08& -2.2--0.45& 0.1  &  0.3 &  BaSTI-IAC & 5 &  120 &(-0.035, 0.037)\\ 
q01b03\_30M\_MG10\_parsec & 13.5-0.02 & -2.2--0.27& 0.1  &  0.3 &  PARSEC & 10 & 4  & (-0.047,0.045)\\ 
q01b03\_43M\_MG5\_parsec &  13.5-0.02 &-2.2--0.27 & 0.1  &  0.3 &  PARSEC & 5 &  43.2 & (-0.035,0.030)\\ 
q01b05\_60M\_MG10 & 13.5-0.02 & -2.2--0.45& 0.1  &  0.5 &  BaSTI-IAC & 10 &  6.9  &(-0.030, 0.030)\\ 
q01b07\_30M\_MG11 &13.5-0.02 & -2.2--0.45& 0.1  &  0.7 &  BaSTI-IAC & 11 & 2.2  & (-0.027, 0.030)\\
q01b07\_81M\_MG5 & 13.5-0.02& -2.2--0.45& 0.1  &  0.7 &  BaSTI-IAC & 5 &  81.3  &(-0.030, 0.030) \\
q04b03\_30M\_MG11 & 13.5-0.02 & -2.2--0.45& 0.4  &  0.3 &  BaSTI-IAC & 11 & 2.2  &(-0.034, 0.038) \\
q06b01\_30M\_MG10 & 13.5-0.02& -2.2--0.45& 0.6  &  0.1 &  BaSTI-IAC & 10 &  3.5  &(-0.039, 0.048)\\
  \hline
  \end{tabular}
  \tablefoot{Columns, from left to right: identification name, which includes information on the millions (M) of stars down to a given limiting magnitude (MG), age and metallicity ranges, the minimum binary mass ratio (q$_{min}$), the binary fraction ($\beta$), the stellar evolution library, the limiting magnitude M$_{G, max}$, the number of stars with a  M$_G\leq 5$ mag and, the shift in colour and magnitude subtracted to the synthetic CMD.}
  \label{mothers}
\end{table*}

\subsection{Simulation of the GCNS observational errors} \label{sec4:ErrorSimulationGCNS}

In the case of the GCNS, since both the photometric errors and those derived from the distance calculation are really small and other sources of error (such as reddening) are negligible, we adopted a simplified error simulation procedure compared to DisPar-$Gaia$, which will be used in future papers of this series.

We considered that the sources of uncertainty affecting the position of an observed star in the CMD were solely the photometric errors and the error in the determination of the distance. In order to implement these observational effects on the mother diagrams, we first assigned to each synthetic star a distance following the global distribution of stellar distances in the GCNS ({\tt dist\_50}). This preliminary step allowed us to, applying the relation between apparent and absolute magnitudes, move our mother CMD to the apparent plane (we verified that extinction is totally negligible in this sample). As shown in \citet{Riello2021DR3_PhotoContent} there is a clear trend of photometric errors with magnitude, with the brighter tail (G$\lesssim6$) presenting larger uncertainties, and a smooth trend to larger uncertainties for fainter stars as well. We found a similar trend for the distance errors, with more distant stars displaying larger uncertainties, which we defined as (dist\_84-dist\_16)/2.0, being 'dist\_84' and 'dist\_16' the 84th and 16th percentiles of the distance PDF (1$\sigma$ upper and lower bounds, assuming a gaussian distribution) provided by \citet{gcns}. In both cases (photometry and distance information), we fitted a 5$^{\rm th}$ degree polynomial ($P_i$) to the run of the error in the parameter $i$ as a function of the value of such parameter, with $i$ corresponding to apparent G$_{BP}$, G$_{RP}$, G or distance. Also, we characterised the running standard deviation of each parameter ($\sigma_i$). Thus, to each star $s$ in the synthetic CMD, with a given set of parameters $i$ (G, G$_{BP}$, G$_{RP}$, distance), we assigned random errors ($\epsilon_i$) following a Gaussian centred at $P_i(i_s)$ and with sigma $\sigma_i(i_s)$, with $i_s$ referring to the value of the parameter $i$ for the star $s$.

Once we have the photometric and distance errors for each synthetic star, we performed a quadratic propagation of uncertainties to translate these observational errors into an error in the absolute magnitude and colour. These errors were simulated by adding to the colour and magnitude of each star a correction following a Gaussian distribution centred at zero and with sigma equal to the propagated error in colour and magnitude. Finally, since the GCNS is considered to be complete down to spectral type M8 \citep[][much fainter than the stars to be used for the SFH calculation]{gcns}, we have not applied any completeness simulation in the synthetic CMD.

\subsection{Parameterising the mother CMD and configuring $Dir$SFH} \label{sec4:parameterizing}

As discussed in Section~\ref{sec3:FitSFH}, $Dir$SFH allows the user to define a number of input parameters that will determine some details of the fit. In this section we will discuss how these parameters have been chosen:

\subsubsection{The arrays of age and metallicity seed points used to define the SSPs} \label{sec4:agemetbins}

The decrease in the isochrone separation toward older ages results into a decrease of the age resolution. Nevertheless, the actual age resolution as a function of age may also depend on the characteristics of the observed CMD, and in particular, on the photometric errors across it. The age and metallicity seed points need to be carefully chosen to optimise the recovery of the age and metallicity information present in the data, while avoiding over-fitting the CMD and keeping manageable computing times. 

For metallicity, and after some testing, we have adopted a typical separation of the seed points of 0.1 dex in [M/H], which is of the order of the typical error in spectroscopic metallicity measurements. 

In order to assess the age precision and accuracy that can be achieved with a high quality {\it Gaia} CMD such as that of the GCNS, we have designed recovery tests based on: i) a composite CMD of four MW open clusters  observed by {\it Gaia}, and ii) seven synthetic {\it clusters} with ages 0.2, 2, 4, 6, 8, 10 and 12 Gyr, a small age range (20 Myr) and a small metallicity range, close to solar metallicity ([M/H]=-0.1 to -0.05), which is the metallicity of most stars in the observed GCNS CMD. These tests consist on testing $Dir$SFH recovery using several arrays of age seed points, which result in corresponding arrays of {\it age bins}, as we will refer, for simplicity, to the difference in age between consecutive age seed points. These tests will allow us to check how the age precision and accuracy vary with different age bins. In Appendix~\ref{clusters} we describe how these datasets have been prepared (membership selection, distance and reddening determination in the case of the open clusters, and ChronoSynth input parameters for the calculation of the synthetic clusters). 

From a set of age seed points similar to the one used in previous works\footnote{The set of age seeds that we used as starting point is the following: Ages(Gyr)= [0.08, 0.195, 0.340, 0.477, 0.606, 0.732, 0.877, 1.057, 1.293, 1.664, 2.190, 2.736, 3.295, 3.852, 4.4585, 5.231, 6.197, 7.235, 8.279, 9.324, 10.368, 11.418, 12.456, 13.5]. It is similar to the set used by \citet{Ruiz-Lara2020Sgr} but the age bins have been optimised by taking into account the typical separation of the isochrones in the populated areas near the turnoff point, as a function of age.}
\citep[e.g.][]{Ruiz-Lara2021LeoI, Ruiz-Lara2020Sgr}, which results in what we will call XL age bins, we created three new sets of age seed points that result in progressively smaller age bins, following the same functional relation between bin size as a function of age as the original set. We will call the four sets XL, L, M and S bins. We have derived the SFH of the composite CMD of the four open clusters and that of the seven synthetic clusters mentioned above, with the four sets of age bins, keeping unchanged all the other parameters involved in the fit. The derived SFHs are presented in Appendix~\ref{clusters}, while we discuss here the main conclusions regarding the precision and accuracy of the age determination as a function of age.

For each real or synthetic cluster we have fitted a 2D gaussian\footnote{Using a program based on the Scikit-learn Python package and the Gaussian Mixture Model code} to the corresponding distribution of ages and metallicities of the stars of the solution CMD in the age-metallicity plane. The square root of each diagonal term of the corresponding covariance matrix (that is, the standard deviations projected in the age and metallicity axis, $\sigma_{age}$ and $\sigma_Z$, respectively) provide a measure of the age or metallicity precision, while the comparison of the 2D fitted gaussian centre with the input mean age or metallicity of the corresponding synthetic cluster gives information on the accuracy of the derived ages or metallicities. 

The metallicity is recovered accurately at all ages, and with a $\sigma_Z$ between 0.05 and 0.10 dex (and thus, of the order of the size of metallicity bins, 0.1 dex), with little dependency of the number of stars in the cluster or the size of the age bins (see discussion in Appendix~\ref{clusters} and Figures~\ref{metallicity_precision} and ~\ref{SC_AMR_projections}).

\begin{figure*}[h!]
    \centering
    \includegraphics[width=0.49\textwidth]{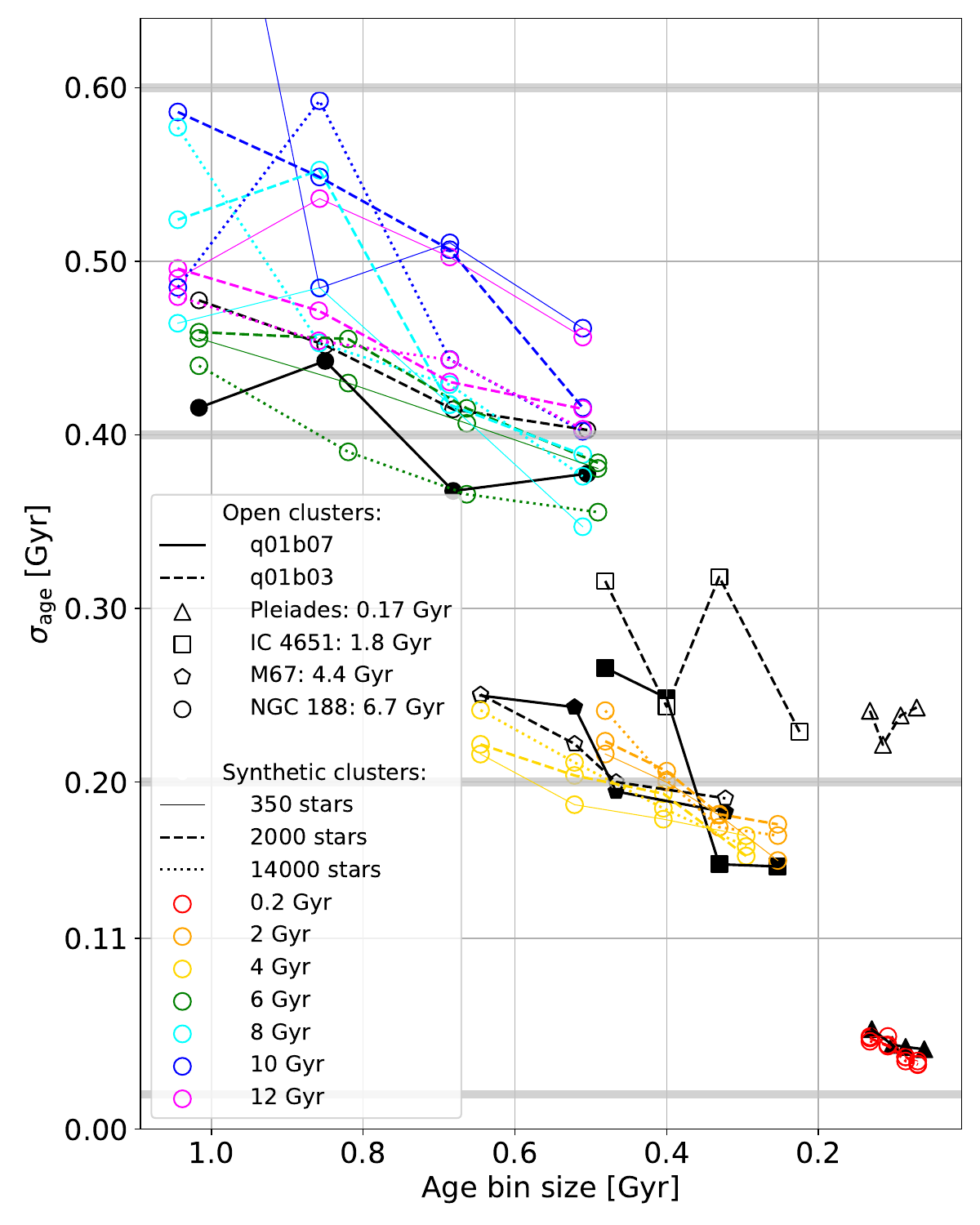} 
    \includegraphics[width=0.5\textwidth]{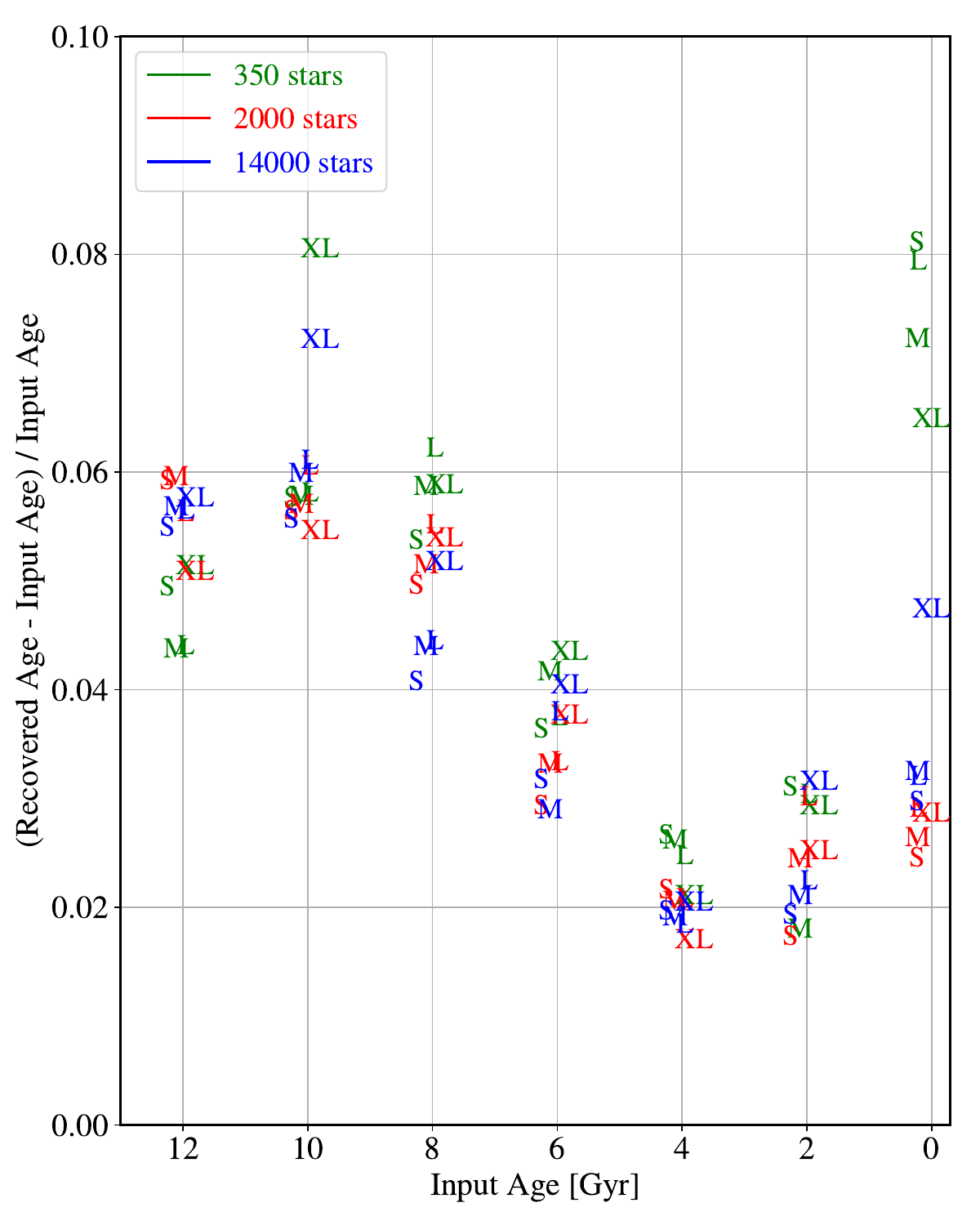}
    \caption{Left panel: precision analysis. $\sigma_{age}$ as a function of the age bin size is displayed for the four open clusters (black symbols) and the seven synthetic clusters (coloured symbols). For each cluster, four points indicating the measured $\sigma_{age}$ for the XL, L, M and S bins are connected with a line. The three sequences of $\sigma_{age}$ for the synthetic clusters correspond to clusters simulated with different number of stars with M$_G$<4.2: $\simeq$ 350, $\simeq$2000 and $\simeq$ 14000. The two sequences of $\sigma_{age}$ for the open clusters refer to solutions obtained with two different mother CMDs with different unresolved binary star characteristics: q01b03\_120M\_MG5 (dashed line and open symbols) and q01b07\_81M\_MG5 (solid line and filled symbols).
    Right panel: accuracy analysis. The relative error of the age determination, as a function of age, is represented. Only the synthetic clusters are displayed. Clusters with different numbers of stars are depicted in different colours as indicated in the labels. The size of the bins used for each measurement is represented with the corresponding letters.} 
    \label{precision_accuracy}
\end{figure*}

The left panel of Figure~\ref{precision_accuracy} displays $\sigma_{age}$ as a function of the age bin size for the four open clusters and the seven synthetic clusters. For each cluster, four points indicating the measured $\sigma_{age}$ for the XL, L, M and S bins are connected with a line. The age bin sizes in the $x$ axis correspond to those at the age of each cluster. In the case of the synthetic clusters, three sequences of $\sigma_{age}$ are displayed, corresponding to simulations with different number of stars brighter than the faint limit of the bundle (M$_G$=4.2): $\simeq$ 350 per cluster (similar to the open clusters), $\simeq$2000, such that the total number of stars in the seven clusters is similar to that in the GCNS CMD, and $\simeq$ 14000, in order to check whether a much larger number of stars leads to a substantially greater precision. In the case of the open clusters, two lines indicate the results of solutions with two different mother CMDs with different unresolved binary star characteristics: q01b03\_120M\_MG5 (dashed line and open symbols) and q01b07\_81M\_MG5 (solid line and filled symbols).

The conclusions that can be extracted from the left panel of Figure~\ref{precision_accuracy} are:
\begin{itemize}
\item The {\bf $\sigma_{age}$ decreases with the size of the age bins}. While this could be expected, the fact that even for the oldest clusters (12-10 Gyr old) the precision is better for smaller age bins is showing that, within this range of bin sizes, we are not hitting a {\it physical} limit imposed by the decreasing separation of the isochrones with increasing age, at least for this CMD with very small photometric and distance errors. 

\item The {\bf precision depends also on age}, with a 'break' between 6-4 Gyr: note that for a very similar age bin size, the age of the 4-2 Gyr old synthetic clusters is recovered with greater precision than that of the 6 Gyr old cluster. In Figure~\ref{CMD_synthetic} it can be seen that the separation of the synthetic clusters starts to decrease faster for the synthetic clusters older than 6 Gyr. Also, the shape of the isochrone changes around this age reflecting the transition from radiative to convective H-burning cores. For ages older than 6 Gyr, the bin size as a function of age is basically kept constant for each XL, L, M and S set. In this range, it can be observed that the age of the 6 Gyr cluster (green) is recovered with slightly better precision than the other older clusters, for which the sequences are quite mixed, indicating little dependence of age precision on age for ages older than 8 Gyr. 

\item The {\bf age precision shows little dependency with the number of stars in the population}. It is remarkable that even with only a few hundred stars, $Dir$SFH is able to determine the age as precisely as with 40x the number of stars. This is an important finding as it shows that this methodology can be used confidently to determine age distributions even for minority populations in the MW.

\item The {\bf age of the open clusters is determined as precisely as that of synthetic clusters of similar age, showing that the BaSTI-IAC models are able to match the data remarkably well}, and that the possible mismatch between the observed populations and those simulated in the mother CMD (as, for example, the characteristics of the binary star population or the IMF) do not affect substantially the recovered age precision.
In fact, in the case of the open clusters, we have recovered the SFH with two mother CMDs with different binary fraction (30\% and 70\%). For M67, the age precision is basically identical with the two binary fractions, while for the other three clusters, a better precision is reached with 70\% binaries. This may reflect different actual binary fractions in each individual cluster. However, except in the case of Pleiades with 30\% binary fraction, the age precision achieved is similar to that reached for the synthetic clusters for which there is a perfect match of the binary fraction between the mock and the mother. 

\item The gray horizontal lines indicate the locus of 10\% precision in age for ages 0.2, 2, 4 and 6 Gyr. The comparison with the sequences of the clusters indicates that, for ages 6-12 Gyr, {\bf even for the XL age bins, the age of the populations is recovered with precision better than 10\%}, a goal that is within the best expectations of Galactic Archaeology, even when including asteroseismology information \citep{Miglio2017_Plato}. \footnote{Note that, in this case, we are referring to the age of a {\it population of stars} rather than a single star.} {\bf For the S bins, the ages of the 2-12 Gyr old clusters are recovered with a precision of the order or better than 5\%}, (with better relative precision for older clusters in each group 2-4 and 6-12 Gyr old), while for the youngest clusters (both synthetic and open) the best resolution achieved is 10\%.
\end{itemize}

The right panel of Figure~\ref{precision_accuracy} displays the relative error of the age determination (a measure of the accuracy), as a function of age. In this case, only the synthetic clusters are displayed, as for them we can compare the recovered age with the mean input age. Clusters with different numbers of stars are depicted in different colours as indicated in the labels, while the age bins used for a particular measurement are represented by the corresponding letters. The $x$ position of the symbols has been shifted slightly for clarity. From this figure, we can conclude that ages are systematically overestimated by a maximum of 6\%. The only instances with a lower accuracy, up to 8\%, are in the case of the XL bins for the 10 Gyr cluster or, in the case with fewer stars, for the youngest cluster. The latter can be easily understood as an effect of the very small number of stars, which may result in an undersampled main sequence turnoff mimicking an older age. For the intermediate-age range (6-2 Gyr), the accuracy is better than 4\%. The figure also shows that neither the size of the bins, nor the number of stars in the population (except for very young ages) are systematically related to the accuracy of the age recovery. 

In Appendix~\ref{clusters} we present this study of the age accuracy and precision in more detail.

\subsubsection{The area in the CMD included in the fit and whether weights are provided within it} \label{sec4:bundle}

\citet{Ruiz-Lara2021LeoI} showed that different bundle strategies had little effect on the resulting SFH of the dwarf galaxy Leo~I. With $Dir$SFH, we consider a single bundle including the whole observed and mother CMD and we have verified that the resulting SFH has little dependence on its exact shape. As mentioned, we consider that the use of a single bundle removes subjectivity, improving the repeatability of the results, and maximizes the information used to compute the SFHs. We have paid special attention to test whether the faint magnitude limit of the bundle would affect the precision of the derived SFH. Inspection of the isochrones in Figure~\ref{description_catalogue} (lower, middle panel) indicates that the best age sensitivity can be expected along the main sequence down to the oldest main sequence turnoff and on the subgiant branch. Thus, in principle, it would be enough to sample the observed and mother CMD down to the magnitude of the oldest turnoff of the more metal rich population included in the mother CMD (as this is the faintest population), that is, M$_G$=4.2. However, it is reasonable to ask whether including a larger portion of the main sequence below the oMSTO could provide useful information and increase the accuracy or precision of the SFH derivation. To check this, we derived the SFH of the synthetic and observed clusters described in (i) using three bundles with different faint M$_G$ limit: 4.2, 4.7 and 5.2 and fitted 2D gaussians to the age-metallicity distribution of each cluster to determine the recovered age and metallicity and the corresponding $\sigma_{age}$ and $\sigma_Z$. No significant difference in the precision or accuracy of the derived ages is observed by changing the faint magnitude of the bundle.
In Appendix~\ref{testing_robustness}, we show a compilation of tests carried out with the goal of assessing the robustness of our SFH recovery. In particular, in Figure~\ref{amr_bundle} we show the SFH of the GCNS derived with the three faint bundle limits. It is clearly seen that the solutions are basically identical. This is an important finding as it implies that there is no gain, as far as the SFH is concerned, in sampling the CMD deeper than the oldest and more metal-rich subgiant branch involved, and thus, it will allow us to reach larger distances in the galaxy than if a fainter magnitude would be necessary. As discussed above, this result can be somewhat expected, as below the oMSTO, stars of different ages are mixed.

We have also tested if using different weights according to the variations in the range of stellar ages across the CMD would produce differences in the resulting SFH (see Sec.~\ref{sec3:FitSFH}, par. ii) and Figure~\ref{weights}). Figure~\ref{amr_wei_uni} compares two solutions calculated with and without weights across the CMD. It can be seen that also in this case the results are basically identical.

\subsubsection{Systematic differences in the {\it Gaia} DR3 magnitude scale and that of stellar evolution models} \label{sec4:shifts}

For each model in Table~\ref{mothers}, the corresponding best shift $(\delta c, \delta m)$ is calculated, and subtracted to each mother CMD before calculating the final SFH. These shifts are also specified in Table~\ref{mothers}. In principle, this shift should be a systematic difference for each stellar evolution library, given a set of bolometric corrections. However, different binary populations in the mother CMD can also lead to small differences in the shifts as they change the overall distribution of stars in colour. For this reason, we have computed the shifts for each mother CMD and listed them in the last column of Table~\ref{mothers}. It can be seen that, for a given binary population and library, the shifts are basically identical, particularly in the case of the BaSTI-IAC library, and they change slightly (for a maximum of 0.01 in magnitude and/or colour) for different binary populations. In general, it appears that models are bluer and fainter than the GCNS by $\simeq 0.03-0.04$ mag. 

\subsubsection{Parametrisation of the unresolved binary population} \label{sec4:binaryparam}

Unresolved binaries in which the component stars are both in the main sequence appear offset to brighter magnitudes and redder colours compared to the single stars main sequence ridge line. Equal-mass binary stars appear offset by -0.75 mag from the locus of a single star of the same mass, while extreme mass ratio binaries locate somewhat to the red and at a similar magnitude in the CMD compared to the most massive star in the pair. Binary stars with intermediate mass ratios populate a continuum of stars between these two extremes. The unresolved binary population can be observed in the GCNS CMD below the oMSTO as a parallel, less populated sequence, above and to the red of the main sequence of single stars. 

\begin{figure}[h!]
    \centering
    \includegraphics[width=0.5\textwidth]{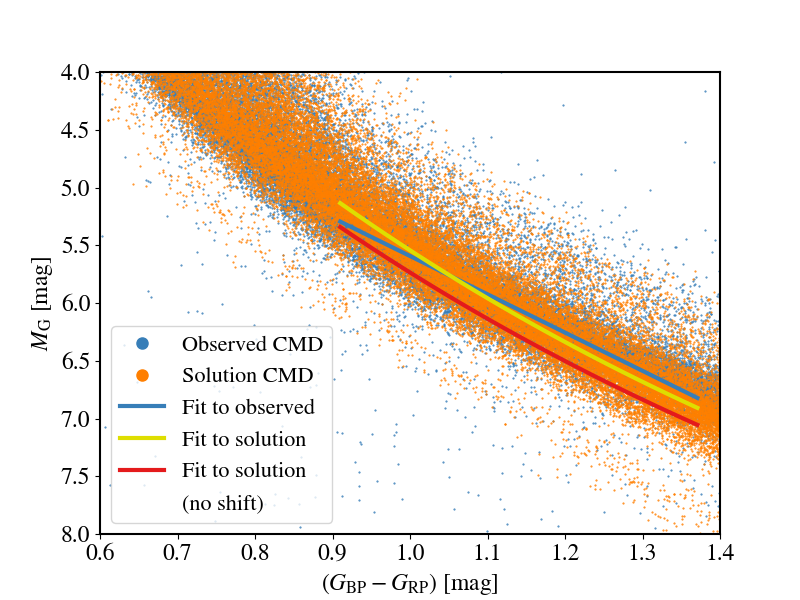}
    \caption{Zoomed region of the main sequence below the oMSTO. Blue dots: CMD of the GCNS. Orange dots: solution CMD obtained from the q01b03\_60M\_deep model. Blue and yellow lines: fit to the main sequence of the GCNS and solution CMD, respectively. Red line: fit to the solution CMD with no shift applied.}
    \label{pol_bin_region}
\end{figure}

Fig.~\ref{pol_bin_region} shows the main sequence of the observational data (blue dots) and the solution from the q01b03\_60M\_MG10 mother CMD (orange dots). The sequence of unresolved binaries can be seen both in the data and in the solution. The magnitude distribution of stars in the main sequence provides valuable information on the characteristics of the binary star population. Comparing the distribution in the observed CMD and that resulting in the best fit CMDs for different $\beta$, and q$_{min}$, (see Table~\ref{mothers}) will allow us to constrain the parameters that result in a good fit of the binary sequence. We will then adopt them to derive the final SFH of the GCNS. 

In order to trace the position of the main sequence main locus (which should approximately correspond to locus of the single stars) as a function of colour, we fitted a Gaussian mixture model (GMM) to the distribution of magnitude of the stars in colour bins of 0.01 mag. and, for each colour bin, we calculated the maximum of the M$_G$ distribution as the peak M$_G$ value of the dominant Gaussian component. We also calculated the colour average for each bin. Thereafter, we performed a polynomial fit to these magnitudes and colours. The blue and yellow line in Fig. \ref{pol_bin_region} represent these fits for the GCNS and of the solution CMD, respectively. The red line represents the fit to the solution CMD with no shift applied. It is reassuring that the shift inferred from the best fit to the bright part of the CMD (above the oMSTO) provides also a good match between the observed and solution main sequence below the oMSTO, which otherwise would be offset from each other (as the blue and red lines are).

We then parametrise the distribution of stars in the main sequence compared to its main locus through $\Delta$G=M$_{G, pol}$-M$_{G,*}$ \citep[see][]{gcns}, such that M$_{G,*}$ refers to the absolute G magnitude of each star and M$_{G, pol}$ is the absolute G magnitude interpolated in the polynomial fit for the colour of the star. We then compute the histogram of the $\Delta$G values (marginalising over the whole colour range).

\begin{figure}[h!]
    \centering
    \includegraphics[width=0.45\textwidth]{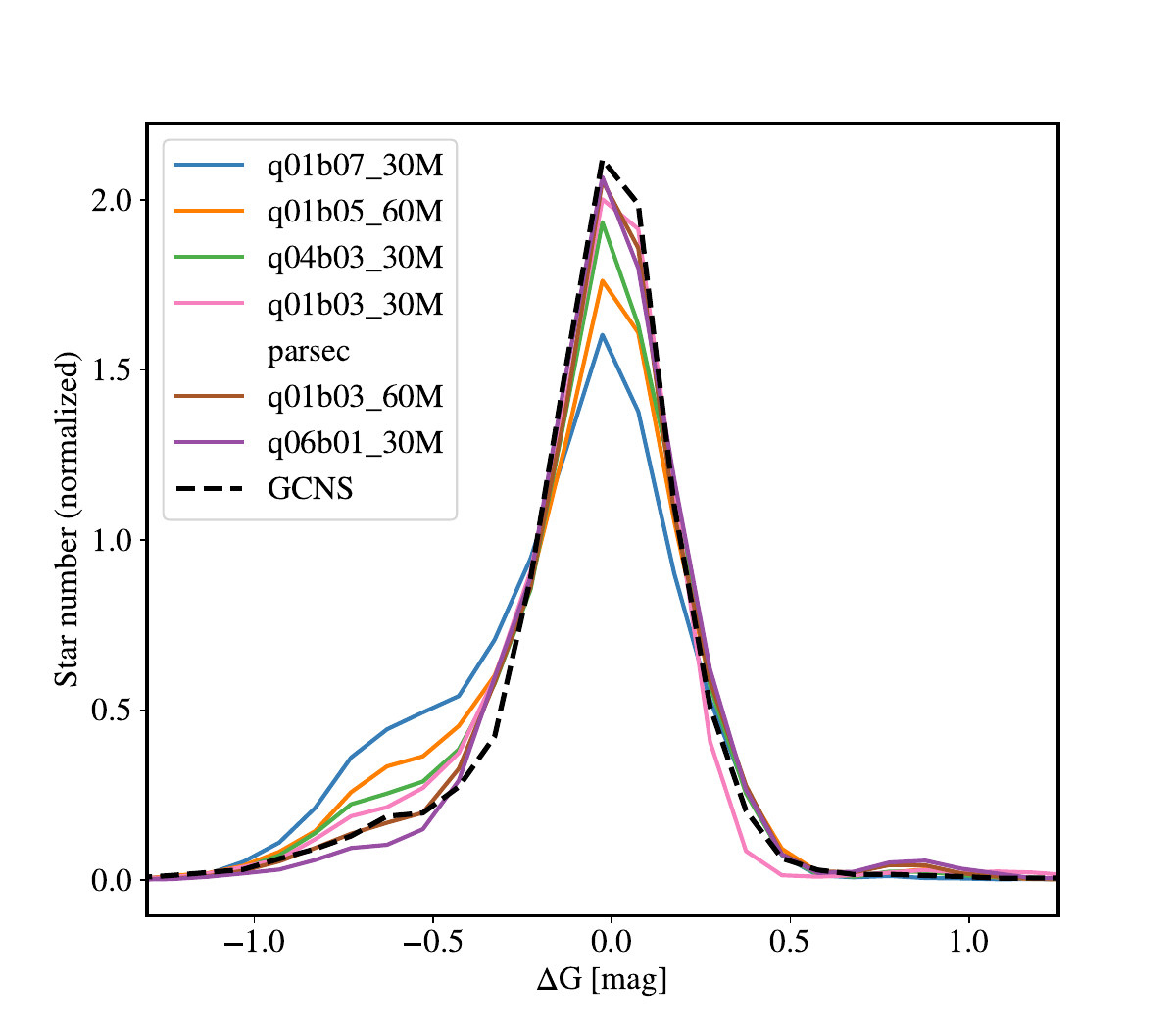}
    \caption{Black dashed line: $\Delta$G histogram of the GCNS. coloured lines: $\Delta$G histograms corresponding to solution CMDs with different $\beta$ and q$_{min}$. All histograms have been normalised to the number of stars in the colour range considered.}  
    \label{deltaG_all}
\end{figure}

Figure~\ref{deltaG_all} depicts, in different colours, the $\Delta$G normalized histograms derived from the solution CMDs for the SFHs derived from the 'deep' mother CMDs (those with M$_{G, max}$=10-11) listed in Table~\ref{mothers}, compared to that of the GCNS (in black, dashed line). Note that all distributions are centred in zero, by definition, and have a main component corresponding to the main sequence of single stars. They also have a secondary bump on the left side of the main component, centred at approximately $\Delta$G = -0.75, which corresponds to the unresolved binary population. Its height depends on the binary fraction $\beta$ and on q$_{min}$, with more prominent bumps for larger $\beta$ in which the components have similar mass (larger q$_{min}$). 

The observed distribution is closely matched by that corresponding to the q01b03\_60M\_MG10 solution while the distribution that differs the most is that of the q01b07\_30M\_MG11 model. The latter contains the largest fraction of binaries among our tests, resulting in a stronger binary bump.  The solutions from q04b03\_30M\_MG11 and q01b05\_60M\_MG10, with intermediate fractions of binaries lie in between\footnote{q04b03\_30M\_MG11 has the same binary fraction as q01b03\_60\_MG10, but the larger q$_{min}$ implies a larger fraction of binaries with similar mass, which are the ones that effectively contribute to the bump.}. The PARSEC model, q01b03\_30Mparsec\_MG10 has a slightly larger bump compared to the equivalent model from BaSTI-IAC and to the observed distribution. It also differs from the observed distribution and from that of the BaSTI-IAC models by the shape of the faint part of the main sequence, with a more abrupt fall to the zero value. Finally, all solutions show a few stars towards positive values of $\Delta$G, which are associated to the presence of synthetic metal-poor stars lying under the main sequence (see Fig.~\ref{pol_bin_region}). The fact that this feature is missing in the observed  GCNS CMD may indicate an underlying subtle mismatch between the empirical data and the theoretical models. In any case, this only affects a minority of stars, so it won't have a substantial effect on the conclusions. Some observed stars are in this region, but they are much more scattered in colour. 

Taking into account these results, we will adopt $\beta$=0.3 with q$_{min}$=0.1 for the final SFH of the GCNS. This value is basically compatible with the results by \citet[][]{2020MNRAS.496.1922B}, who  studied the problem of unresolved binaries in Gaia DR2 data based on the renormalised unit weight error (RUWE) parameter from the Gaia catalogue. They exploit the fact that in the case of stars belonging to an unresolved system, the motions of the centre of light and mass are decoupled, and thus, assuming a single-source for the astrometric model fails. They found that, using this ruwe parameter, they can identify such unresolved systems. As part of their analysis, they study how the binary fraction evolves across the CMD. From their Figure 9, we can see how the average unresolved binary fraction, in the region of the CMD that we analyse, ranges from $\sim$~12$\%$ (faint main sequence) to $\sim$~50$\%$ (bright main sequence), with average values compatible with the 30$\%$ we are finding. A similar result is found by \citet{Penoyre2022_binGCNS} for the GCNS.

\section{Results: the SFH derived from the GCNS.} \label{results}

In this section, we will discuss the deSFH and current age and metallicity distributions of the stars within $100 \, \rm{pc}$ of the Sun. We will focus on the solution obtained with the q01b03\_120M\_MG5 mother CMD, which is the largest synthetic CMD we have computed with the solar-scaled BaSTI-IAC stellar evolution models (120 million stars with M$_G\le 5$), adopting $\beta$=0.3, q$_{min}$=0.1 and a Kroupa IMF \citep{kroupa1993}. We will use a bundle with faint limit M$_G$=4.2, weight of each CMD 'pixel' calculated as the inverse of the variance of the stellar ages in that 'pixel' (see Section~\ref{sec3:FitSFH}), S age bins and 0.1\,dex metallicity bins. A shift of ($\delta c, \delta m$) = (-0.035, 0.04) has been subtracted to the colour and magnitude of the stars in the mother CMD (see Table~\ref{mothers}).

\begin{figure*}[h!]
    \centering    
    \includegraphics[width=1\textwidth]{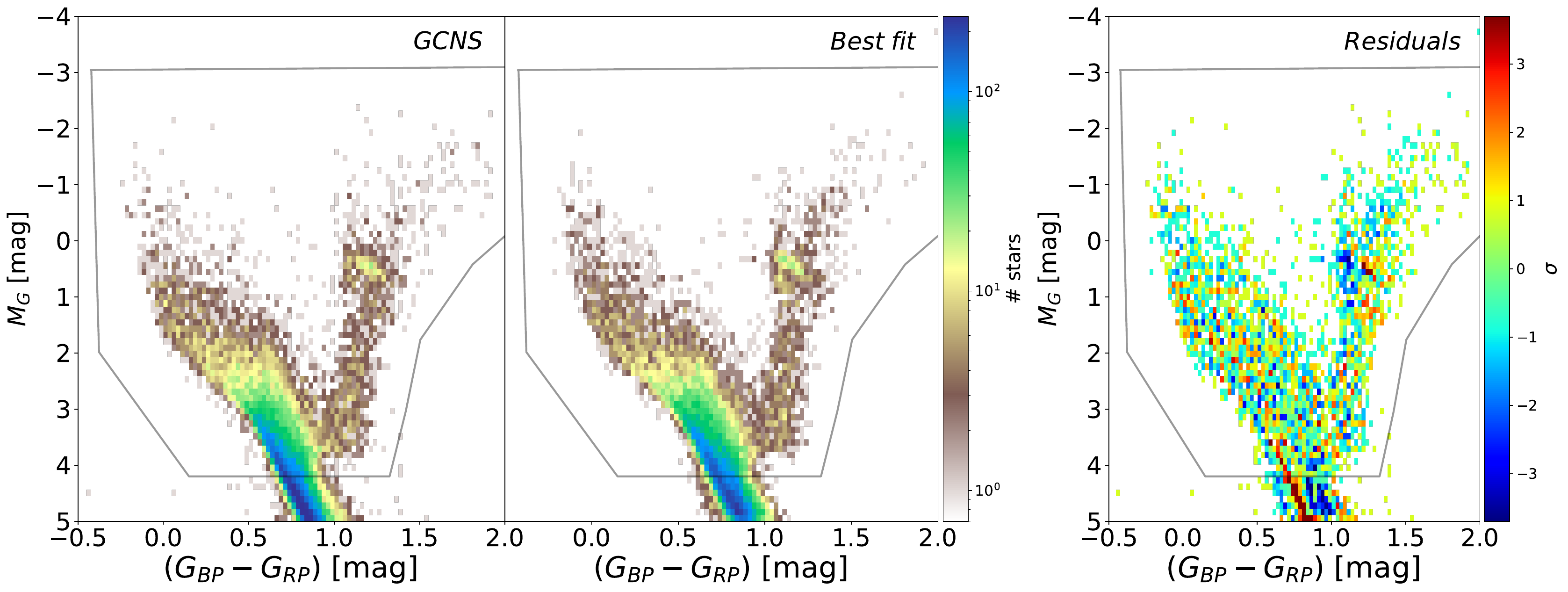}
    \caption{Observed (left) and solution (middle) CMDs of the GCNS. The right panel displays the residuals of the CMD fitting (in $\sigma$ units assuming Poisson errors). The bundle encompassing the stars included in the fit is also plotted in all panels.}  
    \label{cmd_sol_res}
\end{figure*}

Figure~\ref{cmd_sol_res} displays (left and middle panels) the observed and the solution CMD of the GCNS, with the bundle including the stars that have been used for the fit superimposed. The bundle is significantly larger than the area covered by the observed CMD since it has to include the whole mother CMD. Since the latter has a much larger metallicity range than the observed and solution CMD, it covers a larger range in colour (see Figure~\ref{weights}). The right-hand panel shows the residuals of the fit. Note the high quality of the fit, with no significant trends or structures in the residuals, which in most 'pixels' in the CMD are within $\pm 1 \sigma$ with deviations up to $\pm$ 3$\sigma$ in only a few pixels.

\subsection{The deSFH of the solar neighbourhood within 100 pc of the Sun.}  

\begin{figure*}[h!]
    \centering    
    \includegraphics[width=1\textwidth]{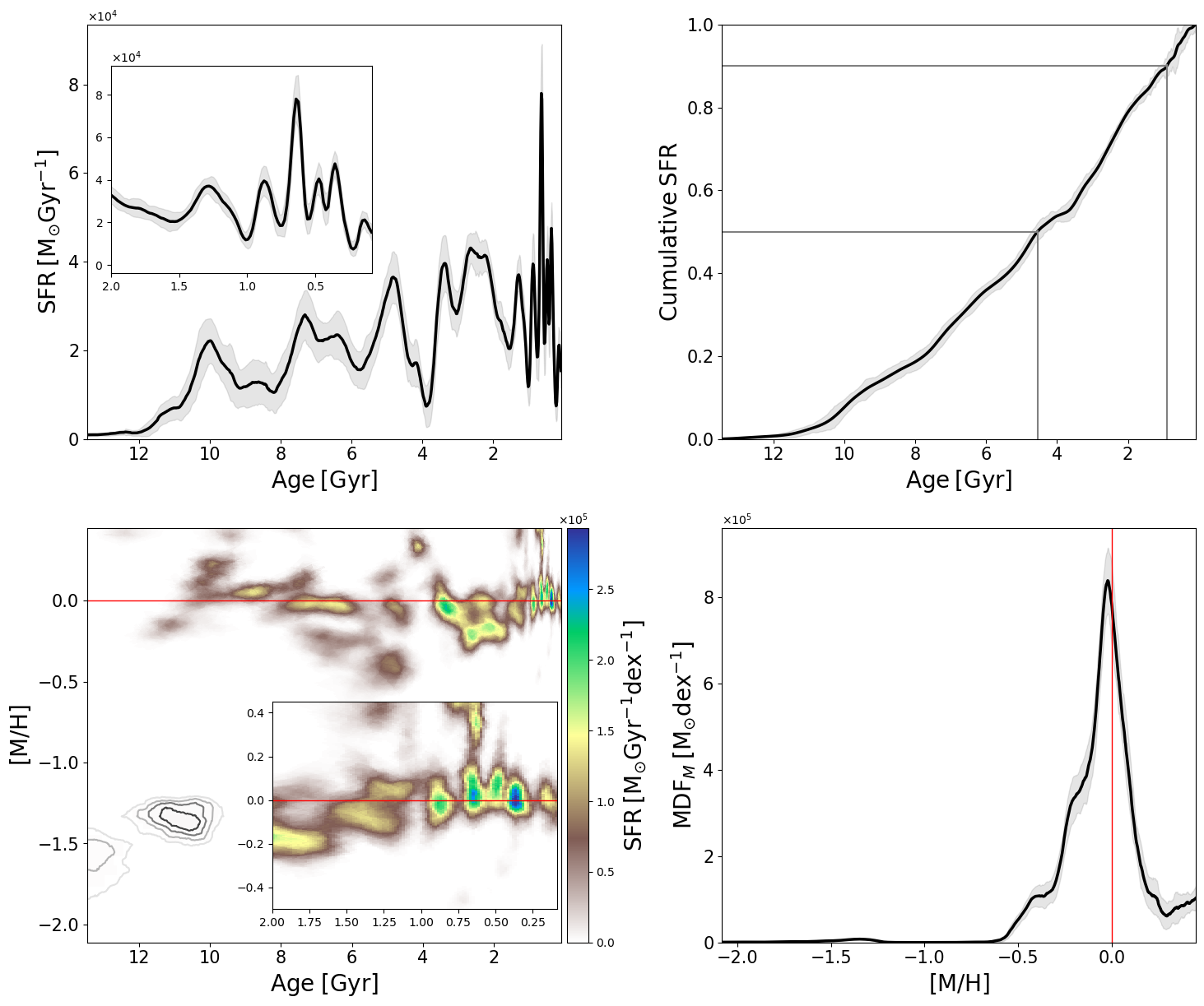}
    \caption{deSFH of the stars within $100 \, \rm{pc}$ of the Sun, derived with the BaSTI-IAC stellar evolution library. Bottom left: Age-metallicity distribution of the mass transformed into stars (astrated mass) as a function of time and [M/H]. The iso-contours indicate the presence of a small amount of star formation at old age and low metallicity. Upper left: star formation rate as a function of  time, SFR(t), integrated over metallicity. In the two left panels, the inset shows and expanded view of the last 2 Gyr.  Lower right: MDF$_M$. Upper right: cumulative distribution of the astrated mass as a function of time. Two lines indicate the 50 and 90 percentiles. In the two lower panels, the red line indicates solar metallicity, [M/H]=0. In all cases the direction of the $x$ axis is such that old age/low metallicity is on the left.}  
    \label{SFH_final}
\end{figure*}

Figure~\ref{SFH_final} displays what we consider our best calculation of the deSFH of the stars currently within 100 pc of the Sun. The bottom left panel shows the age-metallicity distribution of the mass transformed into stars as a function of lookback time (age) and metallicity, [M/H]. Old ages are on the left. The colour bar indicates the star formation rate in units of M$_\sun$\,Gyr$^{-1}$\,dex$^{-1}$. The upper left panel displays the deSFR(t), which is the marginalization over metallicity of the deSFH, in units of M$_\sun$ Gyr$^{-1}$. The lower right panel displays the metallicity distribution function of the mass transformed into stars (MDF$_M$), in units of M$_\sun$ dex$^{-1}$. Finally, the upper right panel shows the cumulative distribution of the mass transformed into stars as a function of lookback time. The lines indicate the 50 and 90 percentiles.

\begin{figure*}[h!]
    \centering
    \includegraphics[width=1\textwidth]{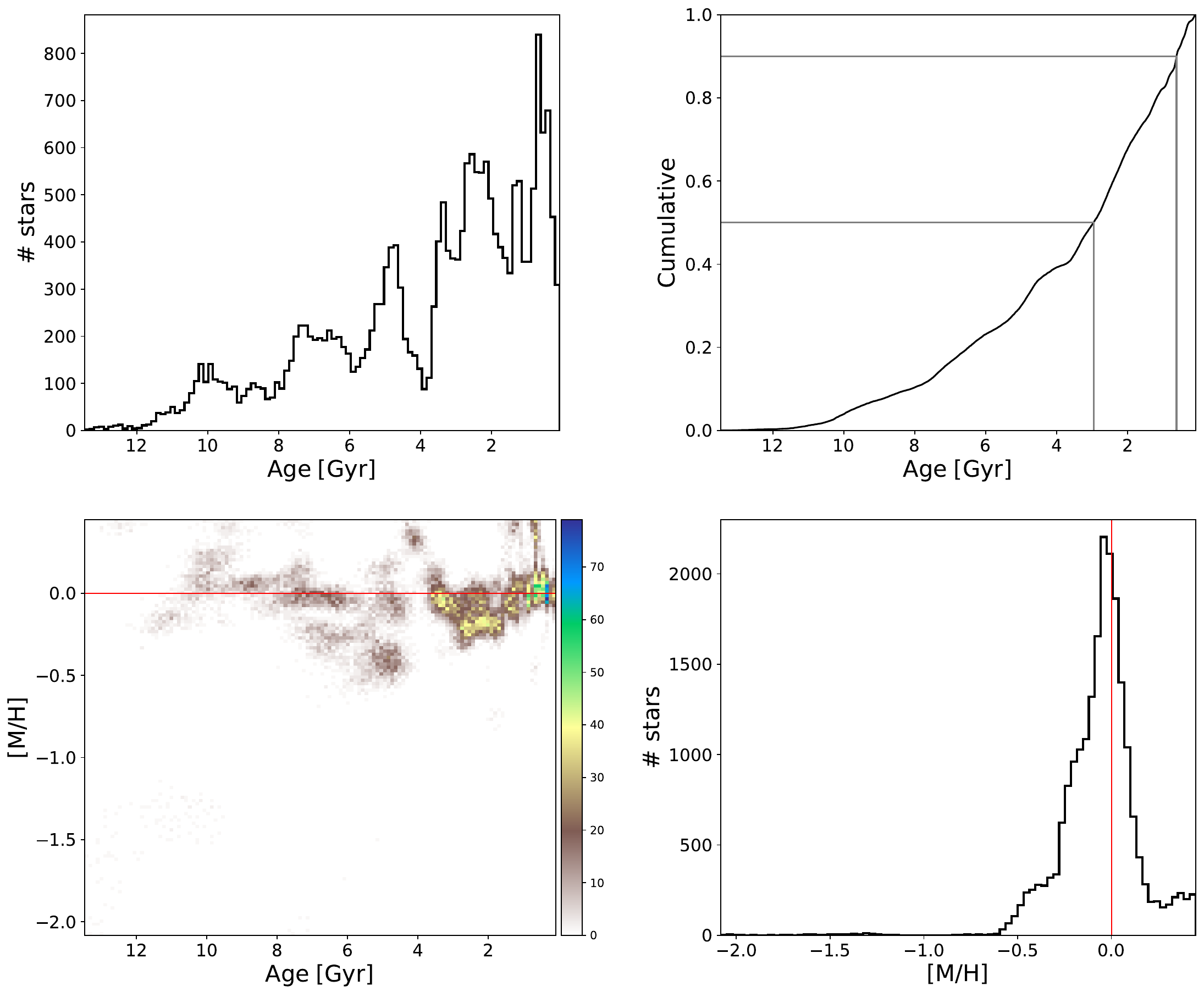}
    \caption{Number of stars with M$_G < 4.2$, as derived from the BaSTI-IAC stellar evolution library. Bottom left: Age-metallicity distribution of alive stars within 100 pc of the Sun. Upper left: number of stars as a function of their age. Lower right panel: MDF$_S$. Upper right: fraction of stars as a function of age. Two lines indicate the 50 and 90 percentiles. In the two lower panels, the red line indicates solar metallicity [M/H]=0. The number are calculated for stars with M$_G < 4.2$.} 
    \label{stellar_final}
\end{figure*}

Figure~\ref{stellar_final} displays an alternative view of the characteristics of the stellar populations in the solar neighbourhood, in terms of the number of stars currently present as a function of their age and metallicity\footnote{In this figure, the number of stars is that {\it inside the bundle}, that is, stars with M$_G < 4.2$. This number could be extrapolated to include the stars with M$_G > 4.2$ using the IMF.}. The bottom left panel shows the distribution in age and metallicity of the stars currently alive. The upper left panel displays the number of stars as a function of their age while the lower right panel presents the stellar metallicity distribution function (MDF$_S$). Finally, the upper right panel shows the fraction of stars as a function of their age. 

Note that the features in Figures~\ref{SFH_final} and \ref{stellar_final} are very similar, the main difference being their relative strength at young and old ages, since a fraction of the mass that has been transformed into stars at any age is not anymore in the form of currently alive stars, and this fraction varies with time. For example, the peak of star formation that can be observed 3 Gyr ago in the SFR(t) plot (Figure \ref{SFH_final}, upper left panel) has approximately twice the intensity of the peak occurred 10 Gyr ago, while the ratio of number of stars is approximately 4:1 (see Figure \ref{stellar_final}, upper left panel).

The most detailed and rich view of the history of the stellar mass in the solar neighbourhood is provided by the panels showing the age-metallicity distribution in Figures~\ref{SFH_final}~and~\ref{stellar_final} (bottom left panels). We remind the reader that, in the case of the deSFH, this is the mass (per unit time and metallicity) that has been transformed into stars {\it somewhere in the Galaxy} to account for the stars that are today in the studied volume. The corresponding panel in Figure~\ref{stellar_final} describes the main features of the stellar populations currently located in the solar neighbourhood. These age-metallicity distributions can be described as follows:

-We measure an age of $\simeq$ 11 Gyr and a metallicity of ${\rm[M/H]}=-0.16$ for the oldest stars. Taking into account the systematic difference between input and recovered ages found in the tests with synthetic clusters (see Figure~\ref{precision_accuracy}), this age could be reduced to about 10.4 Gyr. Thereafter, a sequence of progressively younger and more metal rich stars is observed, up to age $\simeq$ 9.7 Gyr (or 9.2 Gyr considering the possible systematics) and ${\rm[M/H]}=0.25$. 

-Between $\simeq 9.5$ and 6 Gyr ago, two main metallicity sequences, somewhat disjoint in time, exist: at an earlier time, stars have metallicity slightly above solar, while after $\simeq$ 8 Gyr ago, the main population has solar metallicity, and a less prominent population with ${\rm[M/H]}=-0.25$) can be observed. 

-Six Gyr ago a dip in the SFR(t) can be observed, followed by a large scatter in metallicity, with three main populations: one with [M/H]$=-0.4$, another at solar metallicity, and a third one with a very narrow age range (slightly older than 4 Gyr) and super-solar metallicity. The latter one coincides with a star formation gap at the expected solar metallicity, and is followed by a conspicuous break of star formation (see the upper left panel displaying the deSFR(t)). 

-Since 4 Gyr ago, the star formation proceeds at an average higher rate until the present time, even though still showing a bursty behaviour. Between 4 and 2 Gyr ago, populations at different metallicities, solar and below solar (${\rm [M/H]} \simeq -0.2$) coexist. Finally, since 2 Gyr ago, the majority of the stars have solar metallicity, with a small fraction of super-solar stars. 

-Almost no star formation is detected with [M/H]$\lesssim-0.5$, even though the whole metallicity range between [M/H]$=-2.2$ and [M/H]=0.45 was present in the mother CMD used to determine the SFH, and no {\it a priori} chemical evolution law was imposed. Only a very small amount of star formation is detected with ages between 12 and 10 Gyr and metallicity around [M/H]$\simeq-1.5$ (it is barely visible in Figures~\ref{SFH_final}, see contours, and~\ref{stellar_final}). It translates in around one hundred stars in this range of age and metallicity which can be only hinted in the left lower panel of Figure \ref{stellar_final}. We will discuss this population in Section~\ref{discussion}. This paucity of low metallicity stars is not surprising since the GCNS sample is primarily composed of thin disk stars: more than 90\% of the stars that we can classify based on kinematics are found to belong to the thin disk; see Figure~\ref{description_catalogue} and Appendix~\ref{app:GDR3}.

The right lower panels of Figures~\ref{SFH_final}~and~\ref{stellar_final} show that the metallicity distribution function peaks very close to solar metallicity with the great majority of the stars having [M/H] in the range 0.2 to -0.3, a minority population down to metallicity [M/H]$\simeq$ -0.5, and an even less relevant population up to [M/H]$\simeq$0.4. The very small old metal poor population that we mentioned earlier can be noticed as a small peak around ${\rm [M/H]}\simeq -1.3$

Finally the upper right panels of Figures~\ref{SFH_final}~and~\ref{stellar_final} show the fraction of the total mass transformed into stars as a function of age, and the fraction of stars older than a given age, respectively. The lines indicate the 50 and 90 percentiles: it can be seen that 50\% and 90\% of the stellar mass was reached 4.5 and 0.9 Gyr ago, respectively, while 50\% and 90\% of the stars are older than 3 and 0.6 Gyr. These numbers picture an average young solar neighbourhood.

\subsection{Robustness of the solution} \label{robustness}

\begin{figure*}[h!]
    \centering    
    \includegraphics[width=1\textwidth]{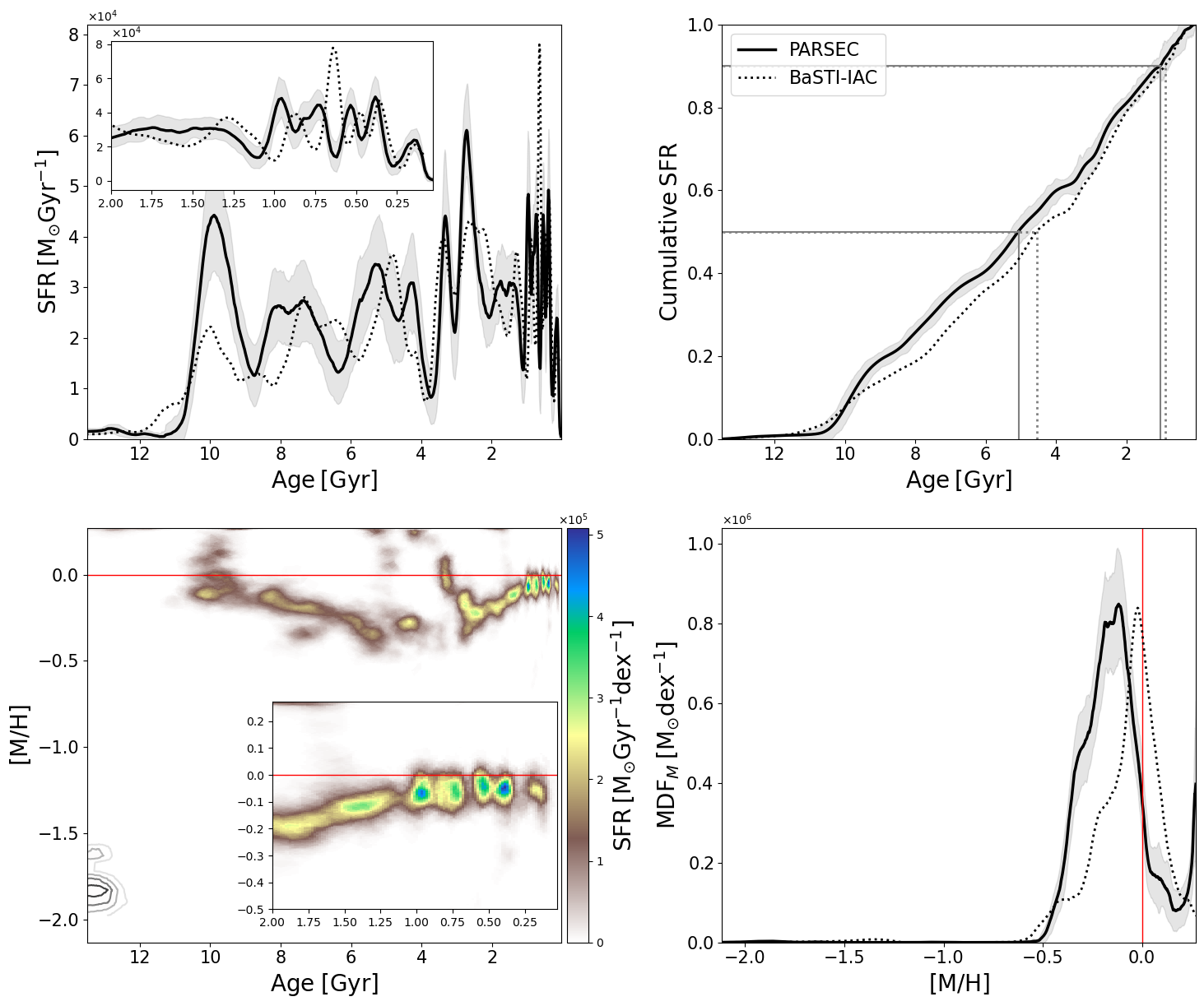}
    \caption{deSFH of the stars within 100 pc of the Sun derived with the PARSEC synthetic CMD. Bottom left: Age-metallicity distribution of the astrated mass as a function of time and [M/H]. The inset shows an expanded view of the last 2 Gyr. The iso-contours indicate the presence of a small amount of star formation at old age and low metallicity. Upper left: star formation rate as a function of  time, SFR(t), integrated over metallicity. Lower right: MDF$_M$. Upper right: cumulative distribution of the astrated mass as a function of time. Two lines indicate the 50 and 90 percentiles. In the two lower panels, the red line indicates solar metallicity [M/H]=0. In all cases the direction of the $x$ axis is such that old age/low metallicity is on the left.}  
    \label{SFH_final_parsec}
\end{figure*}

In Section~\ref{sec4:deriving} we have discussed a number of choices that have been tested, regarding both parameters adopted to calculate the mother CMDs and the input parameters of the fit, in order to determine their impact on the final SFH. These parameters are: the number of stars in the mother CMD, the size of the age bins, the faint magnitude limit of the bundle, the weights across the CMD, and the unresolved binary stars parameterisation. Additionally, a mother CMD has been computed with the PARSEC stellar evolution models. SFHs determined with different choices in each one of these aspects are shown in a number of figures in Appendix~\ref{testing_robustness}. Here, for its relevance, in Figure~\ref{SFH_final_parsec} we show the solution obtained with the PARSEC stellar evolution models.

Concentrating first in the various solutions obtained with the BaSTI-IAC stellar evolution library, we see that the main conclusions regarding the SFH of the GCNS do not change with the different parametrisations which, from oldest to youngest ages, can be summarised as follows (see Figure~\ref{SFH_final}: i) the oldest substantial\footnote{Very minor older populations are also found, with low and high metallicity, [M/H]$\simeq -1.5$, and [M/H]$\simeq$0.3.} stellar populations are measured as $\simeq$11.5 Gyr old; ii) there are several episodes of star formation with decreasing age and increasing metallicity in the subsequent $\simeq$ 2 Gyr, until 9-9.5 Gyr ago; iii) there are hints of splitting of the age-metallicity distribution between 10 and 6 Gyr ago, with a large scatter and different episodes of star formation with different metallicities around $\simeq$5 Gyr ago; iv) there is a sudden decrease in the star formation activity around 4 Gyr followed by an increased average star formation rate, which lasts until the present time; v) the star formation activity is bursty at all ages. 

Most of these features hold in the solution with the PARSEC models, which we show in  Figure~\ref{SFH_final_parsec}, where the BaSTI-IAC solution has been superimposed in thin dashed lines in the panels displaying the deSFR(t), the MDF$_M$, and the cumulative distribution of the mass transformed into stars.  The main differences are that the overall population is retrieved more metal poor with PARSEC (see the MDF$_M$ in the lower right panel of Figure~\ref{SFH_final_parsec}), and that the metallicity range for each given age is narrower, with less obvious metallicity splits. However, the range of metallicity around 10 Gyr ago is also present, as well as an split around 8 Gyr ago. A feature that is hinted in the BaSTI-IAC solutions and is more evident in the PARSEC one is a somewhat decreasing metallicity trend since the oldest age ($\simeq 11 $ Gyr ago) until 4 Gyr ago. At this age, after the same break in the SFH, the metallicity starts to increase to reach solar metallicity at the present time. Comparing the deSFR(t) in the upper left panel of Figure~\ref{SFH_final_parsec}, it can be noticed that the same episodes of star formation are present in both cases, even though the relative importance of some of them may be somewhat different, and age shifts are observed, specially between 9 and 4 Gyr ago. It can also be noticed that no significant star formation before 11 Gyr ago is measured with the PARSEC models, while star formation starts to rise 12 Gyr ago in the BaSTI-IAC solution, with metallicity slightly below solar.  Finally, the shape of the cumulative SFR after 10 Gyr ago indicates a slightly more constant average SFR in the case of the solution based on the PARSEC library, with 50\% and 90\% of the stellar mass reached 5 and 1 Gyr ago - as opposed to 4.5 and 0.9 Gyr for the case of the BaSTI-IAC model library.  

Going back to the comparison of the different SFHs obtained with the BaSTI-IAC models, some differences in the solutions can be associated with changes in some of the parameters: a larger number of stars in the mother CMD (Figure~\ref{amr_mother_nstars}) or smaller age bins (Figure~\ref{amr_tam_bins}) result in sharper features in the SFH. The effect of the age bin size is most evident. For example, while there are hints of the split in the age-metallicity distribution at intermediate-ages regardless of the size of the bins, they are more clearly visible with the smallest (S) bins. The reality of this latter feature, which appears in most of the solutions with different parameters, deserves further attention. In Section~\ref{sfh_clusters} we have discussed an experiment designed to check whether this metallicity split may be real: we derived the SFH for a composite CMD of 14 synthetic clusters of 7 different ages and two metallicities separated by a small metallicity gap (0.15 dex). Figure~\ref{double_metallicity} displays the age-metallicity distribution recovered from the CMD of this composite population. It can be seen that the two metallicities are best separated for ages of 8 Gyr or older, where the metallicity is more precisely recovered (see also Figure~\ref{metallicity_precision}), and of course, they are best separated also for the smallest (S) age bins. For younger ages, the existence of the two metallicities is still hinted in the cluster's age-metallicity distribution, but in a less obvious way. We conclude, thus, that the different metallicity sequences seen in the SFH of the GCNS are real, at least for stars older than 2 Gyr, age when the metallicity distribution gets narrower.  

\subsection{Comparison with spectroscopic metallicity distribution functions.} \label{Comparing_MDF}

The comparison of the MDF$_S$ resulting from the SFH calculation with that obtained from spectroscopy can provide a totally independent, external check of our results, as no assumptions on the age-metallicity distribution or the metallicity distribution function have been made in the SFH derivation.

\begin{figure*}[h!]
    \centering
    \includegraphics[width=1\textwidth]{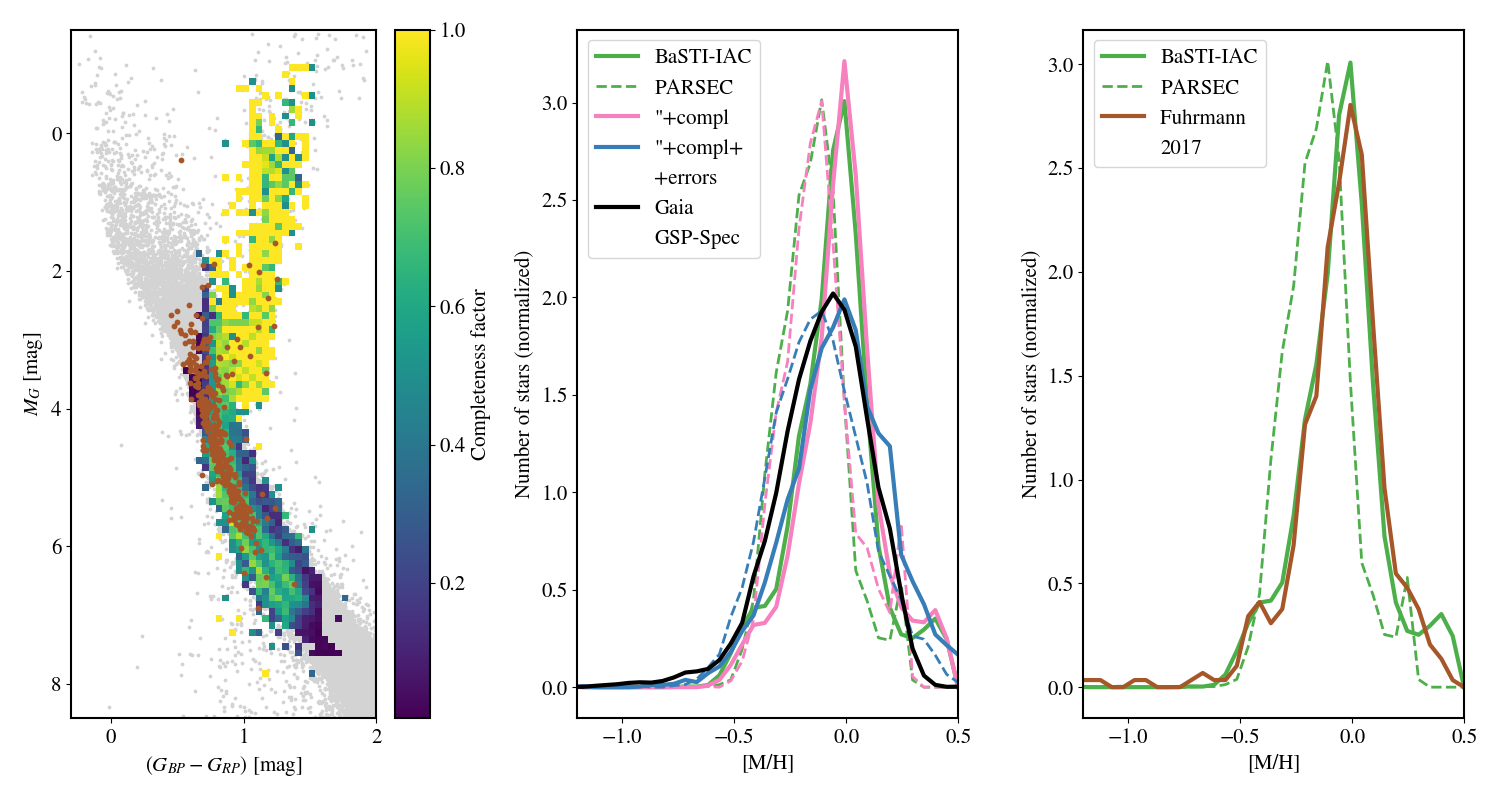}
    \caption{Comparison of the MDF$_S$ derived from CMD-fitting (both with the BaSTI-IAC and PARSEC stellar evolution models, with two spectroscopic MDF. Left panel: completeness in the CMD of the metallicity measurements from $Gaia$ GSP-Spec (blue-yellow histogram) and CMD of the stars measured by \citet[]{Fuhrmann2017_LocalStarPops}, in brown, superimposed on the CMD of the GCNS, in grey. Middle panel: MDF obtained from $Gaia$ GSP-Spec metallicities \citep{RecioBlanco2023_GaiaDR3} for stars in the GCNS. Right panel: MDF derived from \citet{Fuhrmann2017_LocalStarPops} stars within 25~pc of the Sun. }
    \label{MDF_comparison}
\end{figure*}

Figure~\ref{MDF_comparison} shows the comparison of our derived MDF$_S$ with two spectroscopic MDF: that for the metallicities derived by $Gaia$ GSP-Spec for the same volume (middle panel; see Section \ref{data}), and that for the \citet{Fuhrmann2017_LocalStarPops} volume complete sample of stars located within 25 pc of the Sun (right panel) and with T$_{eff} \geq 5300$ K. We show both the MDF$_S$ from our preferred deSFH obtained from the BaSTI-IAC mother CMD displayed in Figure \ref{SFH_final} (solid coloured lines) and that obtained from the PARSEC mother CMD, displayed in Figure \ref{SFH_final_parsec} (dashed coloured lines). 

For the comparison with the {\it Gaia} GSP-Spec sample, we took into account the completeness and typical errors of the spectroscopic metallicity measurements within the GCNS volume. The completeness is represented in the left panel of Figure~\ref{MDF_comparison}, indicated as the number of stars with spectroscopic [Fe/H] divided by the total number of stars in the GCNS, in pixels of ($\delta$ col, $\delta$ mag)= (0.05, 0.1) mag across the CMD. The GCNS CMD is shown in the background as grey dots. It can be seen that there is a sharp cut at $G_{BP}-G_{RP} \simeq 0.75$ corresponding to the T$_{\rm {eff}}=8000$ limit of GSP-Spec \citep{Creevey2023_GaiaDR3, RecioBlanco2023_GaiaDR3}. This cut implies that main sequence stars younger than $\simeq$2 Gyr are not represented in the {\it Gaia} GSP-Spec MDF (MDF$_{Gaia}$ thereafter). The figure also shows that RGB stars have {\it Gaia} metallicity determinations with high completeness, while the completeness is around 50\% for low main sequence stars. Given that, according to our derived age-metallicity distributions, the metallicity range of stars younger than $\simeq$ 2 Gyr is within the limits of the general metallicity range, and that stars up to $\simeq$ 1 Gyr old are present in the RGB, we do not expect that the incompleteness of the spectroscopic metallicity measurements across the CMD will significantly affect the shape of the MDF$_{Gaia}$. In any case, for consistency, we simulated the Gaia GSP-Spec incompleteness and errors as explained below.

In the middle panel of Figure~\ref{MDF_comparison} the MDF$_{Gaia}$ (black line) is compared with our MDF$_S$. The curves are histograms of the number of stars per 0.05 dex bins, normalised to the total number of stars. The green lines corresponds to the metallicity distribution of the whole solution CMD, while the pink lines are the same MDF$_S$  after completeness correction, MDF$_S^C$. To perform this correction, we randomly selected, for each CMD pixel, a fraction of stars from the solution according to the completeness. As expected, the two lines are basically identical, confirming that the metallicity distribution across the CMD is basically uniform. Note that the peak of the MDF$_S$ and the MDF$_{Gaia}$ occur basically at the same metallicity in the case of the BaSTI-IAC solution (solid line), indicating that the metallicity scale of our solution from these models is the same as the GSP-Spec metallicity scale\footnote{Slightly different $(\delta c, \delta m)$ shifts would result in a slightly different peak of the age-metallicity distribution, but it is reassuring that our best estimate of this shift provides a solution that leads to a remarkable agreement with the spectroscopic MDF$_{Gaia}$}. This is not the case for the PARSEC models (dashed line): in this case, the peak metallicity of the solution MDF$_S$ is more metal poor than that of MDF$_{Gaia}$.  The difference in shape of the distributions can be attributed to the different dispersion of the metallicity measurements. In the case of the SFH, the average standard deviation of the metallicity derived from the open clusters (for the S age bins) is $\sigma_{[M/H]}^{OC} \simeq$0.06 dex. The standard deviation of the differences between the GSP-Spec metallicity measurements, based on Table D.1. in \citet{RecioBlanco2023_GaiaDR3}, is around 0.1 - 0.12 dex. In order to simulate this different dispersion in the MDF$_S$, we used a more conservative value of 0.15 dex, that we call spectroscopic sigma. We then add a normal distribution of noise to the metallicity values of the solution, with a standard deviation equal to the square root of the quadratic difference of the spectroscopic sigma minus $\sigma_{[M/H]}^{OC}$. The blue lines are the MDF$_S^C$ after adding this extra dispersion (MDF$_{S,d}^C$). Note that the width of MDF$_{S,d}^C$ are now practically identical to that of MDF$_{Gaia}$. 

The right panel of Figure~\ref{MDF_comparison} displays the comparison of our MDF$_S$ (from BaSTI-IAC, solid green line, and PARSEC, dashed green line) with the MDF obtained by \citet{Fuhrmann2017_LocalStarPops}. It is based in high-resolution spectroscopic data obtained with the FOCES \citep{Pfeiffer1998_FOCES} and BESO \citep{Steiner2006_BESO} échelle spectrographs for the northern and southern samples, respectively. The fainter sources were observed with the FEROS échelle spectrograph \citep{Kaufer1999_FEROS}. This catalogue provides very high quality [Fe/H] and [Mg/Fe] determinations, with typical errors of 0.08 dex in [Fe/H]. Since these errors are similar to our estimated errors in the metallicity determination of the clusters, we do not apply any extra scatter to our MDF$_S$. However, since our metallicity values represent the global metallicity of the stars, [M/H], we have transformed the [Fe/H] values into [M/H] using [Mg/Fe] as a proxy for [$\alpha$/Fe], and following the relation 

\begin{equation}
    \mathrm{[M/H]} = \mathrm{[Fe/H]} + \log\left(0.694\times10^{\mathrm{[\alpha/Fe]}}+0.306\right) \label{1}
\end{equation}

from \citet{Salaris1993_feh_to_mh}, with coefficients slightly different from the original reference in order to take into account the reference solar mixture adopted in the BaSTI-IAC models, based on the heavy element distribution provided by \cite{Caffau2011SolarAbundances}. The agreement between \citet{Fuhrmann2017_LocalStarPops} MDF and our MDF$_S$ derived from the BaSTI-IAC solution is outstanding: the peak metallicity and the width and shape of the distributions are basically identical, including its asymmetry slightly biased to lower metallicities. Additionally, both MDF show a small peak at [M/H] $\simeq -0.4$, which was erased when degrading our metallicity errors to match those in the MDF$_{Gaia}$. In \citet{Fuhrmann2017_LocalStarPops}, most stars with this metallicity belong to the low [Mg/Fe] branch, with ${\rm [Mg/Fe]} \simeq -0.15$, and are consequently classified as Pop I stars. In our solutions, the [M/H]$\simeq -0.4$ stars have ages around 5 Gyr \footnote{See also the position of the 5 Gyr old, [M/H]=-0.5 isochrone compared to the thin and thick disks CMDs in Figure~\ref{description_catalogue}}. This striking agreement of our MDF$_S$ with a high quality spectroscopic MDF supports that the interesting low-metallicity feature in the age-metallicity distribution at intermediate ages (see Figures~\ref{SFH_final}~and~\ref{stellar_final}) and the many structures and splits in our age-metallicity distributions are real, including possibly the bump at the high metallicity end, [M/H]$\geq 0.2$, which is present also in the \citet{Fuhrmann2017_LocalStarPops} MDF. Even though the shape is different from the high metallicity feature in our MDF$_S$, the fraction of stars encompassed in this feature is similar, and thus, this also supports that the high metallicity features that can be observed in our solution may be real. The most striking of those features is the one at age $\simeq$ 4 Gyr and [M/H]=0.4.  

The agreement between MDF$_S$ obtained from the PARSEC solution and the \citet{Fuhrmann2017_LocalStarPops} MDF is clearly worse: the peak metallicity does not coincide and, more importantly, the width and the fine structure of the MDF are in worse agreement. This reinforces our choice of the BaSTI-IAC library as the reference stellar evolution library for this work. 

\subsection{Comparison with literature age-metalliicty distributions}\label{otherAMRs}

\begin{figure}[h!]
    \centering
    \includegraphics[width=0.3\textwidth]{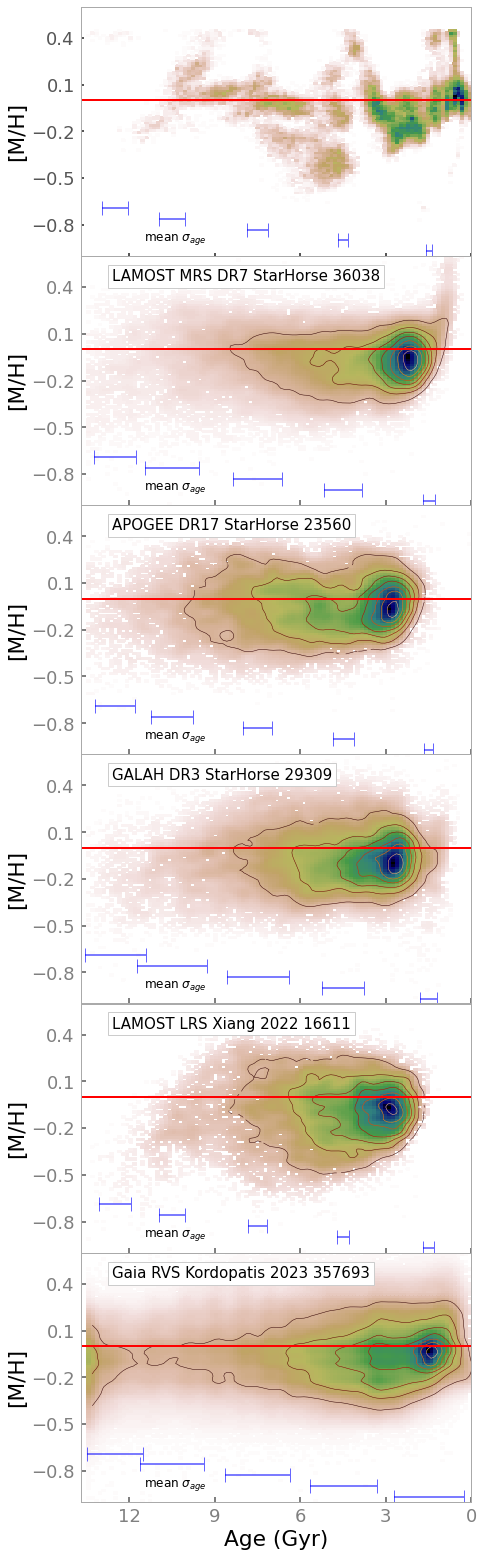}
    \caption{Age-metallicity distributions from the literature. The red line indicates solar metallicity. Panels from top to bottom show results from different sources: $Dir$SFH results within 100 pc obtained in section \ref{results}, with error bars indicating $\sigma_{age}$ (see Figure~\ref{precision_accuracy}); StarHorse results for LAMOST MRS DR7, APOGEE DR17 and GALAH DR3; \citet{Xiang_Rix_2022Natur.603..599X} results for LAMOST LRS DR7; and \citet{Kordopatis2023GaiaAges} results for $Gaia$ GSP-Spec. Each panel shows the number of stars used to derive the age-metallicity distribution, constrained to similar ranges of R$_{g}$, R$_{apo}$ and R$_{peri}$ as the sample presented in this paper. Error bars indicate the mean uncertainty in the age (as provided for each catalogue) within specific ranges (0-3, 3-6, 6-9, 9-12, 12-14 Gyr).}
    \label{AMR_comparison}
\end{figure}

Figure~\ref{AMR_comparison} displays a number of age-metallicity distributions adapted from recent literature data \citep{Xiang_Rix_2022Natur.603..599X, Queiroz2023_starhorse, Kordopatis2023GaiaAges}, as indicated in the labels and on the figure caption. The samples analysed in these works are very heterogeneous not only in terms of the volume covered, but also in the data used (from different spectrocopic survey samples) and the resulting selection functions. They all have in common the use of Bayesian isochrone fitting techniques to derive ages. The precision in the ages determined for individual stars depends on the accuracy of the spectroscopic measurements and the stellar evolutionary state, as highlighted in previous studies \citep{MintsHekker2017UniDAM, SandersDas2018}. Isochrone fitting methods are inherently linked to the choice of stellar evolution models. In Figure~\ref{AMR_comparison}, ages are derived by comparison with the PARSEC models (used in \citealt{Queiroz2023_starhorse} and \citealt{Kordopatis2023GaiaAges}) and the Yonsei-Yale \citep{Demarque2004} models \citep[used in][]{Xiang_Rix_2022Natur.603..599X}, which introduce differences in age scales compared to the BaSTI-IAC models used in the $Dir$SFH determination. In addition, each Bayesian method incorporates different prior assumptions. Furthermore, APOGEE \citep{Majewski2017}, GALAH \citep{DeSilva2015}, LAMOST \citep{Cui2012}, and Gaia RVS \citep{RecioBlanco2023_GaiaDR3} each incorporate specific assumptions in the derivation of spectroscopic [Fe/H] and alpha abundances, which we present in terms of global metallicity using eq.1.

In order to mitigate the effect of the different selection functions and make the Galactic stellar populations represented in the different panels more comparable to those present in the volume of 100 pc radius analysed in this paper (top panel), we have selected, from \citet{Kordopatis2023GaiaAges} only those stars with guiding Radius $7\le R_g\le 9$ and Z${_{\rm max}}\le 500$ pc, similar to the majority of the stars in our sample (see Figure~\ref{orbit_params})\footnote{The orbital parameters represented in Figure~\ref{orbit_params} are taken from \citet{Kordopatis2023GaiaAges}}. Then, we have selected the same stars in the \citet{Xiang_Rix_2022Natur.603..599X} and \citet{Queiroz2023_starhorse} samples. 

The features in the four central panels are very similar and characterised by a basically flat age metallicity relation with metallicity slightly below solar and prominent maximum in the stellar distribution around 3 Gyr ago. The metallicity distribution seems to get slightly narrower around 4 Gyr ago with hints of widening between 6 and 4 Gyr ago. The absence of stars with ages younger than $\simeq$ 2 Gyr is entirely due to selection effects, mainly caused by the T$_{\rm{eff}}$ and logg limits of the stars actually present in these catalogues, and the further limits imposed in the papers for the stars actually used to derive the ages. The bottom panel, corresponding to \citet{Kordopatis2023GaiaAges} ages determined from Gaia RVS data, displays the most numerous stellar sample, and it also shows a high concentration of stars around 3 Gyr ago. However, in this case, a more prominent maximum is seen around 1.5 Gyr ago. This difference with the other panels can be attributed  to the fact that \citep{Xiang_Rix_2022Natur.603..599X} and \citet{Queiroz2023_starhorse} restrict their analysis to subgiant branch stars or main sequence turnoff and subgiant branch stars, respectively, while \citet{Kordopatis2023GaiaAges} don't have this restriction and thus, include in their sample brighter, and thus younger stars, as well as RGB stars. Note however the large errors of this sample, specially at young ages. 

These age-metalliticy distributions provide much less detail, owing to larger age errors, than the one derived in this paper from CMD-fitting, and displayed in the top panel. A detailed comparison of the age-metallicity distributions obtained with both methodologies is out of the scope of this work. However, it is possible to identify the overdensity of 3 Gyr old stars with below solar metallicity with the various star formation events that we resolve between 4 and 2 Gyr ago. The younger overdensity in the \citet{Kordopatis2023GaiaAges} age-metallicity distribution coincides with the star formation episodes that we identify in the last 1 Gyr. Finally, the slightly increased metallicity dispersion between 6 and 4 Gyr ago could be the same that we observe, resolved in three star formation episodes at different metallicities.

This comparison shows the superior ability of CMD fitting to derive age distributions compared to those obtained from the combination of individual ages. This is thanks to the fact that CMD-fitting uses, in addition to the colours and magnitudes of the stars, the extra information on the relative lifetimes of stars of different masses and in different evolutionary stages, provided by stellar evolution theory, and on the relative number of stars of different masses, according to externally derived IMF. It also has the advantage that the required data are only photometry and parallaxes from Gaia, and thus the selection function is easier to control (and in this particular case, the sample can be considered complete). Of course, the important caveat is that ages are not obtained for each individual star, and this precludes the direct association of ages to other important properties, such as chemical abundances or kinematics, which help disentangling different stellar populations in the MW. Both sets of techniques are therefore highly complementary and necessary to advance in our understanding of the evolutionary processes that have shaped the MW.

\section{Discussion: the evolutionary history of the Milky Way thin disk} \label{discussion}

\begin{figure}[h!]
    \centering
    \includegraphics[width=0.5\textwidth]{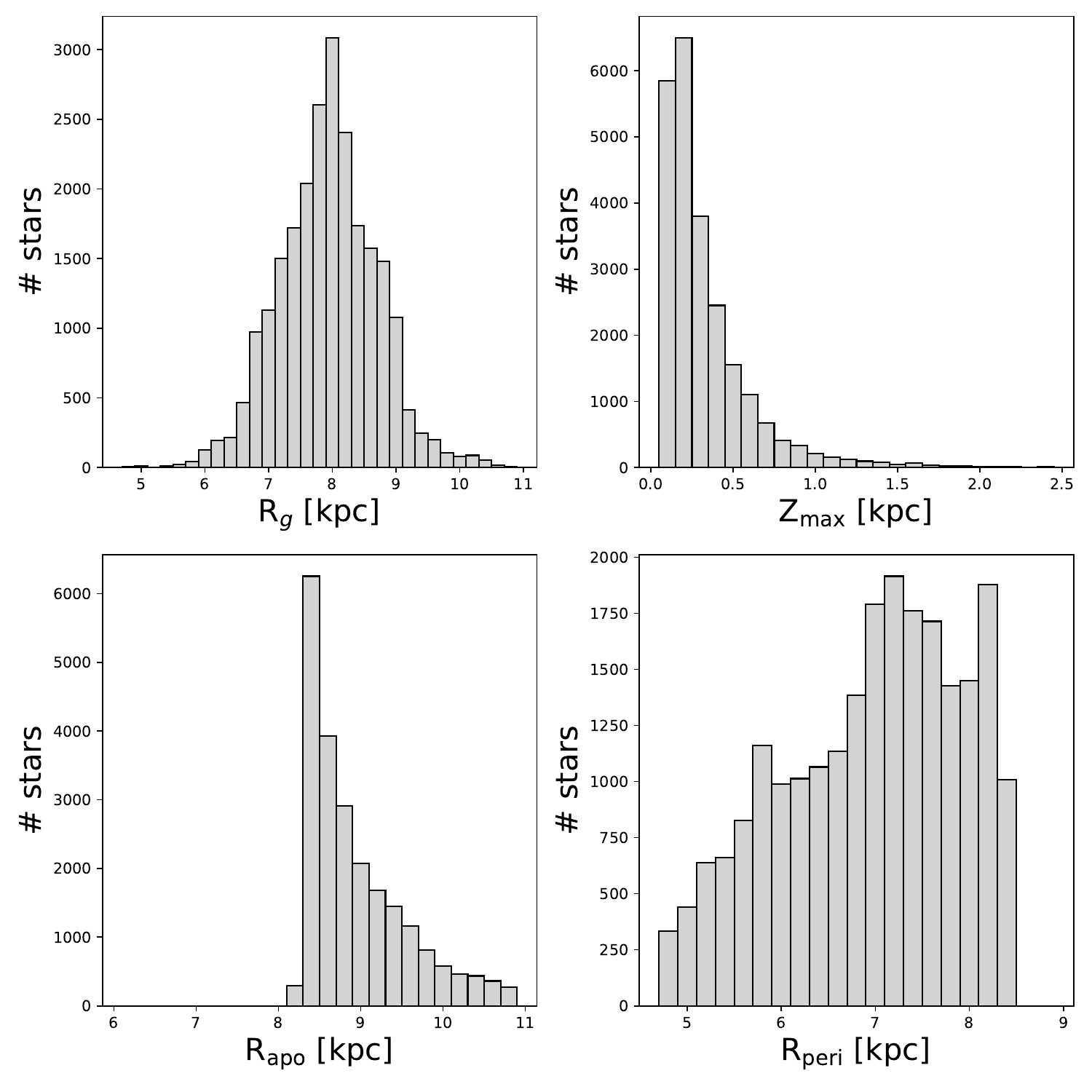}
    \caption{Orbital parameters for stars in the GCNS. Upper left panel: guiding radius (R$_g$); upper right panel: Zmax; lower left panel: apocentric radius (R$_{apo}$); lower right panel: pericentric radius (R$_{peri}$) radius.} 
    \label{orbit_params}
\end{figure}

Figure~\ref{orbit_params} displays the guiding radius (R$_g$, upper left panel), Zmax (upper right panel), apocentric (R$_{apo}$) and pericentric (R$_{peri}$) radius (lower left and right panels, respectively), for stars in the GCNS. We have adopted for these parameters the values provided by \citet{Kordopatis2023GaiaAges}. It can be seen that the majority of the stars currently within 100 pc of the Sun have guiding radius between 7 and 9 kpc and Zmax up to 0.5 kpc. In their apocenters they can travel as far as 11 kpc while in their pericenters they can approach the Galactic centre as close as 5 kpc. We live, therefore, in a quite cosmopolitan place, with a good fraction of the inhabitants possibly born quite far away. As a result, the age-metallicity distribution derived in this paper from the exquisitely observed bubble of 100 pc radius from the Sun contains, in fact, information of a much larger volume. This local stellar population, however, is heavily dominated by thin disk stars, as discussed in Section~\ref{data} and graphically shown in Figure~\ref{description_catalogue}. Indeed, our kinematic selection classifies only $\simeq$ 10\% of the GCNS stars as thick disk. Most of the stars have low [$\alpha$/Fe], and so they would be chemically classified as thin disk as well. The results in this paper, therefore, provide a sharp view of the formation and evolution of the MW thin disk, while still providing hints of the nature of the stellar populations in the thick disk and the halo.

We shall start our discussion in chronological order, by noting the small amount of stars older than 10 Gyr, with [M/H]$\lesssim -1$, which are metallicity values typical of the stellar halo. This population is shown with contours in Figure~\ref{SFH_final}. Based on previous derivations of the SFH of kinematically selected halo stellar populations \citep{Gallart2019Gaia, Massari2023_CARMAI}, the stars with this age-metallicity location would correspond to the GES remnant. Interestingly, we don’t see in our solution signs of a population older than $\simeq11$ Gyr and [M/H] between -0.5 and -0.3 which according to \citet{Gallart2019Gaia} and \citet{Massari2023_CARMAI} would correspond to the red sequence of the halo CMD \citep[see also][]{haywood2018twoseq} also called the Splash \citep{Belokurov2020Splash}. In our solution CMD, for M$_G \leq 4.2$, 96 out of 21357 stars\footnote{When quoting number of stars in a particular population, we refer to stars inside the analysed area of the CMD.} have [M/H]$<-0.6$ and ages older than 9 Gyr. However, when classifying the stars in the GCNS based on their kinematics, only 9 stars end up being classified as halo within this magnitude range. Therefore, most of the stars that produce the signal that we observe at old age and low metallicity must have disc kinematics. This would be in qualitative agreement with the numerous claims in the literature about metal-poor stars in disc-like orbits, possibly tracing an early rotating disk component \citep{Morrison1990, haywood2013, ChibaBeers2000, Carollo2010, sestito20, diMatteo2020FlatRound, Fernandez-Alvar2021_MPdisk, cordoni21,  Mardini2022_AtariDisk, Xiang_Rix_2022Natur.603..599X, BelokurovKravtsov2022Aurora, FeltzingFeuillet2023MWD, Chandra2023_threePhaseMW, Nepal2024arXiv240200561N, Fernandez-Alvar_2024arXiv240202943F} and opens interesting prospects about the possibility of studying the ages of stars in the metal poor disk by analysing larger samples of stars pertaining to different Milky Way volumes. However, owing to the small number of stars in this particular sample, our results on this matter should be considered as preliminary. 

The thin disk is expected to form after the early epoch of frequent mergers and to grow from the smooth accretion of gas \citep{brook2012thinthick, Agertz2021_VINTERGATAN_I, Yu2023_BornThisWay, Grand2017_AurigaDiskFormation}. Since most stars in our sample belong to the thin disk, our results should be able to provide tight constraints regarding the oldest age {\it of the bulk} of the Milky Way thin disk stars.  Our deSFH indicates that these stars have near solar metallicity and are~$\simeq$~11~Gyr old (after taking into account our age precision at old ages, and $\simeq$ 10.4 Gyr old after additionally correcting by age systematics).\footnote{This is actually an upper limit, as the 11 Gyr old population in Figures \ref{SFH_final} and \ref{stellar_final} could actually originate in the stars that we have kinematically classified as thick disk in Section~\ref{data}. However, deriving SFHs for kinematically distinct populations is beyond the scope of this paper, and will be the subject of Fernández-Alvar et al. (2024, in prep).} This age indeed coincides very well with the time of the last major merger experienced by the Milky Way, that of Gaia-Enceladus-Sausage \citep[GES,][]{belokurov2018_sausage, helmi2018} dated to have occurred around 11-10 Gyr ago (\citealt{Gallart2019Gaia}, \citealt{Bonaca2020GES}, \citealt{Montalban2021GES}, although see also \citealt{dolon23}). The high metallicity of the oldest stars in the thin disk must be the result of a rapid initial chemical enrichment in our Galaxy in an epoch of intense star formation activity expected to take place in its inner parts \citep[in an inside-out scenario in which star formation is more efficient in the innermost regions, e.g.][]{Larson1976FormDisk, chiappini1997, Chiappini2001, Haywood2018InnerDisk, Matteucci2021A&ARv}, and it is followed by a trend of slightly decreasing metallicity with time. Inverted age-metallicity relations have been inferred by other works \citep[][ see also Section \ref{otherAMRs}]{Sahlholdt2022MNRAS.510.4669S, Xiang_Rix_2022Natur.603..599X, Imig2023TaleTwoDiskApogee}, and different explanations have been proposed. The most natural one could be that this trend is the result of more steady star formation fuelled by gas accretion while the radius of the MW disk progressively grows \citep{Chiappini2001, minchev2014AA, Ciuca2021}. In this scenario, the combined effects of gas dilution and metal enrichment by ongoing star formation may be the origin of the observed trend \citep{Weinberg2017} for the main populations formed between $\simeq$ 11 and 6 Gyr ago. The effects of radial migration \citep{Grand2015_RadialMigration, Minchev2018BirthRadii, Okalidis2022} and satellite infall \citep[][see below]{Grand2018_AurigaChemDistinctDisks,Agertz2021_VINTERGATAN_I, Lu2022_satelliteInfall} are expected to also play a role. 

A dramatic event in the history of the Milky Way must have happened around 6 Gyr ago, when a dip in the star formation activity is observed, followed by a clear disruption of the previous relatively steady age-metallicity trend. This event may be the first infall of the Sagittarius dwarf galaxy (Sgr), thought to have occurred around this epoch \citep{law2010, laporte2018}, and seen to have possibly caused repeated bursts of star formation coinciding with its pericentric passages \citep{Ruiz-Lara2020Sgr}. Cosmological simulations find that accretion of large satellites is usually accompanied by the accretion of gas (both from the gas in the satellite itself \footnote{Sagittarius must have been gas-rich at the time of its accretion, as star formation close to the centre of the remnant galaxy has been seen to proceed for many Gyr after its infall, possibly coinciding with the subsequent pericentric passages \citep[see][]{Siegel2007_SgrSFH}.} and from cosmological cold flows, \citealt{RuizLara2016_satelliteAccretion}, \citealt{Grand2020_dualOriginThickDiskGES},  \citealt{Agertz2021_VINTERGATAN_I}), and result in a drastic chemical and kinematic reorganisation of the disk, accompanied by disk growth. For example, \citet{Grand2018_AurigaChemDistinctDisks} found that the accretion of a gas-rich satellite about 6 Gyr ago supplied fresh, metal-poor gas which reinvigorated star formation in the outer disc. This resulted in a resumption of disc growth and distinct features in the age-metallicity distribution. Another example is the work of \citet{Agertz2021_VINTERGATAN_I}, who find that a satellite accretion occurring at $z\simeq1.5$ ($\simeq$ 9 Gyr ago) results in the rapid formation of an extended disk which is more metal poor than the existing disk when the satellite was accreted, leading to a feature in the age-metallicity distribution of their simulated disk that is highly reminiscent of the low and intermediate metallicity population that we infer around 5 Gyr ago. What is missing in their predicted age-metallicity distribution is the high-metallicity population that we infer in ours. In our case, this may indicate that the star formation triggered by this accretion event may have reached the inner galaxy as well, where star formation would take place in a more metal enriched environment. Stellar migration to the solar radius could possibly be enhanced at this time by a growing bar \citep{Nepal2024_YoungBar} which formation could have been triggered by the interaction with Sgr itself (e.g. \citealt{Purcell2011NaturSgr}; however see also \citealt{Merrow2023_barGESarXiv} for claims that the GES could be linked to the bar formation). Thus, both populations with metallicity clearly different (higher and lower) than solar, observed between 6 and 4 Gyr ago, could be present in the analysed volume owing to radial migration (from the inner and outer disk, respectively). It is interesting to note that the three populations follow an age sequence, with the most metal poor population being the oldest, and the more metal rich, the youngest. Additionally, the number of stars in them decreases from the more metal-poor to the most metal-rich population (with $\simeq 1000, 700$ and 400 stars respectively). This could indicate a propagation of the star formation, possibly with decreasing intensity, from the outer to the inner disk over a period of 1.5-2 Gyr.

\begin{figure}[h!]
    \centering
    \includegraphics[width=0.4\textwidth]{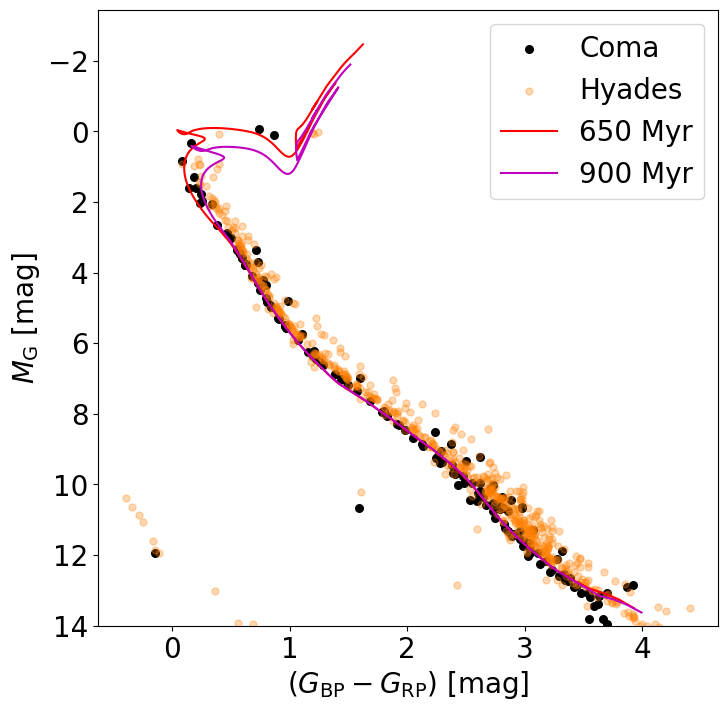}
    \caption{CMD of the two open clusters contained within 100 pc: Hyades and Coma Berenices. The superimposed isochrones correspond to the age of the two main star formation episodes in the last 1 Gyr, occurred $\simeq$ 900 and 650 Myr ago}
    \label{twoclusters}
\end{figure}

After the striking dip observed 4 Gyr ago in the deSFH (Figure~\ref{SFH_final}) and the age-metallicity distribution (Figure~\ref{stellar_final}), star formation in the extended solar neighbourhood starts again and proceeds at an average higher rate than ever before, and in an episodic manner. The stars formed in the first episode after the 6-4 Gyr old break have metallicity very close to (or just slightly below) solar, but a further metallicity decrease is observed between 3 and 1.5 Gyr ago, with most of the stars having [M/H] $\simeq -0.2$ and only a minor population with solar metallicity. This age can be linked to another pericentric passage of Sgr \citep[][with exact timing depending on the models]{Purcell2011NaturSgr, laporte2018}, and even to a moment in which the MW starts to experience the combined influence of both Sgr and the LMC \citep{Besla2007FirstInfallMC, laporte2018, Vasiliev2023Galax}. Therefore, the same mechanisms discussed above, with possibly less intensity owing to a stripped, less massive Sgr may be at play here. Finally in the most recent 2 Gyr, the deSFH exhibits a notable bursty pattern (which is observable due to enhanced age resolution in this time range) and a discernible upward trend in metallicity, with the majority of stars having metallicity very close to solar. This pattern can be attributed to both the gradual increase in local metallicity and the limited radial migration that likely occurred within this relatively short time span, with the majority of stars in the analysed volume born in close proximity to their current locations \citep{MinchevChiappiniMartig2013_solarVicinity}. An exception to this is the significant $\simeq$ 600 Myr old, metal-rich population, containing around 250 stars (see inset in the lower left panel of Figure~\ref{SFH_final}), which presence in the solar neighbourhood must again be attributed to radial migration. Finally, the ages of the two strongest episodes of star formation that we derive in the last 1 Gyr, occurring $\simeq$ 650 and 900 Myr ago, coincide very approximately with the ages of the two open clusters contained within 100 pc, Hyades and Coma Berenices, respectively \citep[see Figure~\ref{twoclusters} and][]{gcns}. However, the presence of the stars belonging to these clusters in our sample cannot account for the whole strength of the bursts, which contain $\simeq$ 800 stars each, while the number of stars in the cluster CMD within the same magnitude range (M$_G \le 5$) is an order of magnitude smaller.

\section{Summary and conclusions} \label{summary}

This paper presents for the first time a detailed dynamically evolved SFH (deSFH) within 100 pc of the Sun, as well as the distribution of age and metallicity of the stars in the same Milky Way volume. They have been derived with a new set of codes to perform CMD fitting, called CMDft.Gaia.  In this section we will present our main conclusions both regarding the stellar content of the solar neighbourhood and the performance of CMDft.Gaia.

{\subsection{The stellar populations within 100pc of the Sun}}

The main conclusions regarding the evolutionary history of the disk near the solar radius, obtained from the deSFH and age and metallicity distributions of the GCNS, which contains mostly thin disk stars inhabiting a range of galactocentric radius between at least 7 and 9 kpc, and Zmax<500pc, can be summarised as follows: 

-The bulk of the stellar population contained in the analysed volume form an age-metallicity sequence, with metallicities around solar. The oldest stars in this main population have an age of about 11 Gyr (or 10.4 Gyr considering systematic effects) and a metallicity of ${\rm[M/H]}=-0.16$, increasing to ${\rm[M/H]}=0.25$ within the next Gyr. 

-Our solution requires around 100 stars older than 10 Gyr and with metallicities around ${\rm[M/H]}=-1.5$. This number is one order of magnitude larger than the stars that we classify as halo based on their kinematics, and thus we could be dating the population of the metal poor disk stars found in many previous works.

-Between  $\simeq 9.5$ and 6 Gyr ago, two distinct metallicity sequences emerged. In the earlier phase, stars exhibited metallicity slightly above solar. Subsequently, around 8 Gyr years ago, the primary population has solar metallicity, with a less prominent group with ${\rm[M/H]}\simeq-0.25$).

-Stars between 6 and 4 Gyr old are clumped in three well separated populations with little difference in age but large metallicity differences: the oldest with [M/H]$=-0.4$, the next with solar metallicity, and the youngest with a very narrow age range $\simeq 4$ Gyr old and supersolar metallicity. The latter one coincides with a star formation gap at the expected solar metallicity, and is followed by a conspicuous break of star formation. 

-Stars younger than 4 Gyr are more numerous (per age range) compared to older ages, and form clumps as a function of age and metallicity: between 4 and 2 Gyr ago, populations with a range of metallicities, including solar and sub-solar (${\rm [M/H]} \simeq -0.2$), coexisted. However, since 2 Gyr ago, the majority of stars have exhibit solar metallicity.

-The stellar metallicity distribution derived with CMDft.Gaia is in excellent agreement with spectroscopically derived ones. In particular, it compares extremely well (even in its details) with that obtained by \citet{Fuhrmann2021_solarTypeStars} with their high-resolution, high signal-to-noise, complete sample of stars within 25 pc of the Sun. The peak metallicity and width of the two distributions are virtually identical, and the same two lower metallicity components, with ${\rm [M/H]} \simeq -0.2$ and $\simeq -0.5$, are observed. Interestingly, our age-metallicity distributions disclose that these are intermediate-age populations, including the ${\rm [M/H]} \simeq -0.5$ component, with metallicity typical of the thick disk, which here we infer to be $\simeq$ 5 Gyr old.

The picture that emerges from the results in this paper is thus that of a thin disk that begins to form the bulk of its stars $\simeq$ 11-10 Gyr ago, from gas chemically enriched in an earlier epoch of high star formation rate, possibly triggered by mergers (\citealt{Gallart2019Gaia}, see also \citealt{orkney22}, \citealt{BelokurovKravtsov2022Aurora}). Several Gyr of star formation with slightly decreasing metallicity follow, possibly the result of steady star formation fuelled by gas accretion while the radius of the Milky Way disk progressively grows. Then, around 6 Gyr ago, some dramatic event seems to be required to cause the break in the star formation activity after which three stellar populations with very different metallicities are observed. This event is likely the first infall of the Sgr dwarf galaxy, and we hypothesise, in agreement with results of cosmological simulations, that it was accompanied by a substantial amount of gas accretion which led to an important chemical and kinematic reorganisation of the disk, accompanied by further disk growth. Stellar migration to the solar radius could possibly be enhanced at this time by a growing bar which formation could have been triggered by the interaction with Sgr itself, and thus the presence of stellar populations with such different metallicity. The age sequence observed from the most metal poor to the most metal rich population  within 6-4 Gyr ago could indicate some kind of propagation of the star formation from the outer disk to the inner galaxy. The final, more vigorous period of star formation since 4 Gyr ago may as well hold some relation with further Sgr pericentric passages (as hinted by some gas dilution between 3 and 1.5 Gyr ago) and converges to a mainly local population of metallicity very close to solar. 

{\subsection {The performance of CMDft.Gaia:}}

CMDft.Gaia is a suite of procedures that includes i) the computation of synthetic CMDs in the {\it Gaia} bands with ChronoSynth; ii) the simulation in the synthetic CMDs of the observational errors and completeness affecting the observed CMD; and iii) the derivation of the SFH itself using $Dir$SFH. 

Salient characteristics of CMDft.Gaia are:

i) the synthetic mother CMDs computed with ChronoSynth have a virtually continuous age and metallicity distribution between desired upper and lower limits of these parameters, and following parameterised IMF and binary population characteristics. 
ii) the final deSFH is derived by averaging numerous SFHs generated using different sets of SSPs. These SSPs are obtained by dividing, as a function of age and metallicity, using a Dirichlet-Voronoi tessellation, a mother CMD featuring a vast number of stars with continuous age and metallicity distribution and a realistic simulation of the observational errors; iii) no assumptions are made on the age-metallicity relation, on the metallicity distribution or on the functional form of the star formation rate as a function of time. This results in robust and detailed information on the evolution of the age and metallicity of the stellar populations present in the analyzed volume over cosmic time.

We have used synthetic populations of real open clusters observed by Gaia, and synthetic clusters in a wider range of age and metallicity, to test the accuracy and precision of $Dir$SFH when determining the ages and metallicities of stellar populations. Additionally, we have tested the effect of the variation of some of the input parameters on the derived deSFH of the GCNS. These tests allow us to reach the following conclusions on the performance of CMDft.Gaia:

-The age of a stellar population with a narrow age and metallicity range is determined with a {\it precision of 10\% or better}, with little dependency on the number of stars in the test stellar population, the size of the age bins, the age, or whether the test population is observed (open clusters) or synthetic.

-The ages of synthetic clusters were systematically overestimated by a maximum of 6\% (and as little as 2-4\% in the age range 1-6 Gyr), except in very special cases, and again with little dependency on the size of the population or the size of the age bins.

-The main features of the GCNS deSFH show little dependency on various input choices for the fit, such as the number of stars in the mother CMD, the size of the age bins, the faint magnitude limit of the bundle, the weights across the CMD, and the unresolved binary stars parametrisation (for sensible choices of this parametrisation). The largest differences occur for solutions obtained with mother CMDs calculated with different stellar evolution libraries, such as the BaSTI-IAC (the library we used for most tests) or the PARSEC stellar evolution libraries. The level of agreement between the derived MDF for each library and spectroscopic MDFs make us prefer the solution with the BaSTI-IAC models, and this is the one we choose for describing the stellar content of the GCNS.

\begin{acknowledgements}
We would like to thank C.B. Brook, R.J.J. Grand, D. Kawata, I. Pérez and A. Recio-Blanco for enlightening discussions. CG, EFA, GB, GT, SC, and TRL acknowledge support from the Agencia Estatal de Investigación del Ministerio de Ciencia e Innovación (AEI-MCINN) under grant “At the forefront of Galactic Archaeology: evolution of the luminous and dark matter components of the Milky Way and Local Group dwarf galaxies in the Gaia era” with reference PID2020-118778GB-I00/10.13039/501100011033. SC, CG and EV  also acknowledge financial support from the Spanish Ministry of Science, Innovation and University (MICIN) through the Spanish State Research Agency, under Severo Ochoa Centres of Excellence Programme 2020-2023 (CEX2019-000920-S). CC acknowledges support from the Fundación Jesús Serra and the Instituto de Astrofísica de Canarias under the Visiting Researcher Programme 2022-2024 agreed between both institutions. EFA also acknowledges support from the "María Zambrano" fellowship from the Universidad de La Laguna. SC acknowledges financial support from PRIN-MIUR-22: CHRONOS: adjusting the clock(s) to unveil the CHRONO-chemo-dynamical Structure of the Galaxy” (PI: S. Cassisi) finanziato dall'Unione Europea - Next Generation EU, and Theory grant INAF 2023 (PI: S. Cassisi). TRL acknowledges support from Juan de la Cierva fellowship (IJC2020-043742-I). AH is grateful for financial support from a Spinoza prize (NWO). EV acknowledges support from the Severo Ochoa program through CEX2019-000920-S, during a visit to IAC. 

This work has made use of data from the European Space Agency (ESA) mission {\it Gaia} (\url{https://www.cosmos.esa.int/gaia}), processed by the {\it Gaia} Data Processing and Analysis Consortium (DPAC, \url{https://www.cosmos.esa.int/web/gaia/dpac/consortium}). Funding for the DPAC
has been provided by national institutions, in particular the institutions participating in the {\it Gaia} Multilateral Agreement.

\end{acknowledgements}

\bibliographystyle{aa}

\bibliography{bib_gaia.bib}

\begin{appendix} 

\section{Complementing the GCNS with Gaia DR3} \label{app:GDR3}

We complemented the original GCNS catalogue with {\it Gaia} DR3 line-of-sight velocities for 174221 stars, as well as metallicity ([M/H]) and [$\alpha$/Fe] abundance measurements for 23629 stars derived from the Radial Velocity Spectrometer by the General Stellar Parametriser-spectroscopy module, \textit{GSP-Spec} \citep[as explained in][]{RecioBlanco2023_GaiaDR3}. This allowed us to explore the stellar population characteristics in terms of kinematics and chemical abundances, which are summarised in Figure~\ref{description_catalogue}.

To select the sample with chemical information, we applied quality cuts based on the flags provided by \citet{RecioBlanco2023_GaiaDR3} and encoded in the {\tt flags\_gspspec} parameter. We selected only stars with bits 1 to 6 in {\tt flags\_gspspec} equal to 0 or 1, bit 7 equal to 0, 1 or 2, bit 8 equal to 0 or 1, and bits 9 to 13 equal to 0. This way we avoid the inclusion of possible erroneous abundance estimates due to biases in $T_{\rm eff}$, $\log{g}$ and [M/H] induced by bad line broadening models, large uncertainties because of flux noise, extrapolation of the parametrisation and other issues, as explained in section 6 and table 2 in \citet{RecioBlanco2023_GaiaDR3}.

Following the suggestions in the same paper we also calibrated [M/H] with the polynomial coefficients in their table 3, and [$\alpha$/Fe] with the polynomial relation as a function of $t=T_{\rm eff}$/5750 in their table 4. We restrict our sample to calibrated $\log g$ values (based on the relation in table 2) between 0 and 4.75 because we detected suspicious [$\alpha$/Fe] overabundances in stars with $\log g$ > 4.75. This is probably due to the vicinity to the $\log g$ = 5 limit of the grid.   

In order to assign probabilities of membership to the thin disc, thick disc and stellar halo, we first derived Galactic space velocities, U,V, W \footnote{As a reminder, this is an heliocentric reference frame, with the x- direction pointing towards the Galactic centre, y- being measured in the direction of Galactic rotation and the z- direction points towards the North Galactic Pole; the velocity components in this reference frame are typically indicated as U, V, W.}. We then transformed them into Galactocentric velocities considering the Sun position with respect to the Galactic centre at 8.178 kpc \citep{gravity19}, the Sun motion with respect to the Local Standard of Rest (LSR) as ($U_\sun$, $V_\sun$, $W_\sun$) = (11.1,12.24,7.25) km\ $\rm s^{-1}$ \citep{Schonrich2010} and a V-velocity for the LSR of 236.26 km/s \citep{reid19}. Then, we adopted the procedure by \citet{Bensby2003A&A} to calculate  the probability for each star to be a member of the thin, the thick disc or the halo. Each Galactic component is assumed to follow a Gaussian distribution for each component of the velocity vector, with the parameters shown in Tab.~\ref{kine}. In terms of number density laws, we assumed an exponential profile both in the radial direction (R) and in the vertical, $z$, direction for the discs, and  a power law ($r^{-2.5}$) for the stellar halo, respectively (r in this case is the spherical 3D Galactocentric distance). It is assumed that 90\% of the stars in the considered volume belong to the thin disc, and 8\% and 2\% to the thick disc and halo, respectively. A star is assigned to a given Galactic component if it has a probability higher than 70\% of belonging to that component. For more details on the kinematic selection of the stars belonging to the different Galactic components, we refer the reader to Fernández-Alvar et al. (2024, in prep). This returned samples of 146108 thin disc stars, 13153 thick disc stars, and 415 halo stars. 

\begin{table*}
 \centering
 \caption{Kinematic parameters}
 \centering
 \begin{tabular}{lrcccrcrcr}
\hline \hline
Population & Relative density & Scale-height & Scale-length & <U> & <V> & <W> & $\sigma_{U}$ & $\sigma_{V}$ & $\sigma_{W}$ \\
 &  & (kpc) & (kpc) & ($\rm km\ s^{-1}$) & ($\rm km\ s^{-1}$) & ($\rm km\ s^{-1}$) & ($\rm km\ s^{-1}$) & ($\rm km\ s^{-1}$) & ($\rm km\ s^{-1}$) \\ 
  \hline\hline
  Thin disc & 0.9 & 0.3 & 2.6 & 0 & 236 & 0 & 39 & 20 & 20 
 \\

  Thick disc & 0.08 & 0.9 & 2.0 & 0 & 206 & 0 & 63 & 39 & 39 
 \\

  Halo & 0.02 & -- & -- & 0 & 0 & 0 & 141 & 106 & 94 
\\
\hline
  \end{tabular}  
\tablefoot{We assumed that the thin and thick disks follow gaussian velocity distributions, with means and standard deviations from \citet{Soubiran2003}, and a halo component with velocity dispersion following \citet{ChibaBeers2000}. We considered the density fractions as in \citet{ramirez13}, scale-heights, scale-lengths, and the halo power law are based on \citet{Bland-Hawthorn_Gerhard2016}. }
  \label{kine}
\end{table*}

\section{Empirical constraints on the $(G_{BP}-G_{RP})$ colour for the fainter portion of the BaSTI-IAC isochrones} \label{app:lowmscor}

\begin{figure}[h!]
\centering
\includegraphics[width=0.4\textwidth]{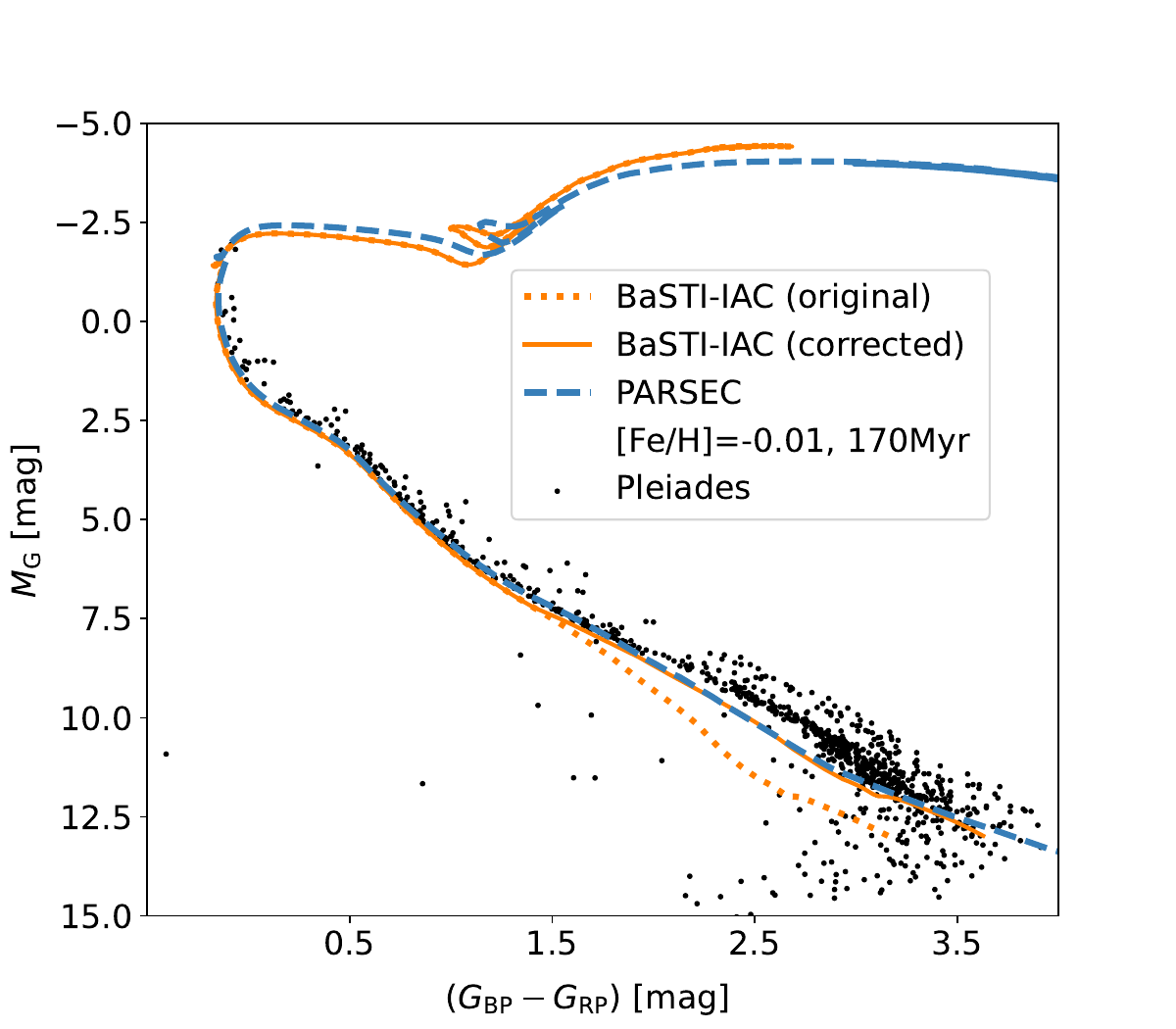}
\caption{{\it Gaia} DR3 Pleiades CMD with 170 Myr isochrones superimposed: BaSTI-IAC isochrones (orange dotted line: uncorrected; orange solid line: corrected adopting \citet{mann} empirical $T_{eff} - (G_{BP}-G_{RP})$ relation) and PARSEC isochrone (blue dashed line).}
\label{pleiades}
\end{figure}

The extraordinary depth and precision of the {\it Gaia} photometry for nearby samples of field stars and open clusters offer an excellent test-bed to assess the accuracy of stellar evolution models and of the bolometric corrections needed to transfer the theoretical framework from the H-R diagram to the {\it Gaia} photometric plane.
Figure \ref{pleiades} displays the comparison between isochrones of 170 Myr from BaSTI-IAC \citep{Hidalgo2018} and PARSEC \citep{Bressan2012} with the {\it Gaia} DR3 CMD of one of the best-studied nearby open clusters: the Pleiades \citep[][see Appendix \ref{clusters} for a description of the procedure used for the membership selection, distance and reddening determination for this cluster as well as for all clusters studied in the present work]{dias:21,soderblom:09}.

The PARSEC and the BaSTI-IAC isochrones shown in this figure have been transferred in the {\it Gaia} photometric plane by adopting their own bolometric correction tabulations -- but see the discussion about this topic in Sect.~\ref{sec3:Synthetic_CMD_Computation}.

\begin{figure}[h!]
\centering
\includegraphics[width=0.4\textwidth]{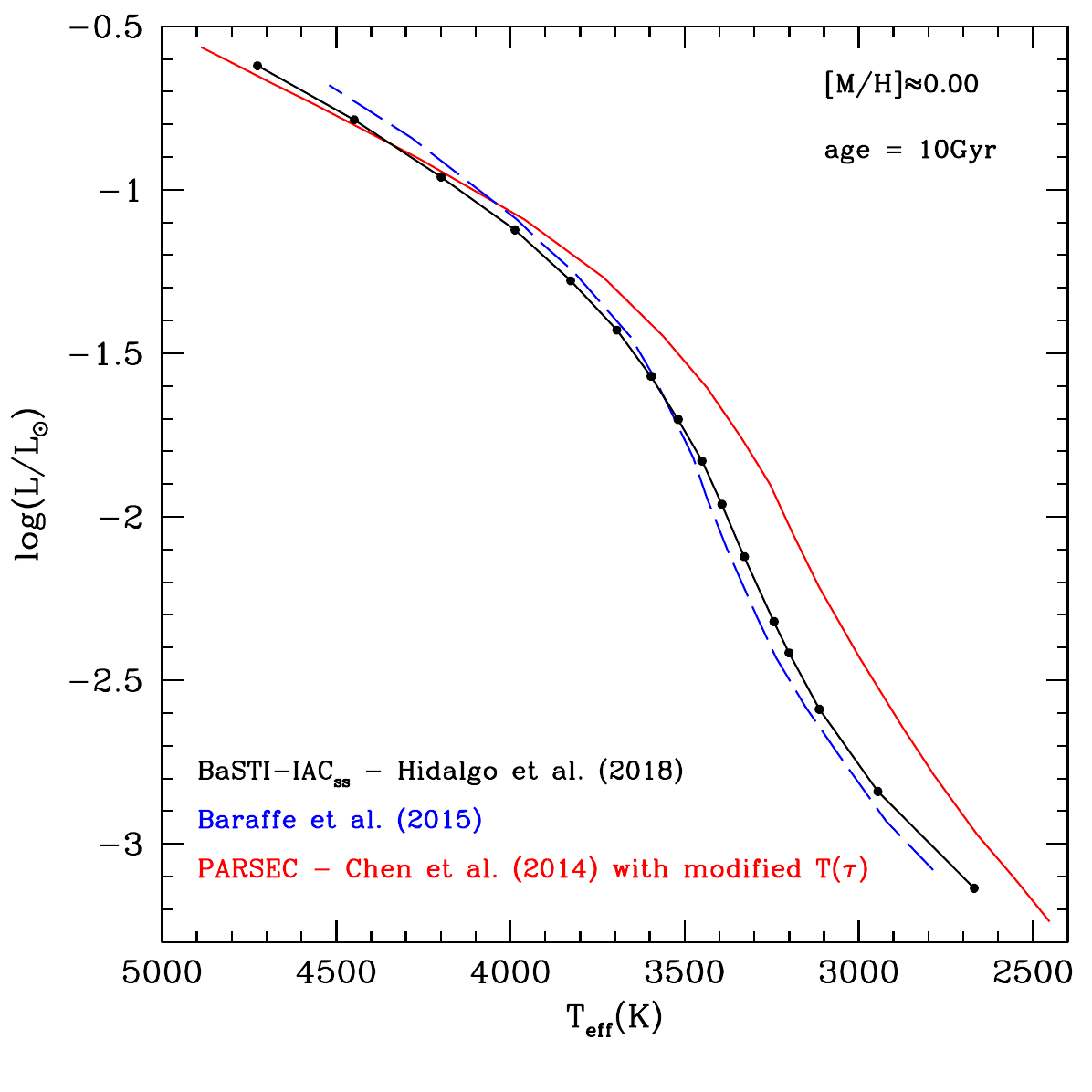}
\caption{The main sequence locus in the H-R diagram for BaSTI-IAC and PARSEC VLM stellar models at solar metallicity for an age of 10 Gyr. The dot-dashed line represents the models provided by \cite{baraffe:15}.}
\label{vlmcomp}
\end{figure}

Fig.~\ref{pleiades} reveals that the two sets of isochrones show a quite different behaviour concerning the faintest portion of the main sequence locus. In particular for $M_G>7$, the BaSTI-IAC isochrone (orange dotted line) is systematically bluer than the observed sequence, with a shift in colour increasing with increasing magnitude up to $\sim0.5$ mag at $M_G\simeq12$. On the other hand, the PARSEC isochrone (blue dashed line), although slightly bluer than the empirical sequence, provides a quite better match to the data. In order to understand the reason for such a behaviour we compared the two sets of isochrones in the theoretical plane and we discovered that, despite the fact that the two stellar models libraries are based on similar physical inputs concerning equation of state, opacity and outer boundary conditions, the effective temperature scale of the very low-mass stars ($M<0.45M_\odot$) in the PARSEC database is significantly cooler than that predicted by the corresponding BaSTI-IAC models, with the difference increasing with decreasing stellar mass. This is shown in Fig.~\ref{vlmcomp}, that discloses the comparison, at solar metallicity, between very low mass (VLM) stellar models from the BaSTI-IAC and the PARSEC libraries. For the sake of comparison, in the same figure we also show the main sequence locus for the VLM models by \cite{baraffe:15} that represent the state-of-the-art for the VLM stellar regime. 

The BaSTI-IAC models are in very good agreement with those by \cite{baraffe:15}; some differences appear only for masses larger than about $\sim0.65M_\odot$ and are due to the adoption of different values for the mixing length parameter. On the other hand the PARSEC models are in good agreement with the other two model libraries for masses larger than about $0.6M_\odot$, and become progressively cooler with decreasing mass. The reason for such behaviour is that, as discussed by \cite{chen:14}, in the last release of the PARSEC models, the outer boundary conditions adopted for computing stellar models in the VLM stellar regime, have been tuned in order to improve the match between these models and the empirical Mass - Radius relationship.  In our belief, this approach is unsafe due to the fact that, for a given stellar mass, the radius of a VLM star can be hugely affected by the presence of a strong magnetic field as usually occurs in M dwarfs \citep{chabrier2000}. 

Being aware of the significant shortcomings that still affect the synthetic spectra computations, we decided to follow a different methodological approach to try to improve the agreement between the observed sequence for VLM stars and the BaSTI-IAC models in the {\it Gaia} DR3 photometric bands: we compared the $T_{eff} - (G_{BP}-G_{RP})$ relationship predicted by the BaSTI-IAC VLM stellar models with the empirical one provided by \citet{mann}, which is based on accurate spectra for a large sample of field M dwarfs with reliable estimate of their effective temperature. We found that the original BaSTI-IAC $T_{\rm{eff}} - (G_{BP}-G_{RP})$ relationship predicts, at a given effective temperature a bluer colour with respect the analogue relation by \cite{mann}: the $\Delta(G_{BP}-G_{RP})$ colour\footnote{The bolometric corrections provided by \cite{mann} have been evaluated by adopting the {\it Gaia} DR1 passbands; therefore by using the BaSTI-IAC isochrones in both {\it Gaia} DR1 and DR3 photometric systems we have estimated the impact ($\sim-0.06$~mag) on the $\Delta(G_{BP}-G_{RP})$ due to the use of the more updated {\it Gaia} passbands} is negligible for $T_{eff}\approx4000$~K, of the order of $\sim0.2$~mag at 3600~K, and of about 0.5~mag at $T_{eff}=3000$~K. Therefore we decided to apply a $T_{eff}-$ dependent correction to the $(G_{BP}-G_{RP})$ of the original BaSTI-IAC isochrones in the VLM star regime in order to reproduce the empirical colour of field M dwarfs. 

The \lq{corrected}\rq\ BaSTI-IAC isochrone is also shown in Fig.~\ref{pleiades} (orange solid line): note that the \lq{corrected}\rq\ BaSTI-IAC isochrone are in quite better agreement with the observed main sequence locus with respect the original BaSTI-IAC one.

\section{Tests with open and synthetic clusters.} \label{clusters}

We have carried out a number of tests regarding the performance of CMDft.Gaia, and the effects on the SFH derivation of various parameters, using composite CMDs of four open clusters observed by {\it Gaia}, and several set of synthetic clusters computed with the BaSTI-IAC stellar evolution library.  In this Appendix we discuss the creation of these mock datasets, and the results of a representative number of tests.

\subsection{Open clusters: member selection, distances, reddening} \label{open_clusters}

We have chosen four open clusters for these tests: one young cluster (Pleiades), one of intermediate-age (IC~4651) and two old clusters (M~67 and NGC~188). The membership selection is performed using the same mixture modelling approach as for globular clusters in \citet{VasilievBaumgardt2021}. Specifically, the distribution of cluster and field stars in the 3d space of parallax and two proper motion components is represented by Gaussians (one for cluster members, two for field stars), which are convolved with individual star's measurement uncertainties when computing the likelihood of membership. Additionally, the spatial distribution of cluster members is assumed to follow a Plummer profile, while field stars are assumed to be uniformly distributed on the sky within the considered region (2--3 times the half-light radius). The parameters of these distributions are optimised to maximise the total likelihood of the entire dataset, after which the membership probability of each star is computed as the ratio of likelihoods of it belonging to the cluster vs.\ the field population. If the cluster's mean proper motion is well separated from the bulk of field stars (as in the case of M~67 and Pleiades), the probability is very close to either 0 or 1, while for NGC~188 and especially IC~4651 the selection is still quite clean for bright stars but becomes less certain beyond $G\approx 18$. Notably, we do not use photometric information in the classification, making it an ideal tool to test the level of agreement with theoretical isochrones.

The CMD of the clusters have been transferred to the absolute magnitude plane using the values of distance and E(B-V) listed in Table~\ref{OC_properties}. The distance for each cluster has been obtained by inverting the average parallax \citep[adopting the parallax zero point of -17 $\mu as$ recommended by][]{Lindegren2021_parallaxBias} of member stars with good photometric and astrometric measurements, by imposing cuts of {\tt parallax\_over\_error > 5}, $G<17$ and the quality parameter cuts adopted by \citet{GriggioBedin2022_M37}. The E(B-V) value was obtained by interpolating in the \citet{lallement2018} 3D reddening map at the position of each cluster. Finally, temperature dependent extinction corrections were derived for each star using the transformations in \citet{casagrandevandenberg2018}.

\begin{figure}
\centering
\includegraphics[width=0.5\textwidth]{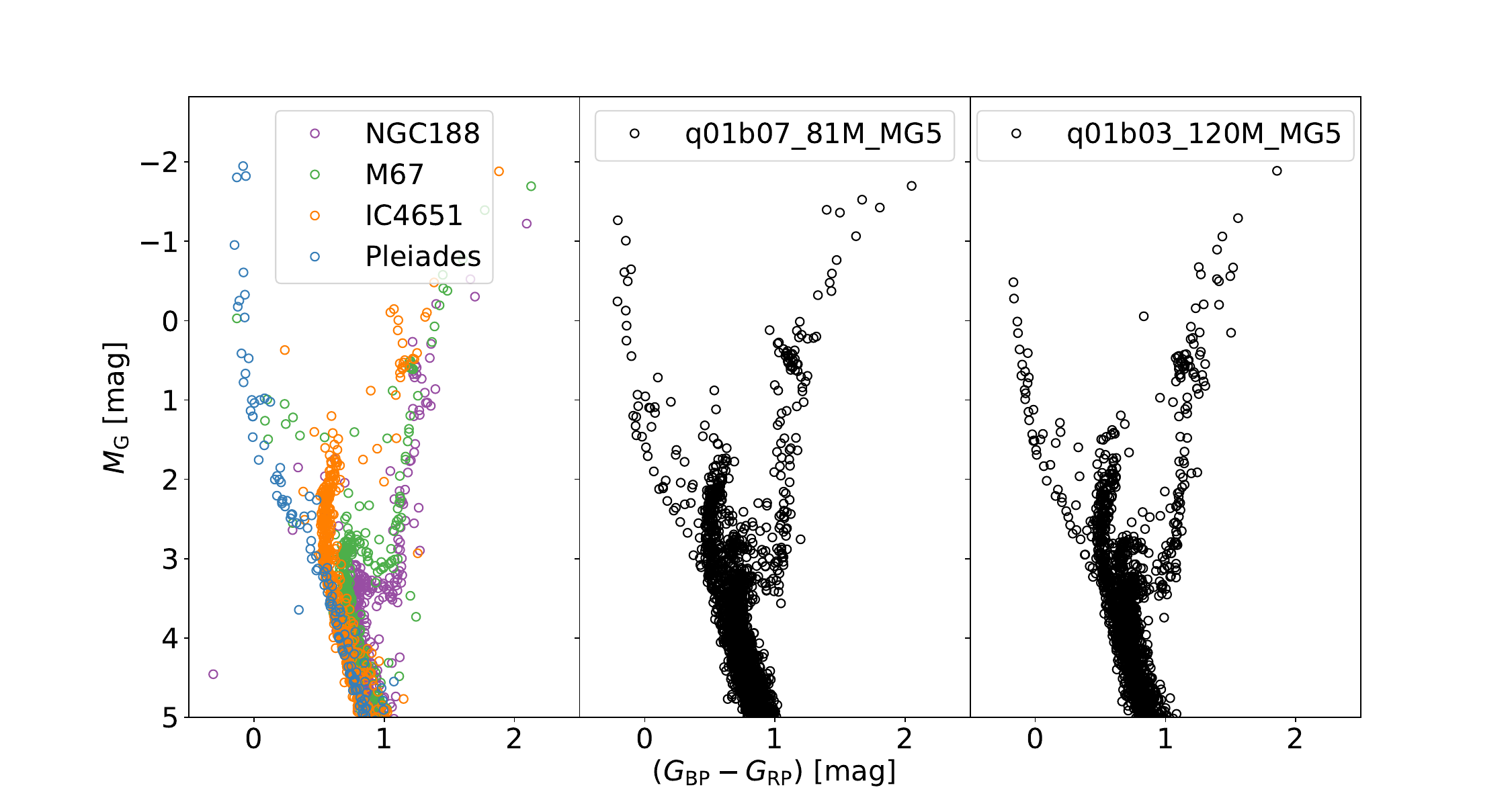}
\caption{CMD of the four open clusters (left panel) and solution CMDs with q01b07\_81M\_MG5 (middle panel) and q01b03\_120M\_MG5 (right panel).}
\label{CMD_open}
\end{figure}

The resulting composite CMD in the absolute plane of the four open clusters is displayed in the left panel of Figure~\ref{CMD_open}. The SFH has been derived with two mother CMDs: q01b03\_120M\_MG5 and q01b07\_81M\_MG5 (see Table~\ref{mothers}) and with the same bundle and error simulation in the mother CMD as for the GCNS. The shift with respect to the mother CMD that provides the best fit to the GCNS CMD has been used in both cases, namely (-0.036, 0.04). SFHs with the four age bins, XL, L, M and S have been computed. The solution CMD for both mother CMDs with the S bins is shown in the middle and right panels of Figure~\ref{CMD_open}. Note how the observed sequences are well reproduced, particularly in the case of the solution with q01b07\_81M\_MG5, as a spurious sequence of stars appears above the subgiant branch corresponding to M67 in the solution with q01b03\_120M\_MG5. As we will see in Section \ref{sfh_clusters}, for this composite population of open clusters,  the solution with the mother CMD computed with a larger binary fractions seems to provide a better solution. A fraction of unresolved binaries of 70\% is broadly consistent with that found by \citet{Malofeeva2023_binaries} for a sample of open clusters.

\begin{table*}
  \caption{Open Cluster parameters}
  
 \centering
\begin{tabular}{lclcc}
  \hline
  \centering
Name & Distance $\pm \sigma$ & E(B-V) & [Fe/H] & Nº stars \\
  & (pc) & & (dex) & (M$_G<4.2$) \\
  \hline\hline
  Pleiades & 135$\pm$2 & 0.043 & -0.01$\pm$0.05 & 82
 \\
  IC4651 & 926$\pm$40 & 0.129 & 0.12$\pm$0.04 & 357
\\
    M67 & 860$\pm$36 & 0.030 & 0.03$\pm$0.05 & 349
\\
    NGC188 & 1853$\pm$108 & 0.079 & 0.11$\pm$0.04 & 345
 \\
  \hline
\end{tabular}
\tablefoot{Adopted open cluster distance, reddening and spectroscopic metallicity \citep[from the HQS sample of][]{Netopil2016_OCmet}. See text for details.}
  \label{OC_properties}
\end{table*}

\subsection{Synthetic clusters}\label{synthetic_clusters}

We have created two composite populations of synthetic clusters computed with ChronoSynth adopting 30\% of binaries with q$_{min}$=0.1, and a Kroupa IMF. The first of them contains seven synthetic clusters with ages 0.2, 2, 4, 6, 8, 10 and 12 Gyr, each with a small age range (20 Myr) and with metallicity between [M/H]=-0.1 and -0.05 dex. Three CMDs with different number of stars with M$_G<4.2$ have been extracted from this composite population, so that the individual synthetic clusters have: $\simeq$ 350 stars per cluster (similar to the open clusters described in \ref{open_clusters}), $\simeq$2000 stars, such that the total number of stars in the seven clusters is similar to that in the GCNS CMD, and $\simeq$ 14000 stars, in order to check whether a much larger number of stars in the input data leads to a substantially better recovery of the input population parameters. The second composite population includes the former one and seven additional clusters with the same ages, and higher metallicities between [M/H]=0.1 and 0.15. This population contains a similar number of stars as in the GCNS (some 1000 stars per cluster) and is aimed at exploring the ability of $Dir$SFH to discriminate populations of different metallicity separated by a very small metallicity gap, in this case 0.15 dex. Photometric and distance errors typical of the GCNS have been simulated. 

\begin{figure}[h!]
    \centering
    \includegraphics[width=0.45\textwidth]{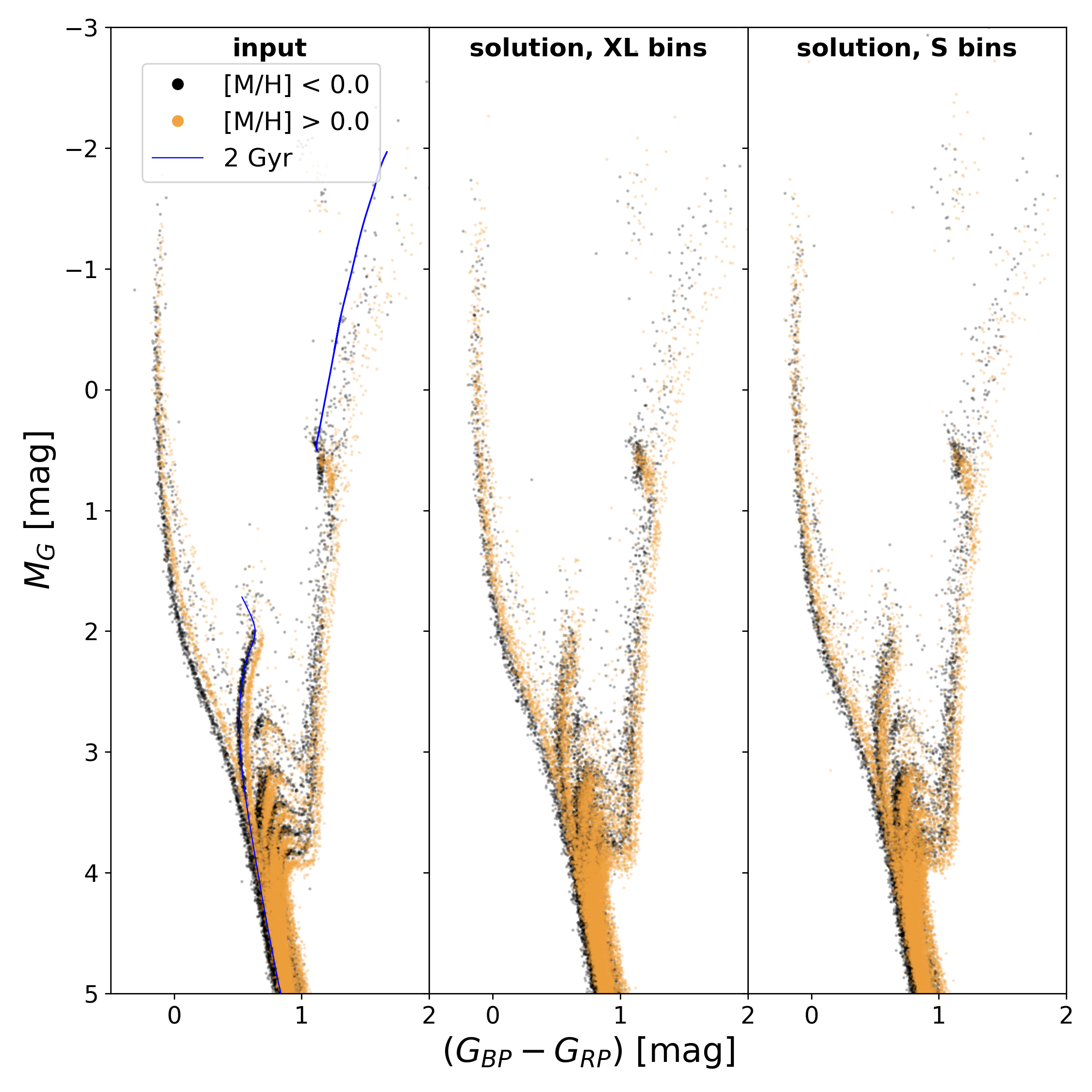} 
    \caption{Left panel: composite CMD of the set of fourteen synthetic clusters with two metallicity sequences separated by 0.15 dex (metallicities between [M/H]=-0.1 and -0.05 dex and between [M/H]=0.1 and 0.15 dex. An isochrone of age 2 Gyr has been superimposed as reference, to show how the binary stars depart from the locus of single stars. The synthetic stars in the low and high metallicity sequences are displayed in black and orange, respectively. Middle and right panels: solution CMDs derived with the XL and S bins, respectively. Note the more scattered sequences in the solution with the XL bins.} 
    \label{CMD_synthetic}
\end{figure}

The left panel of Figure \ref{CMD_synthetic} displays the CMD of the composite population containing the two metallicity sequences. The sample with the intermediate number of stars, similar to that of the GCNS is displayed. This CMD shows how the two metallicities at each age are slightly separated, with the more metal rich population redder in the main sequence and in the RGB, and fainter at the end of the subgiant branch and in the red clump. The binary stars are clearly seen to the red of the younger main sequences, and broadening the lower composite main sequence. The stars scattered around the subgiant branches are also binary stars. Note how the binaries modify the main sequence turnoff of the intermediate-age populations: for example, in the 2 Gyr old sequence, the typical shape of the overall contraction region (marking the transition from the main sequence to the subgiant branch stage) gets blurred by binary stars bluer and/or brighter than the single star sequence (an isochrone of the lower metallicity and 2 Gyr has been superimposed as a reference). The middle and right panels of Figure~\ref{CMD_synthetic} display the solution CMD with bins XL and S, respectively. In both cases all the sequences are more blurred than in the original input CMD, even though they are still distinguishable even in the case of the larger XL age bins. Note the subtle effect of the bin size in the resulting solution CMDs.

\subsection{SFH of open and synthetic clusters} \label{sfh_clusters}

\begin{figure*}[h!]
    \centering
    \includegraphics[width=0.45\textwidth]{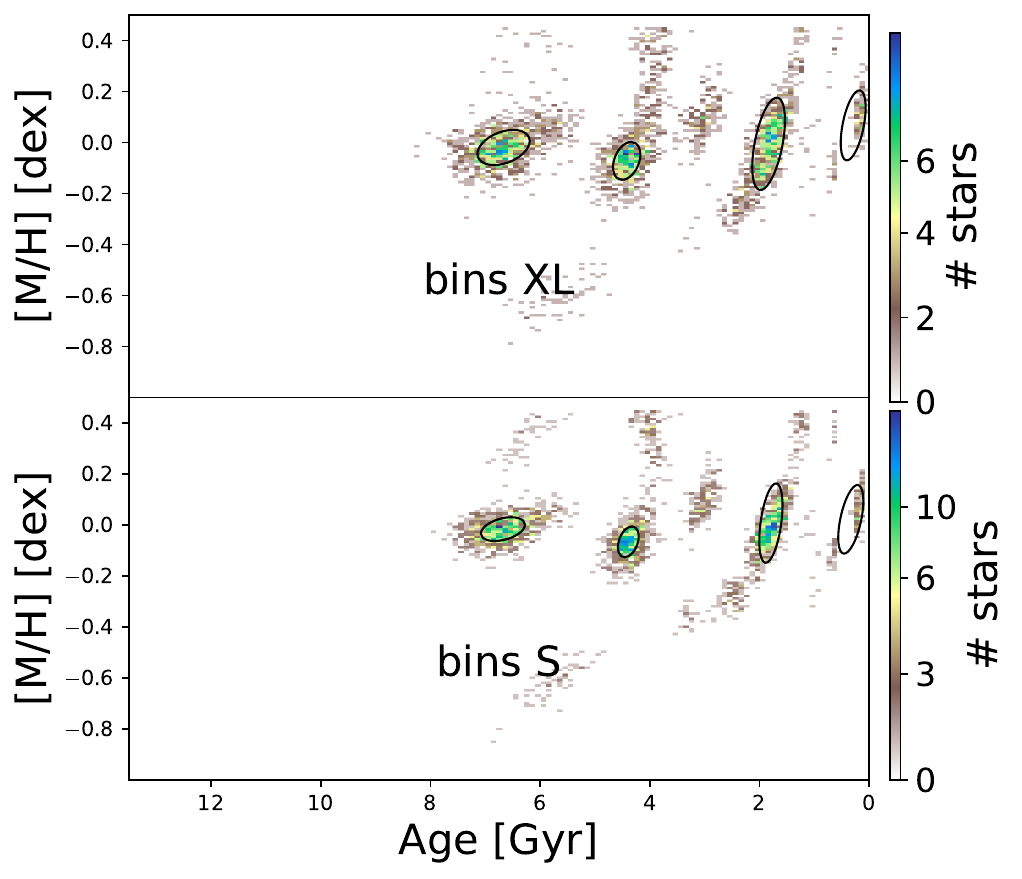} 
    \includegraphics[width=0.45\textwidth]{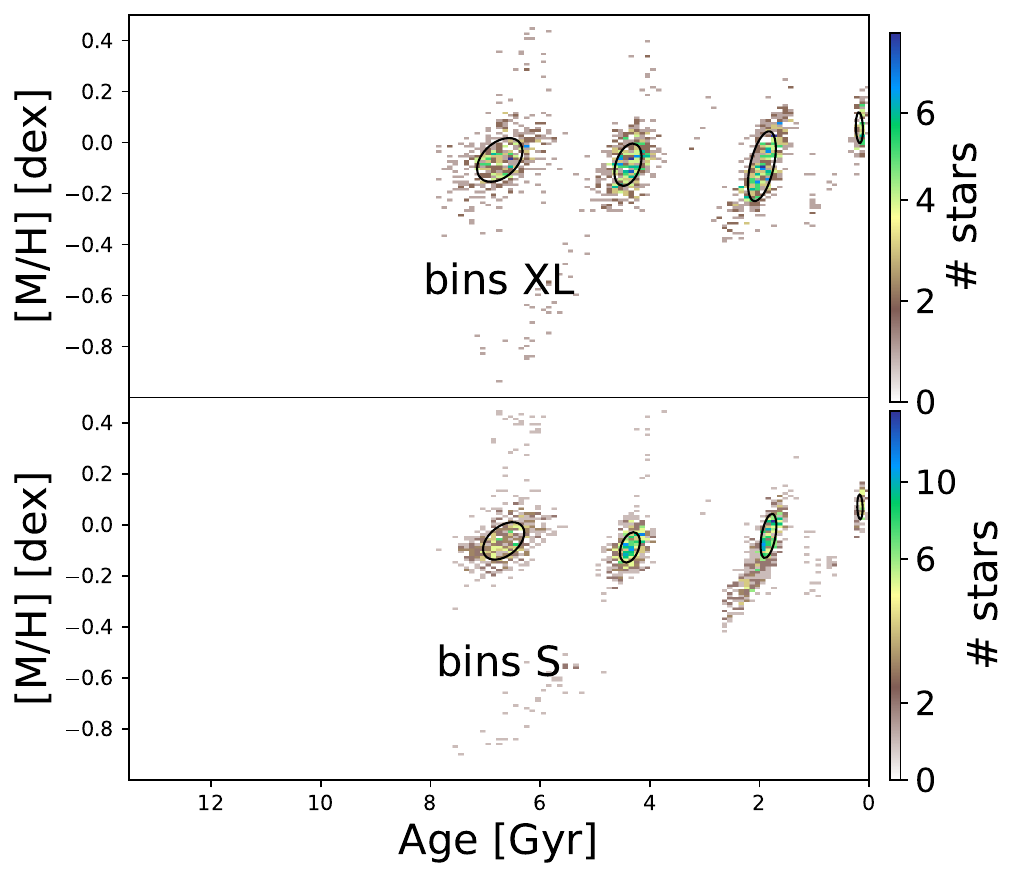}
    \caption{Age-metallicity distribution obtained from the composite CMD of the open clusters. Left panels: solution from q01b03\_120M\_MG5. Right panels: solution with q01b07\_81M\_MG5. The results for the XL bins (upper panels) and S bins (lower panels) are shown. Best fit ellipses with 1-$sigma$ axes are overlaid (eight ellipses in total have been fit, and only those with weight larger than 5\% are displayed).}
    \label{OC_AMR}
\end{figure*}

\begin{figure*}[h!]
    \centering
    \includegraphics[width=0.45\textwidth]{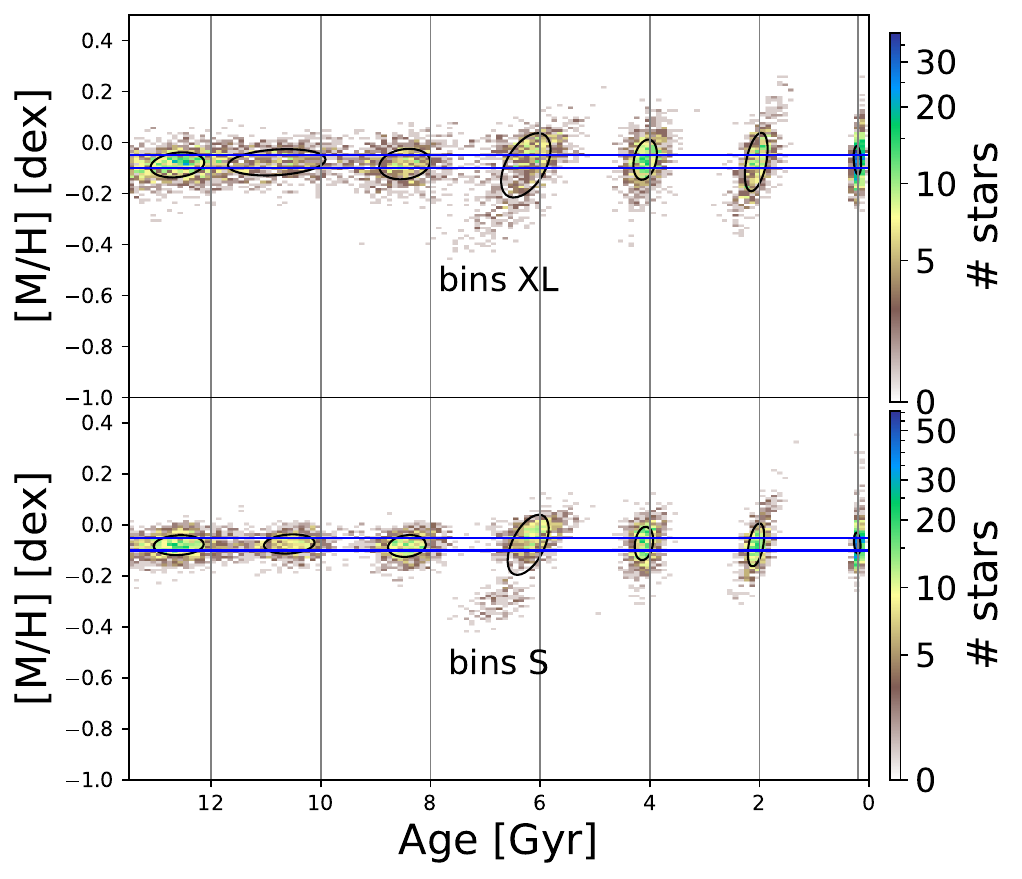} 
    \includegraphics[width=0.45\textwidth]{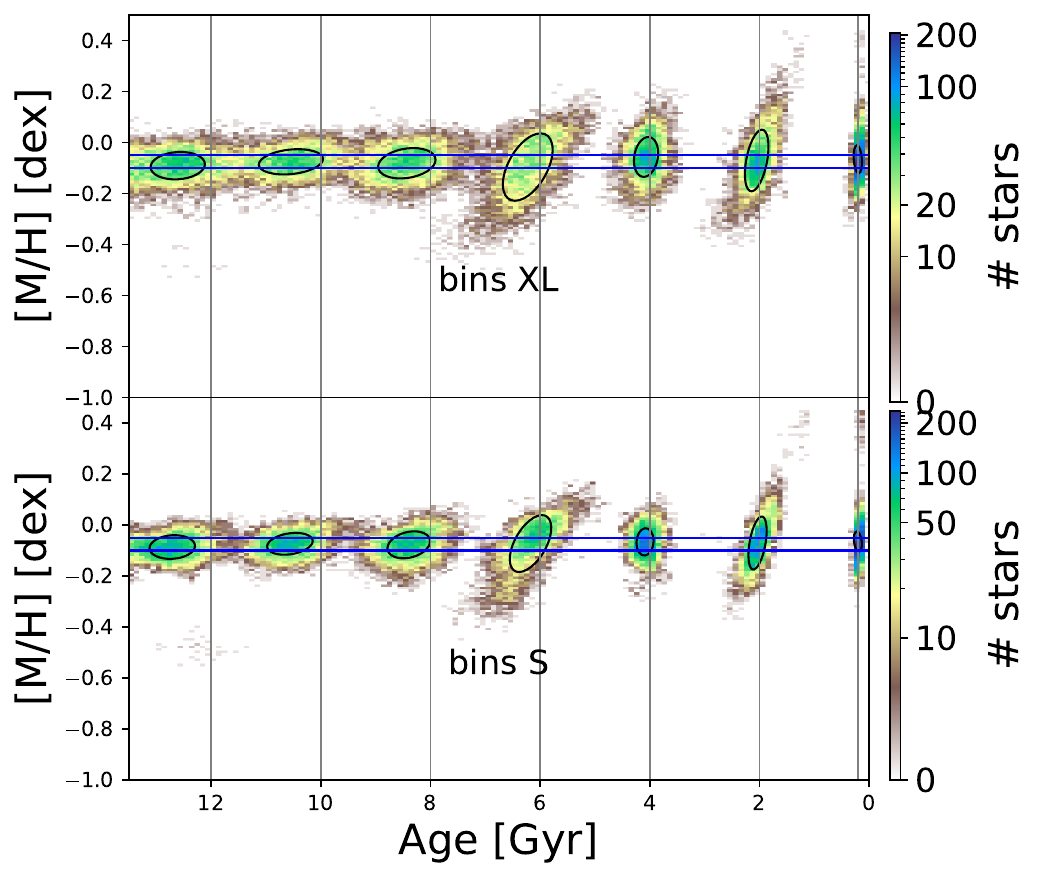}
    \caption{Age-metallicity distribution obtained from the composite CMD of synthetic clusters (sets with single metallicity sequence).  Left panels: result from the mock with the small number of stars . Right panels: results from the mock with intermediate number of stars. The results for the XL bins (upper panels) and S bins (lower panels) are shown.  Two horizontal lines mark the metallicity range of the input population, and vertical lines indicate the input ages. Best fit ellipses with 1-$sigma$ axes are overlaid. } 
    \label{SC_AMR}
\end{figure*}

Figure \ref{OC_AMR} shows two solutions (in the age-metallicity plane) for the composite CMD of the open clusters. The left panel displays the solution obtained with the q01b03\_120M\_MG5 mother CMD while the right panel corresponds to the solution with q01b07\_81M\_MG5. The results for the XL bins (upper panels) and S bins (lower panels) are shown. Note that the four populations have been more cleanly recovered with the q01b07\_81M\_MG5 mother.

Figure \ref{SC_AMR} displays the solution in the same plane for the composite CMD of synthetic clusters (sets with single metallicity sequence). The result from the mock with the small number of stars is shown in the left panel, while the one for the mock with intermediate number of stars is displayed in the right plane. The results for the XL bins (upper panels) and S bins (lower panels) are shown. 

In all cases, best fit ellipses with 1-$\sigma$ axes are overlaid. In the case of the synthetic clusters, two horizontal lines marking the metallicity range of the input population, and vertical lines indicating the input age are included in the plot. The comparison of the recovered age-metallicity relation with the reference lines allows us to visually evaluate the precision and accuracy of the results. It is remarkable that, even with only $\simeq$ 350 stars per cluster (left panel), $Dir$SFH is able to separate the 7 clusters in age, even though the separation becomes marginal for the 8-12 Gyr old clusters, specially with the XL bins. 
\begin{figure}[h!]
    \centering
    \includegraphics[width=0.4\textwidth]{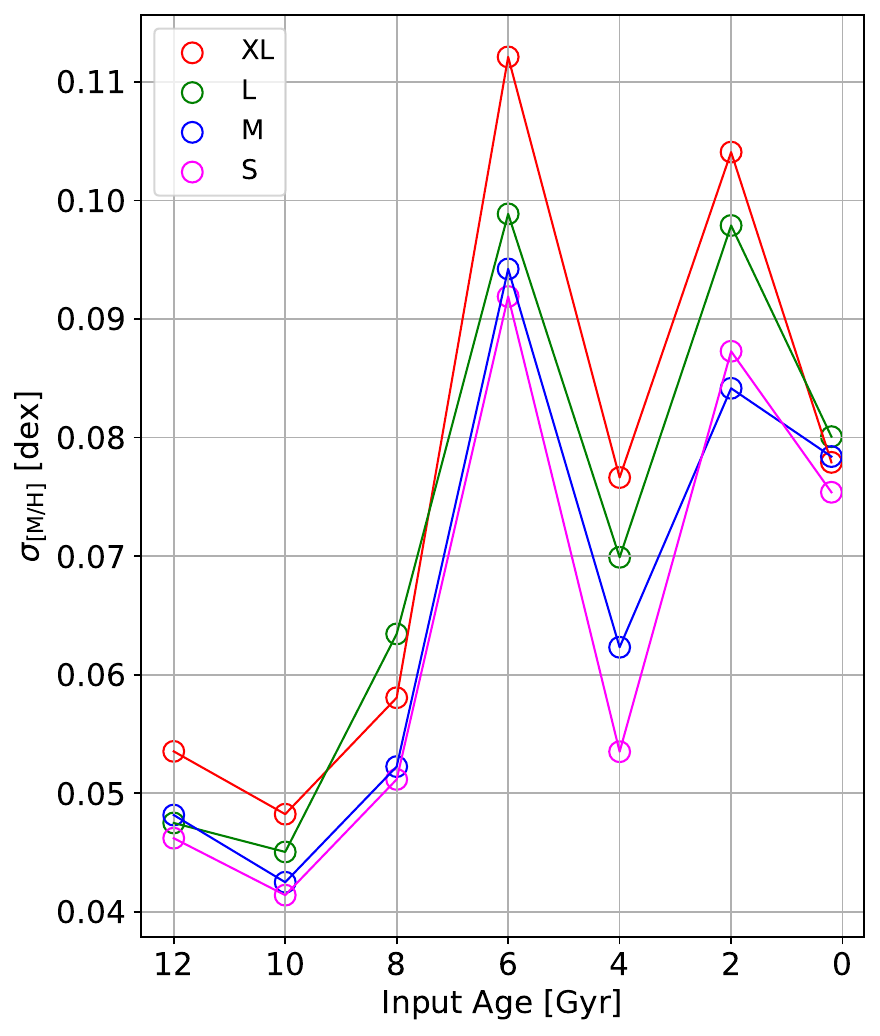} 
    \caption{Analysis of the precision of the metallicity recovery. The $\sigma_{met}$ is show as a function of the age. While the metallicity bins are kept constant (0.1 dex), the effective size of the SSPs is also affected by the size of the age bins. Results for different age bins are shown in different colours as indicated in the labels} 
    \label{metallicity_precision}
\end{figure}

Visual inspection of Figure \ref{SC_AMR} already discloses that the metallicity value is accurately recovered at all ages with a precision that depends on the age, with the metallicity of the three older clusters recovered more precisely than that of the four younger clusters (see Figure \ref{metallicity_precision}). Additionally, in the clusters younger than 7 Gyr (and specially in the 2 and 6 Gyr old clusters), the angle of the ellipse indicates the occurrence of age-metallicity degeneracy, more prominent for these ages than for other age intervals. This is possibly due to the characteristic distribution of stars in the CMD near the turnoff in this age range, with the typical ‘hook’ shape that is reproduced by the models when overshooting is accounted for (see Figure~\ref{CMD_synthetic}), making the degeneracy more difficult to break. This occurrence is also clearly seen in the age-metallicity relation of the open clusters in Figure D.3, in which a larger scatter of a small number of stars toward lower and higher metallicity (and older and younger age, respectively) can be appreciated.

\begin{figure*}[h!]
    \centering
    \includegraphics[width=0.49\textwidth]{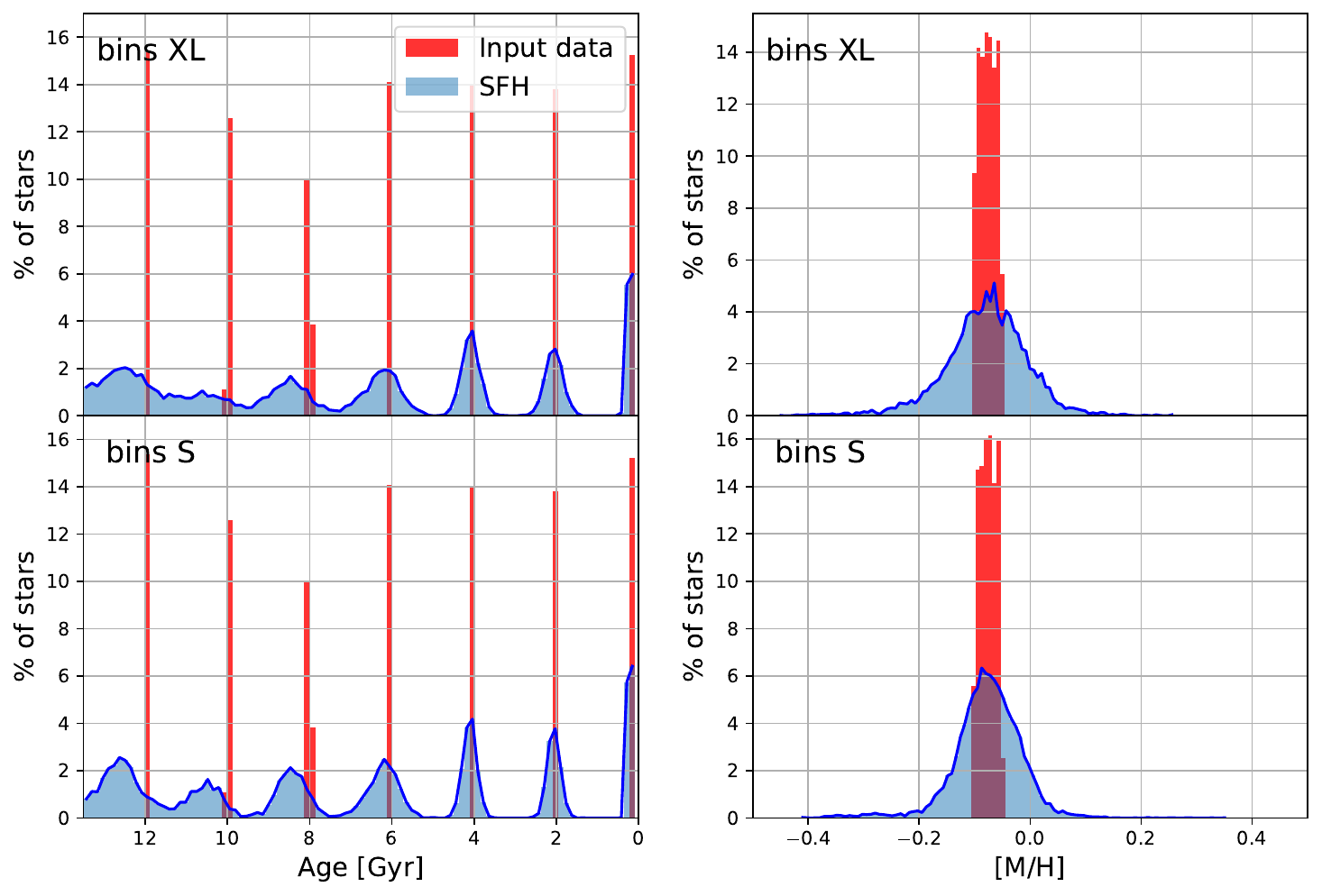} 
    \includegraphics[width=0.49\textwidth]{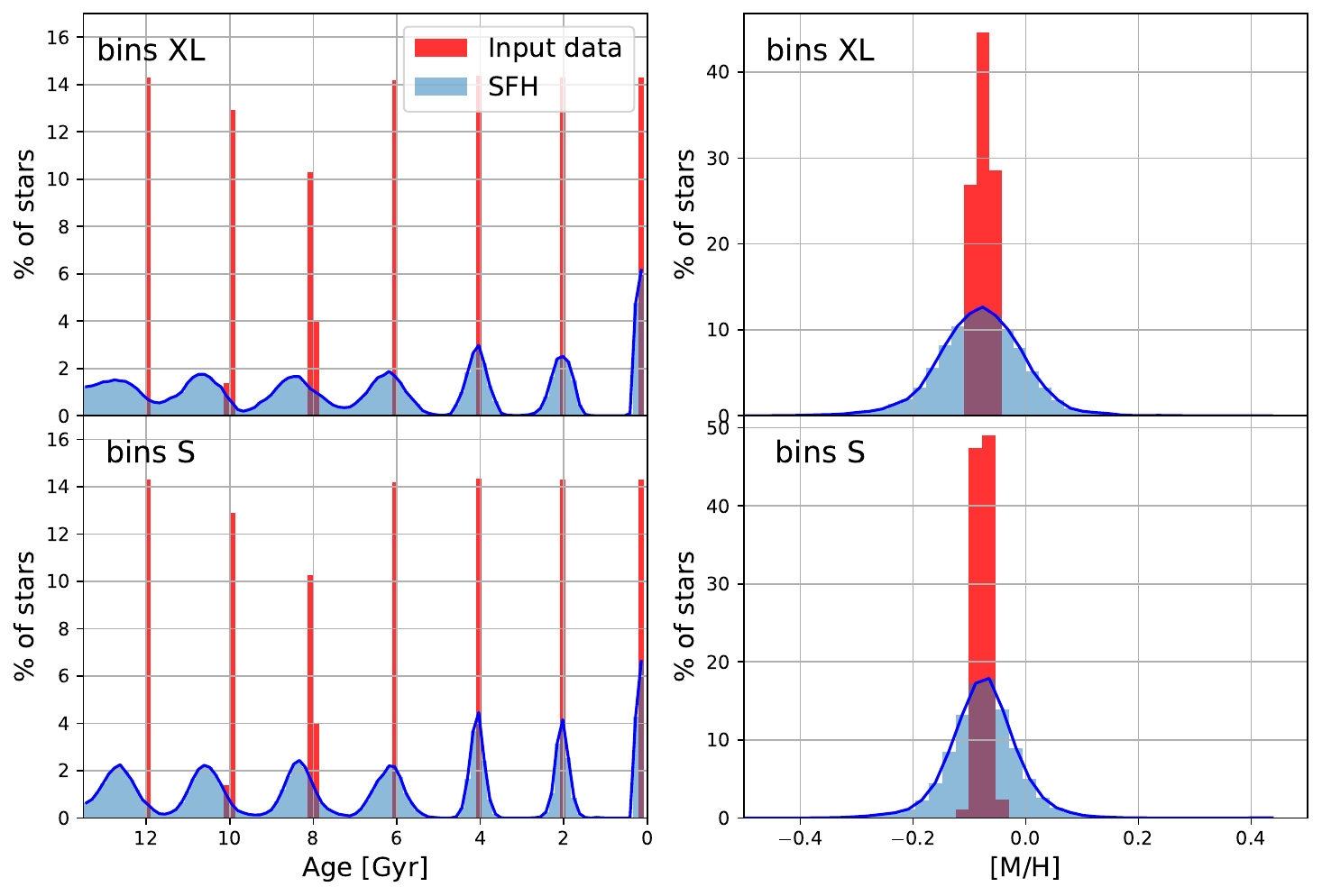}
    \caption{Normalised histograms of percentage of stars as a function of age and metallicity for the synthetic clusters. Left panel: set with $\simeq$ 350 stars per cluster. Right panel: set with $\simeq$ 14000 stars per cluster. The upper and lower panels display the results with the XL and S age bins, respectively.} 
    \label{SC_AMR_projections}
\end{figure*}

The fact that the populations are recovered systematically older (see Figure~\ref{precision_accuracy}) is also evident in Figure \ref{SC_AMR} for clusters older than 6 Gyr, from the mismatch between the vertical lines indicating the input ages of the synthetic clusters, and the centre of the ellipses.  Figure \ref{SC_AMR_projections} presents the same information in the form of normalised histograms of percentage of stars as a function of age and metallicity for an easier visualisation. The broadening of the age and metallicity sequences (smaller for the S bins) and the systematic shifts in the recovered age, while the metallicity is recovered with no systematic shifts can be clearly appreciated. It is also worth remarking that $Dir$SFH is very successful in delimiting the metallicity of the populations: even though the mother CMD used to calculate the SFH contains stars in the whole metallicity range, the stars in the solution are well confined in a narrow metallicity range.

\begin{figure}[h!]
    \centering
    \includegraphics[width=0.45\textwidth]{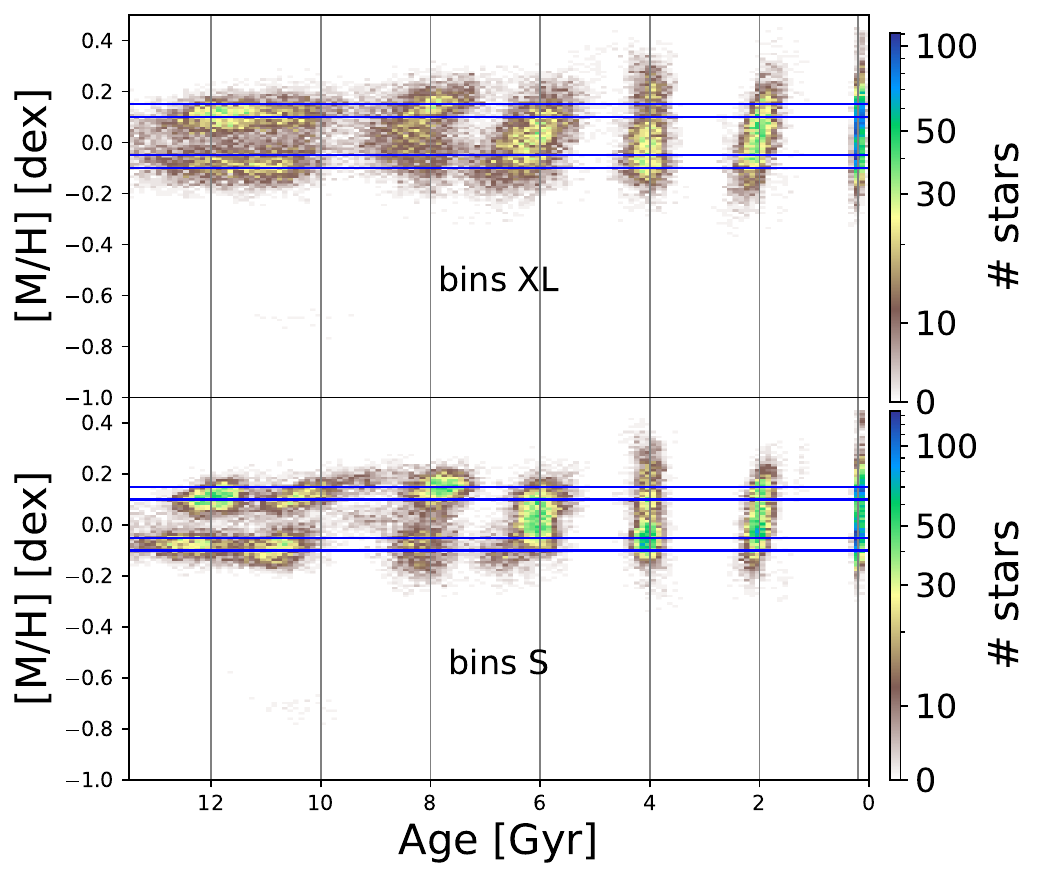} 
    \caption{Age-metallicity distribution for the composite synthetic cluster CMD with two metallicity sequences. Results with the XL and S age bins are shown in the upper and lower panels, respectively} 
    \label{double_metallicity}
\end{figure}

Figure \ref{double_metallicity} shows the age-metallicity distribution retrieved for the composite synthetic cluster CMD with two metallicity sequences. Note that both the XL and S age bins are able to discriminate the double metallicity sequence for ages equal or older than 8 Gyr, even though, in the case of the XL bins the different populations are much less defined than in the case of the S bins. For younger ages the S bins are necessary for a marginal metallicity separation of the different clusters. This result however indicates that our CMD-fitting methodology should be able to discriminate small metallicity differences in populations of the same age, specially at old ages.

\section{Robustness of the solutions}
\label{testing_robustness}

In this Appendix we display a number of age-metallicity distributions derived varying some parameters of the fit, or those used to compute the mother CMDs.

\begin{figure}[h!]
    \centering
    \includegraphics[width=0.5\textwidth]{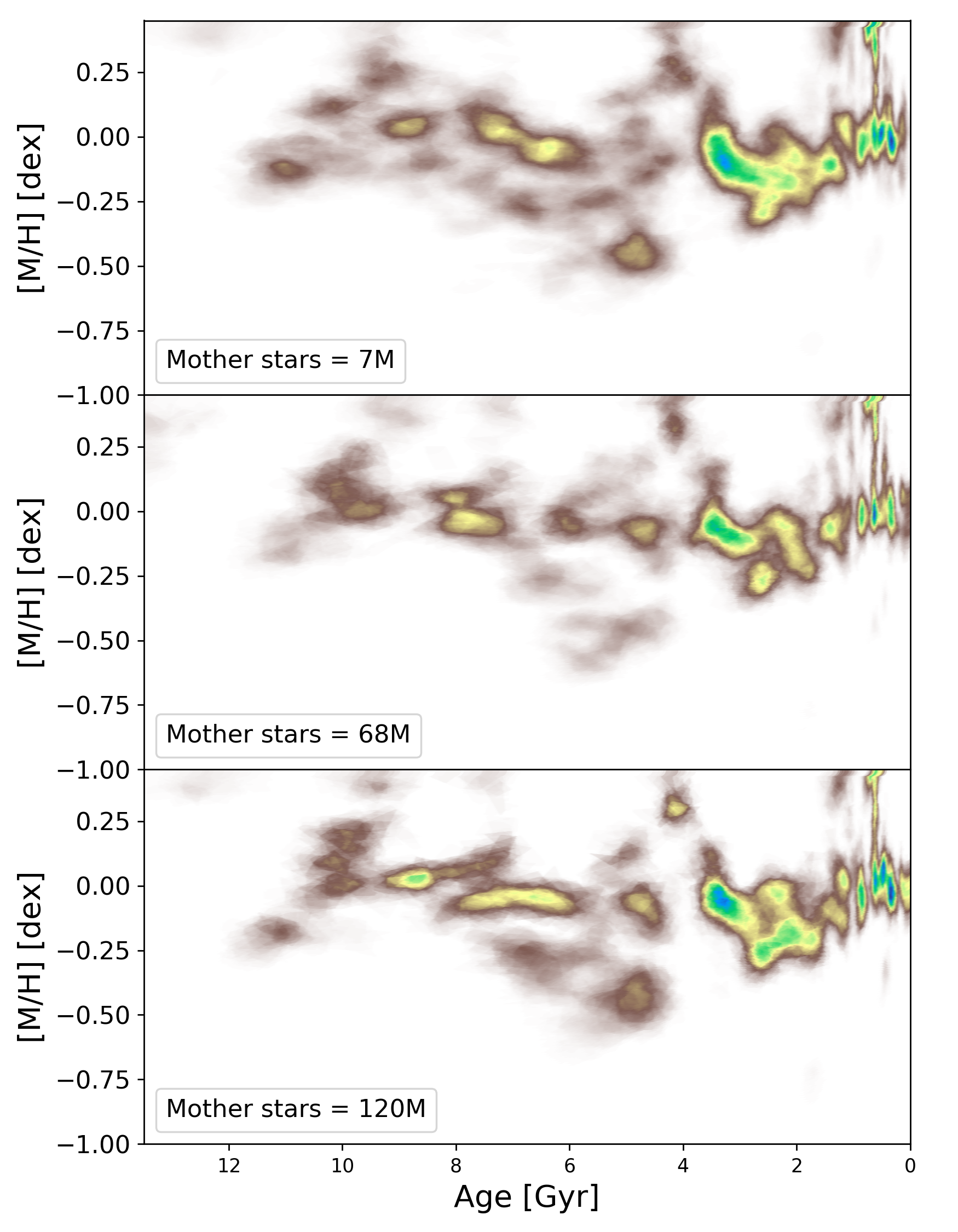} 
    \caption{Age-metallicity relation of the mass transformed into stars as a function of time and [M/H] derived from three mother CMDs with different number of stars down to M$_G=5$, calculated from the BaSTI stellar evolution library adopting q$_{min}$=0.1, $\beta$=0.3 and a Kroupa IMF. They are the mother CMDs named q01b03\_60M\_deep, q01b03\_112M\_MG6 and q01b03\_120M\_MG5 in Table~\ref{mothers}.} 
    \label{amr_mother_nstars}
\end{figure}

Figure~\ref{amr_mother_nstars}  displays age-metallicity distributions derived from three mother CMDs computed with identical parameters (indicated in the caption) except for the number of stars, as indicated in the figure labels. Note the similarity of the features in all SFHs. The 7M solution, however, seems to be more noisy that the other two, indicating that a large number of stars in the mother CMD is important in order to avoid spurious features in the SFH.

\begin{figure}[h!]
    \centering
    \includegraphics[width=0.5\textwidth]{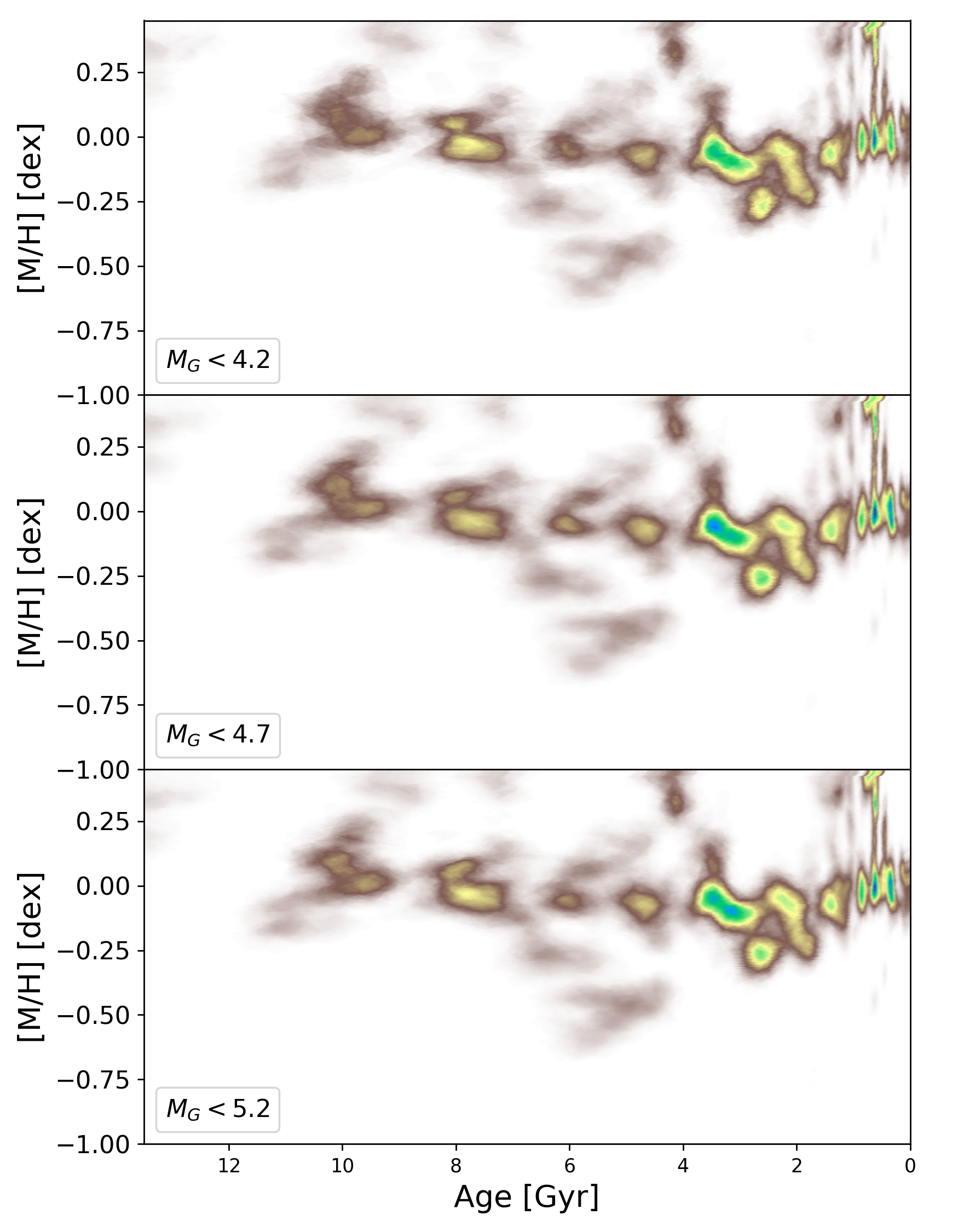} 
    \caption{Age-metallicity relation of the mass transformed into stars as a function of time and [M/H] derived from the same mother CMD (q01b03\_112M\_MG6) but three different faint magnitude limit of the bundle, as indicated in the labels. } 
    \label{amr_bundle}
\end{figure}

Figure \ref{amr_bundle} displays three age-metallicity distributions derived from the same mother CMD (see caption), for different faint magnitude limit of the bundle (see labels). The features in the SFH are basically identical in all three cases, indicating that there is no clear gain of age and metallicity information when including in the CMD fit main sequence stars with magnitude fainter than those in the subgiant branch of the oldest and more metal rich population present in the mother CMD (in our case M$_G<4.2$).

\begin{figure}[h!]
    \centering
    \includegraphics[width=0.5\textwidth]{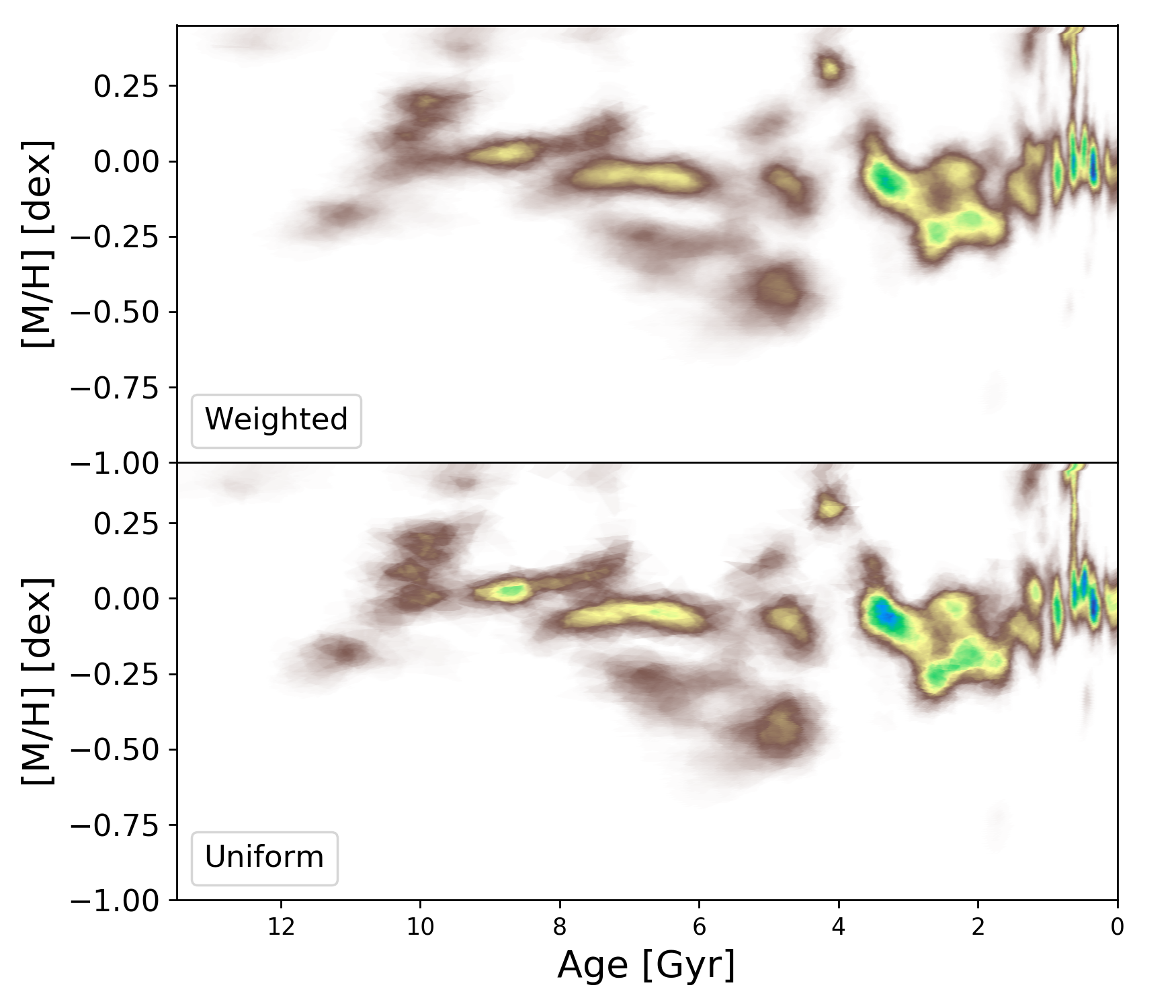} 
    \caption{Age-metallicity distribution of the mass transformed into stars as a function of time and [M/H] derived from the same mother CMD (q01b03\_120M\_MG5) but adopting different weights across the CMD for the fit: weights calculated as the inverse of the variance of the stellar ages across the CMD, or uniform (no) weights.} 
    \label{amr_wei_uni}
\end{figure}

Figure \ref{amr_wei_uni} displays two age-metallicity distributions derived from the same mother CMD applying or not weights across the CMD for the fit (see caption for details). The differences between the two SFHs are minimal.

\begin{figure}[h!]
    \centering
    \includegraphics[width=0.5\textwidth]{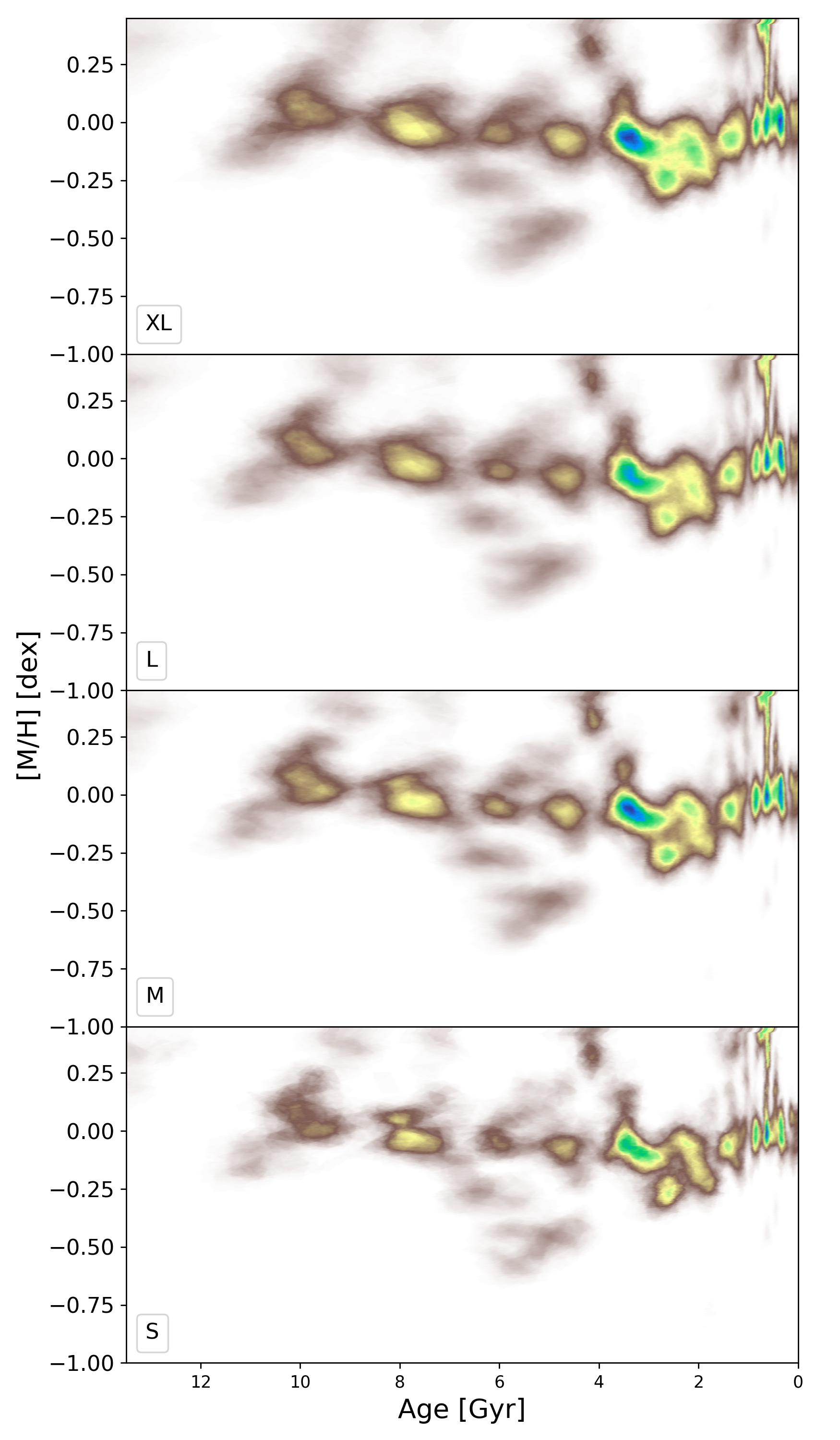} 
    \caption{Age-metallicity distribution of the mass transformed into stars as a function of time and [M/H] derived from the same mother CMD (q01b03\_112M\_MG6) but the four different age bin sizes (XL, L, M and S) tested in this work. In all case, the metallicity bin size is 0.1 dex.} 
    \label{amr_tam_bins}
\end{figure}

Figure \ref{amr_tam_bins} displays four age-metallicity distributions derived from the same mother CMD (q01b03\_112M\_MG6), with the four different age bins sizes  (XL, L, M, S) tested in this paper. Note how the features in the age-metallicity distributions sharpen as the age bins used in the fit are smaller. The main features, however, are the same in all four. In particular, the split in the metallicity of the two oldest events of star formation is already hinted in the age-metallicity distributions derived with the XL bins, but is more clearly visible in the age-metallicity distributions derived with the S bins.

\begin{figure}[h!]
    \centering
    \includegraphics[width=0.5\textwidth]{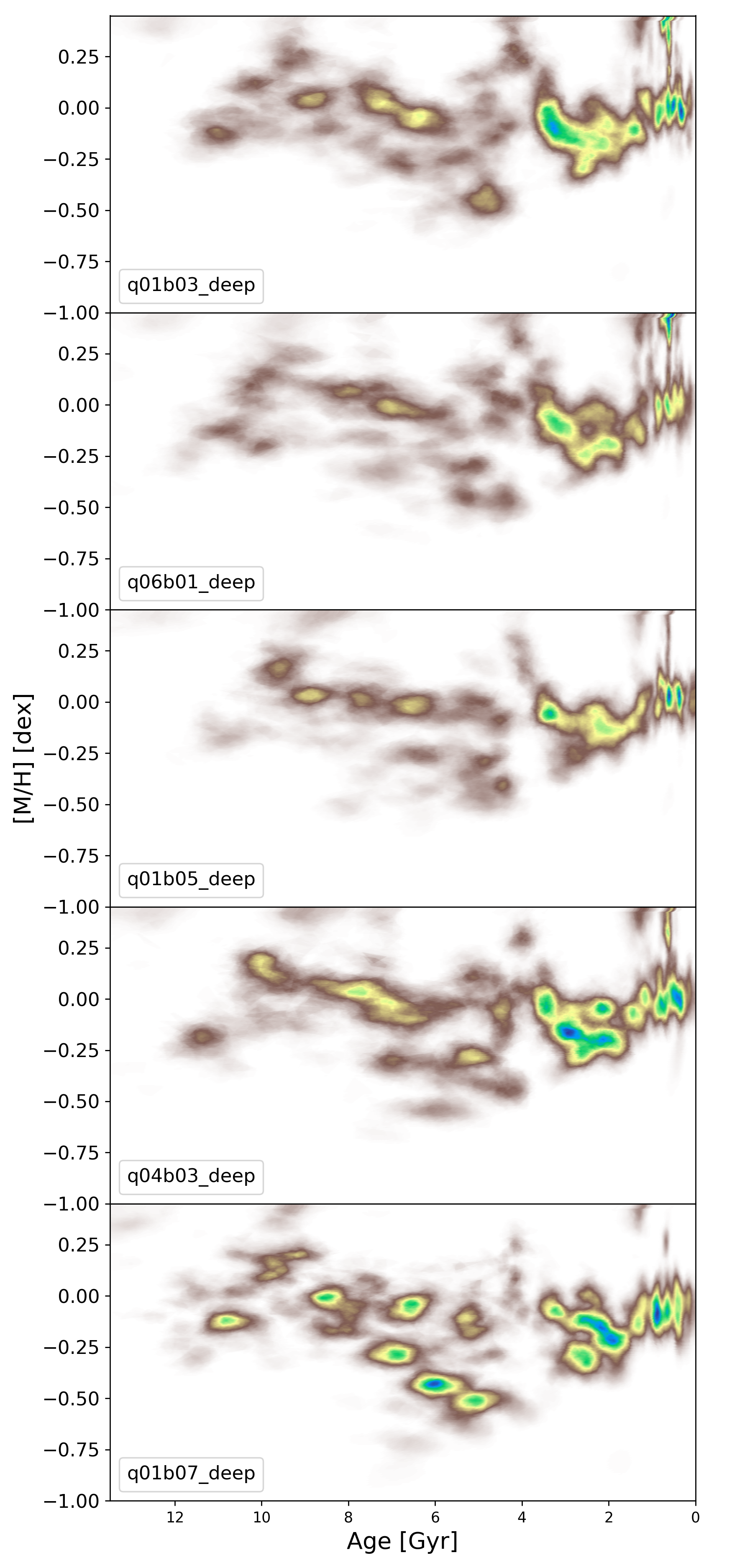} 
    \caption{Age-metallicity distribution of the mass transformed into stars as a function of time and [M/H] derived from five mother CMDs computing adopting five different parametrisations of the binary star population, as indicated in the labels (see Table \ref{mothers} for details).} 
    \label{amr_binarias}
\end{figure}

Figure \ref{amr_binarias} displays five age-metallicity distributions derived from the five deep mother CMDs computed with different binary parametrisations, but otherwise identical parameters of the mother CMDs and the SFH fit. In all cases, the mother CMDs have a small number of stars in the area of the CMD used for the fit  ($\simeq$ 2M-7M) which, as seen in Figure \ref{amr_mother_nstars} can lead to more noisy solutions that in the case of mother CMDs with a larger number of stars available for the fit. Note that, except for q01b07\_deep (the mother CMD that results in the highest discrepancy in the distribution of stars in the binary sequence compared to the single stars main sequence, see Figure~\ref{deltaG_all}), the resulting age-metallicity distributions have very similar features. The presence of a more conspicuous metal poor populations at intermediate age in the case of q01b07\_deep (short for q01b07\_81M\_MG5; similar abbreviations have been used in the other panels) can be expected from the fact that an excessive number of binaries produces a red population in the CMD that is compensated in the solution by introducing a larger amount of metal poor stars, which are bluer, in the solution.

\begin{figure}[h!]
    \centering
    \includegraphics[width=0.5\textwidth]{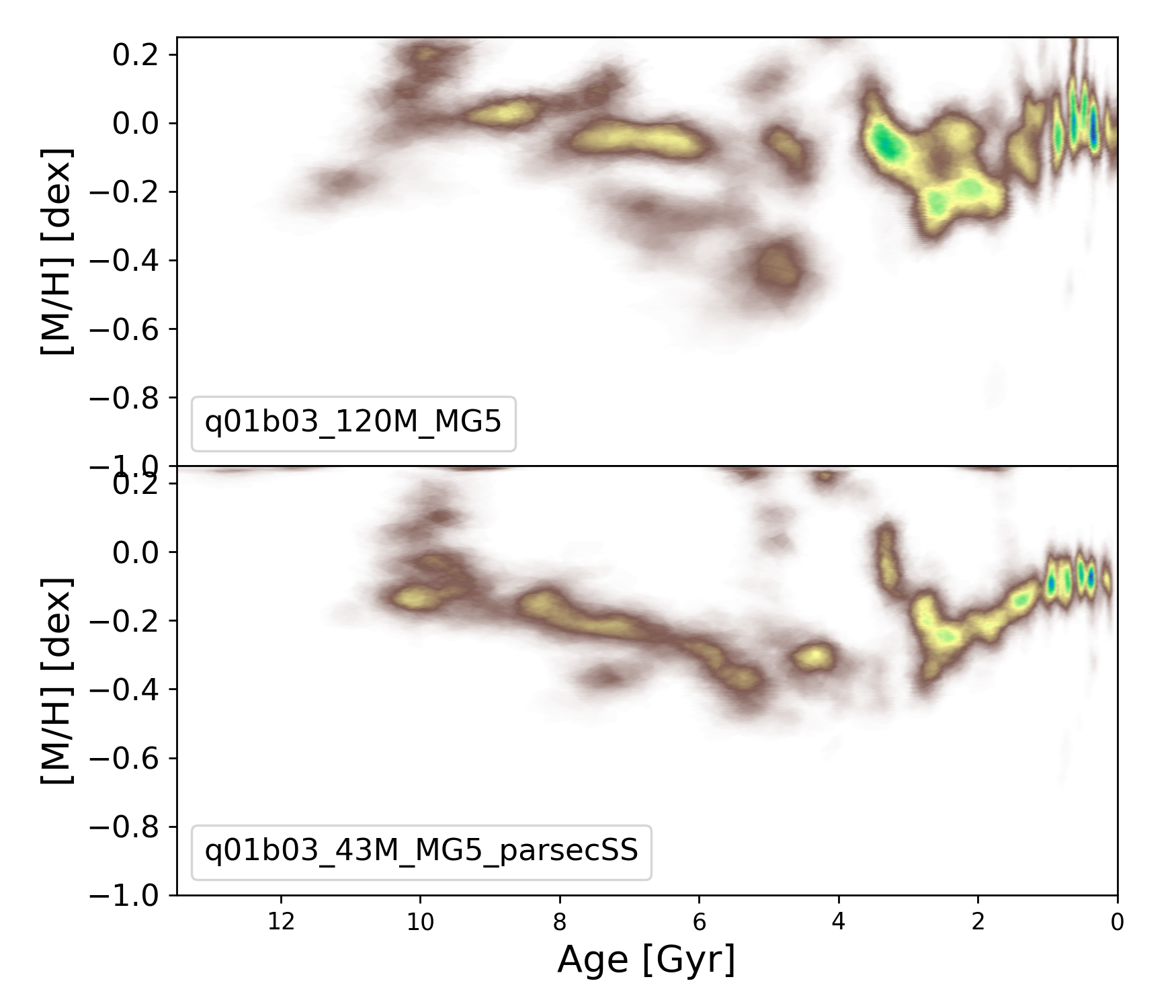} 
    \caption{Age-metallicity distribution of the mass transformed into stars as a function of time and [M/H] derived from the BaSTI mother CMD adopted for our final solution, q01b03\_112M\_MG5, and a PARSEC mother CMD with the same binary stars parametrisation (q01b03\_43Mparsec\_MG5), as indicated in the labels (see Table \ref{mothers} for details).} 
    \label{amr_binarias1}
\end{figure}

\end{appendix}

\end{document}